\documentclass[amsmath, amssymb, preprintnumbers, showpacs, showkeys, aps,prb,superscriptaddress,twocolumn]{revtex4-1}

\pdfoutput=1 
\usepackage{amssymb}
\usepackage{comment}
\usepackage[utf8]{inputenc}
\usepackage{graphicx}
\usepackage{amsmath}
\usepackage{csquotes}
\usepackage{xcolor}
\usepackage{placeins}
\usepackage[pdftex,colorlinks, linkcolor={red!50!black},
    citecolor={blue!50!black},
    urlcolor={blue!80!black}]{hyperref}

\usepackage[globalcitecopy]{bibunits}
\defaultbibliography{bibliography}
\defaultbibliographystyle{naturemag}

\def\bs#1{\boldsymbol{#1}}	    

\DeclareMathAlphabet{\mathbbold}{U}{bbold}{m}{n}

\def\Tr{\mathrm{Tr}}

\begin{document}
\begin{bibunit}

\title{Exceptional Topological Insulators}

\author{M.~Michael~Denner}
\email{michael.denner@physik.uzh.ch}
\affiliation{Department of Physics, University of Zurich, Winterthurerstrasse 190, 8057 Zurich, Switzerland}
\author{Anastasiia~Skurativska}
\affiliation{Department of Physics, University of Zurich, Winterthurerstrasse 190, 8057 Zurich, Switzerland}
\author{Frank~Schindler}
\affiliation{Department of Physics, University of Zurich, Winterthurerstrasse 190, 8057 Zurich, Switzerland}
\affiliation{Princeton Center for Theoretical Science, Princeton University, Princeton, NJ 08544, USA}
\author{Mark~H.~Fischer}
\affiliation{Department of Physics, University of Zurich, Winterthurerstrasse 190, 8057 Zurich, Switzerland}
\author{Ronny Thomale}
\affiliation{Institut für Theoretische Physik und Astrophysik, Universität Würzburg, 97074 Würzburg, Germany}
\author{Tom\'{a}\v{s} Bzdu\v{s}ek}
\affiliation{Condensed Matter Theory Group, Paul Scherrer Institute, 5232 Villigen PSI, Switzerland}
\affiliation{Department of Physics, University of Zurich, Winterthurerstrasse 190, 8057 Zurich, Switzerland}
\author{Titus Neupert}
\affiliation{Department of Physics, University of Zurich, Winterthurerstrasse 190, 8057 Zurich, Switzerland}
\date{\today}
\begin{abstract}
We introduce the exceptional topological insulator (ETI), a non-Hermitian topological state of matter that features exotic non-Hermitian surface states which can only exist within the three-dimensional topological bulk embedding. We show how this phase can evolve from a Weyl semimetal or Hermitian three-dimensional topological insulator close to criticality when quasiparticles acquire a finite lifetime. The ETI does not require any symmetry to be stabilized. It is characterized by a bulk energy point gap, and exhibits robust surface states that cover the bulk gap as a single sheet of complex eigenvalues or with a single exceptional point. The ETI can be induced universally in gapless solid-state systems, thereby setting a paradigm for non-Hermitian topological matter.

\end{abstract}

\maketitle
{\section{Introduction}}
Since their theoretical conception~\cite{FuKaneMele:2007,MooreBalents:2007} and experimental discovery~\cite{ChenBi2Te3:2009,HasanBiSe:2009}, three-dimensional topological insulators (3D TIs) have become the focal point for research on topological quantum matter. Their key feature are conducting surface states resembling a single species of gapless Dirac electrons, which are protected against surface perturbation as long as time-reversal and charge-conservation symmetry are preserved~\cite{Hasan:2010}. Transcending the realm of quantum matter, the TI phase has since been realized in many different settings including meta-materials, such as photonic and phononic crystals~\cite{Parappurath:2020,Yang:2019,Malzard:2015,Regensburger:2012}.

Most of such meta-material platforms are accidentally or tunably lossy, such that their effective Hamiltonian description involves non-Hermitian terms due to the lack of energy conservation~\cite{Ganainy:2018}. The same holds for interacting electronic quantum systems in which quasiparticles attain a finite lifetime, as manifested in a complex self-energy~\cite{Kozii:2017,Yoshida:2018,Shen:2018,Aguado:2019}. Starting from the initial classification of topological matter based on Hermitian Hamiltonians, the study of systems with non-negligible loss and gain calls for an extension to non-Hermitian topological matter. At this early stage of the field, several principles have been uncovered: (i) non-Hermitian systems have stable band degeneracies in two dimensions (2D), called exceptional points~\cite{Heiss:2004,Berry:2004,Schen:2018} (Fig.~\ref{fig:pbc_w3D}a). (ii) Two different types of gaps have to be distinguished when eigenvalues are complex -- line gaps, which can be adiabatically transformed into a Hermitian system, and point gaps, where this is not the case~\cite{Kawabata:2019}. (iii) The topological bulk-boundary correspondence may break down for non-Hermitian systems due to the skin effect, which leads to dramatic shifts in the spectrum for open versus periodic boundary conditions as well as to a piling up of bulk states at the boundary~\cite{Xiong:2018,Wang:2018,Kunst:2018,PhysRevB.99.201103,bbc-skin,Weidemann311,Slager:2020,Torres:2018}. (iv) The structure of topological invariants becomes more intricate, as complex-valued energy eigenvalues can themselves acquire a winding number~\cite{Zhang:2020,Ueda:2018}.
 
The key property of the single Dirac electron on the 3D TI surface is that it represents an anomaly: in purely 2D such a state can neither be regularized on a lattice (implied by the fermion doubling theorem) nor in the continuum (implied by gauge symmetry)~\cite{Nielsen:1981a,Redlich:2084}. 
Our search for a non-Hermitian analogue of the 3D TI thus adopts a perspective of reverse-engineering: what could the anomalous non-Hermitian surface states be which necessitate a 3D topological bulk embedding?
Two options come to mind: (1) a 2D band structure with a single exceptional point~\cite{Schnyder:2019:1}; and (2) a single band with eigenvalues $E(k_x,k_y)=k_x+\mathrm{i}k_y$, which represents a vortex~\cite{Kunst:2019}, without the otherwise required antivortex.
In this work, we introduce exceptional topological insulators (ETIs) as the paradigmatic class of 3D non-Hermitian topological systems. We here extend the notion of `insulator' to the case of point-gapped non-Hermitian systems, irrespective of their transport properties. Under periodic boundary conditions, ETIs have a point gap in the spectrum, while in the presence of a boundary they support one of the above two types of surface states. We show that surface manipulations can interpolate between the cases (1) and (2). In particular, we demonstrate that our ETI models do not exhibit a non-Hermitian skin effect, such that the surface states are not overshadowed by a collapse of the point gap. In contrast to the conventional 3D TI, the surface states of an ETI do not require time-reversal symmetry for their protection and may therefore generically occur in non-Hermitian systems.\\

{\section{Results}}
\noindent
\textbf{Model.} We formulate a microscopic electronic quantum model for an ETI. Our results, however, hold independently from this setting, and readily carry over to systems with other degrees of freedom. 

Consider a tight-binding model on a cubic lattice with an $s$ and a $p$ orbital at each site, each of which can be filled with spin $\uparrow$ and $\downarrow$ electrons (Fig.~\ref{fig:pbc_w3D}b). Let the Pauli matrices $\sigma_\mu$ and $\tau_\mu$ act on the spin and orbital degrees of freedom, respectively, with $\mu=0,x,y,z$ and the 0-th Pauli matrix as the 2$\times$2 identity matrix. The Bloch Hamiltonian is defined as

\begin{eqnarray}
    H(\mathbf{{k}})&=&
    \left(\sum_{j=x,y,z}\!\!\!\cos k_j-M\right)\!\!\tau_z\sigma_0
    +\lambda\!\!\!\sum_{j=x,y,z}\sin k_j \, \tau_x\sigma_j \nonumber
    \\
    &\phantom{=}&+\,[\sin \alpha \, \tau_0+\cos \alpha \, \tau_z](\mathbf{{B}}\cdot\bs{\sigma})  
    +\mathrm{i}\delta\,\tau_x\sigma_0.
    \label{eq: Herm Hamiltonian}
\end{eqnarray}
For $\mathbf{{B}}=\delta=0$, $H(\mathbf{{k}})$ is a well-known Hamiltonian of a conventional 3D TI if $1<|M|<3$ with phase transitions towards trivial insulators at $|M|=1$ and $|M|=3$ (see Fig.~\ref{fig:pbc_w3D}c). The parameter $M$ controls the band inversion between $s$ and $p$ orbitals, while $\lambda$ represents the spin-orbit coupling. 
Furthermore, $\mathbf{{B}}$ represents a Zeeman field, which we take to be $\mathbf{{B}}=(B,B,B)^\top$ throughout, and $\alpha$ accounts for a possible imbalance between the $g$-factors of the $s$ and $p$ orbitals. (For $\alpha=\pi/2$ the $g$-factors are the same, for $\alpha=0$ they have opposite sign.) The term proportional to $\delta$ introduces the non-Hermiticity. We provide a physical motivation for its specific form below.

Our regime of interest is close to the phase transition between the topological and trivial insulator phase (Fig.~\ref{fig:pbc_w3D}d), where the low-energy physics of the model is described by a 3D Dirac equation for $\delta=B=0$. For concreteness, we choose parameters $M=3$ and $\lambda=1$ throughout, and then consider a finite $\delta$ {(see Supplementary Note 2 for a full phase diagram)}. Assuming periodic boundary conditions (PBC), we observe that $\delta$ opens a series of \emph{point} gaps in the bulk complex spectrum of $H(\mathbf{{k}})$ (Fig.~\ref{fig:pbc_w3D}e), while the \emph{line} gap pertaining to the Hermitian 3D TI phase is closed for $|M-3|<\delta$ (Fig.~\ref{fig:pbc_w3D}f). We posit that the constructed point-gapped Hamiltonian exhibits the phenomenology of an ETI in the presence of open boundary conditions (OBC). The role of $B$ will become clear once we study its surface states below; for now we only require $B$ to be small enough to not close the point gap at zero energy.\\

\noindent
\textbf{Topological Invariants.} Information about the spectrum with OBC can be inferred from topological invariants computed from bulk states with PBC. While the palette of available topological invariants depends on the symmetry~\cite{Kawabata:2019}, here we only consider point-gap invariants that remain in the absence of symmetry.
 
Specifically, in 3D there are three non-Hermitian weak integer winding numbers $w_{\mathrm{1D},j}$ tied to specific directions in momentum space~\cite{Sato-PRL:2020}. In addition, there is an intrinsically 3D integer invariant~\cite{Kawabata:2019,Ueda:2018,Das:2019} $w_{3\textrm{D}}$. While non-vanishing $w_{\mathrm{1D},j}$ have been related to the collapse of the point gap under OBC (non-Hermitian skin effect)~\cite{Zhang:2020, Sato-PRL:2020}, the physical significance of $w_{3\textrm{D}}$ has previously not been clarified.

In the following, we demonstrate the bulk-boundary correspondence for a system with nonzero $w_{\mathrm{3D}}$ that does not suffer from a skin effect, and find that it exactly corresponds to the ETI as characterized above. The necessary condition for the absence of the skin effect,   $w_{\mathrm{1D},1}=w_{\mathrm{1D},2}=w_{\mathrm{1D},3}=0$, holds for the Hamiltonian in Eq.~\eqref{eq: Herm Hamiltonian}. The values of $w_{\mathrm{3D}}$ in the point gaps of the Hamiltonian are indicated in Fig.~\ref{fig:pbc_w3D},~e and~f.\\

\begin{figure}
    \includegraphics[width=\linewidth]{Fig_pbc.pdf}
    \caption{\label{fig:pbc_w3D} \textsf{\textbf{Constructing an ETI from a Hermitian 3D TI.} \textbf{{a}}~Schematic band structure near an exceptional point (green). \textbf{{b}}~Real space cubic lattice model of the 3D TI with $s$ ($p$) orbitals depicted in blue (red). \textbf{{c}}~Bulk spectrum of the Hermitian 3D TI in the non-trivial phase ($1<M < 3$) along one momentum direction $k$ with a superimposed Dirac surface state (blue). \textbf{{d}}~Bulk spectrum of the Hermitian 3D TI at the transition point ($M = 3$) along one momentum direction $k$. \textbf{{e}}~Bulk spectrum in the complex plane $[\text{Re}(E), \text{Im}(E)]$ of the non-Hermitian model in Eq.~(\ref{eq: Herm Hamiltonian}) under PBC for all momenta $k_x, k_y, k_z$ at zero magnetic field ($B = 0$, $M = 2.3$, $\lambda=1$ and $\delta=0.5$). As $|M-3|>\delta$, a line gap opens along the imaginary axis, with six hole regions with corresponding point-gap invariants $w_{\mathrm{3D}}$. \textbf{{f}}~The bulk spectrum for $M=3$ and $\delta=1$ for the non-Hermitian model shows five hole regions, with corresponding point-gap invariants $w_{\mathrm{3D}}$.}}
\end{figure}

\noindent
\textbf{Surface States.} We study the model~\eqref{eq: Herm Hamiltonian} in the presence of an open boundary. Owing to the Zeeman term, the model is symmetric under $2\pi/3$-rotations around the $(111)$-axis, implying the OBC spectra for $x$, $y$, and~$z$ termination are equivalent. We thus consider OBC with $N$ layers in $z$ direction and PBC in $x$ and $y$ directions (the `slab geometry'). We set $\delta = \lambda = 1$ and $M=3$ throughout the discussion.

We begin with the case $B=0$, when the problem is analytically tractable~\cite{Terrier:2020} at $k_x=k_y=0$ (see {Supplementary Note 1 and 2}), and the characteristic polynomial $E^4(E^2-1)^{2(N-1)}$ has four roots at $E=0$. However, we find only two linearly independent eigenstates at this energy (one localized at either surface), indicating that the Hamiltonian is defective at this point. Solving for the dispersion of the zero-energy states perturbatively in $k=(k_x^2+k_y^2)^{1/2}$, we find $E_N(k_x,k_y)=\pm\sqrt{\pm\mathrm{i}}\,2^{N/2}\sqrt{k}$. In the thermodynamic limit $N\to\infty$, this leads to an infinitely steep set of eigenvalue branches, which is why we call $k_x=k_y=0$ an \emph{infernal point}. A similar exceptional point with an order equivalent to system size was found in Ref.~\onlinecite{Terrier:2020}, although using a two band model (see also {Supplementary Note 6}).

\begin{figure*}[t]
    \includegraphics[width=\linewidth]{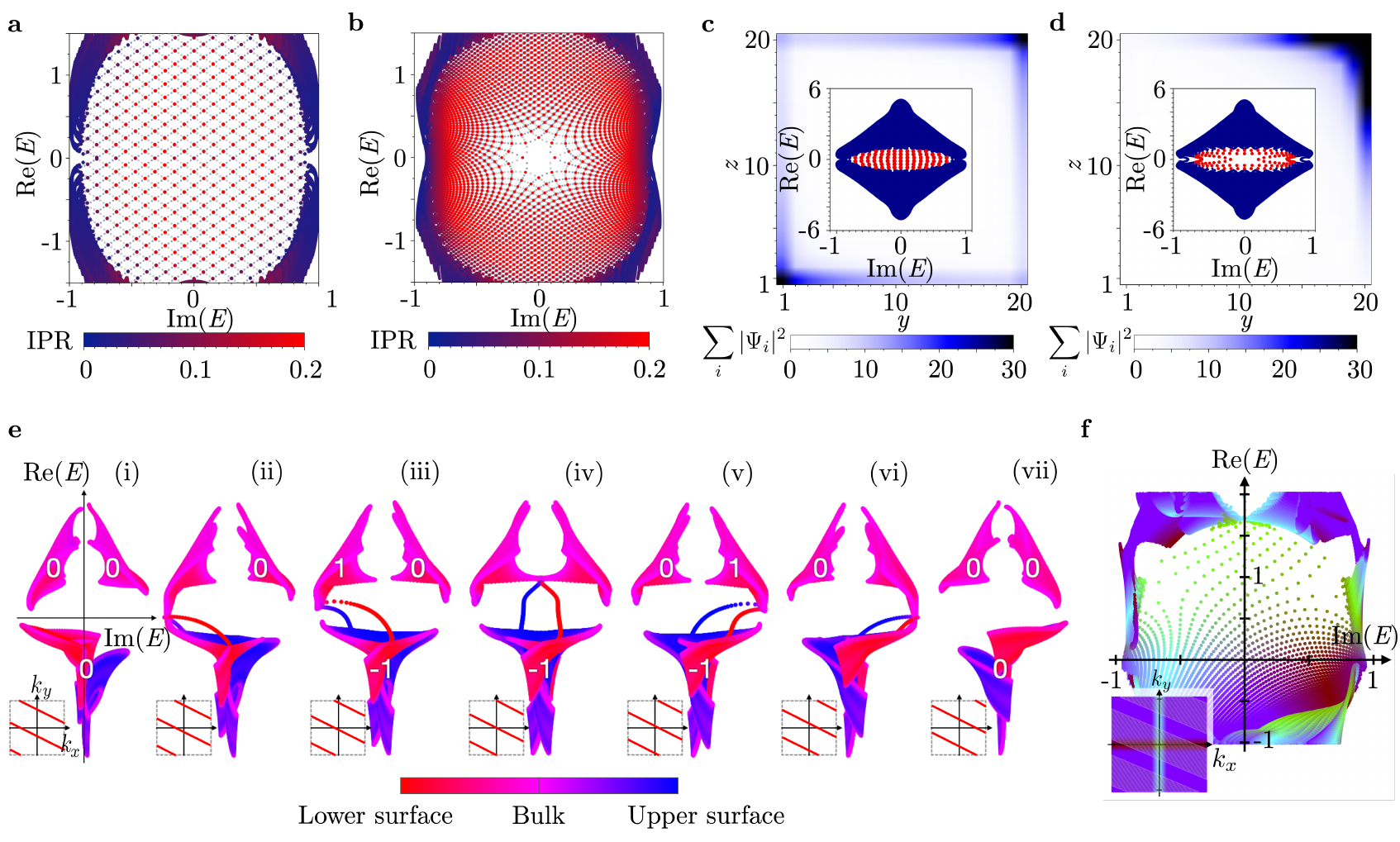}
    \caption{\label{fig:obc_2_loc} \textsf{\textbf{Open boundary condition (OBC) spectra of the exceptional topological insulator.} \textbf{{a}}~and~\textbf{{b}}~Complex-energy spectrum for the central point gap [cf.~Fig.~\ref{fig:pbc_w3D}f] for OBC in $z$ direction and for a momentum resolution of $\delta k=2\pi/800$  in  both $k_{x,y}$-directions ($B = 0.2$, $\delta = 1$). Red tones highlight states of large surface localization, with the light grey mesh of finer resolution $\delta k=  2\pi/2400$ indicating the band energies along lines parallel to either the $k_x$ or $k_y$ axis. {Concretely, we compute the inverse participation ratio (IPR), which sums the wavefunction over the lattice sites along the finite dimension $z$. A completely delocalized (bulk) state corresponds to vanishing IPR.} A single surface band covers the point gap with a regular grid for $\alpha = \pi/2$ (a), whereas the presence of an exceptional point at the origin for $\alpha = 0$ is revealed by a $2\pi$ disclination (b) (see also {Supplementary Note 3}). \textbf{{c}}~and~\textbf{{d}}~Summed density profile $\sum_i |\psi_i|^2$ of the eigenstates in the central point gap for open boundaries in $z$ and $y$ direction indicated with color scale ($B = 0.2$, $\delta = 1$), showing a surface skin effect. The insets display the corresponding energy spectrum in the complex plane, with the selected states highlighted in red. The single surface band for $\alpha = \pi/2$ shows a boundary localization in two opposite corners (c), whereas the exceptional point for $\alpha = 0$ localizes all surface states in one corner (d). \textbf{{e}}~Chern number (white numbers) evolution for a series of cuts in the surface Brillouin zone to illustrate the chiral charge pumping around the point gap for $\alpha = 0.9$ ($B = 0.5$, $\delta = 1$). The employed colormap highlights the localization of the respective eigenstates. \textbf{{f}}~Combination of cuts from panel e, zoomed onto the central point gap, highlighting how the single-sheet covering in panels a,b relates to the chiral edge states in e. The inset indicates the employed color scheme for momenta in the surface BZ as well as the cuts used for diagonalization (shaded area). Only bulk states and states localized on the upper surface are displayed  ($\alpha = 0.9$, $B = 0.5$, $\delta = 1$).
    }}
\end{figure*}

We argue however that the infernal point exhibits a {fine-tuned}, rather than the {generic}, surface-state structure of an ETI. Specifically, we find the surface spectrum to be regularized by small perturbations, such as the finite Zeeman term in Eq.~\eqref{eq: Herm Hamiltonian}, which we expect to be generally present in a physical realization. {Note that $B\neq0$ breaks the isotropy of the model by selecting the (1,1,1)-direction. While the OBC spectra in $x$, $y$, and $z$ direction are equivalent, a surface termination perpendicular to the (1,1,1)-direction still exhibits the infernal point.} We first study the effect of the Zeeman term with $\alpha=0$ and $\alpha=\pi/2$ separately, and then discuss the transition between the two.

For $\alpha=\pi/2$ (Fig.~\ref{fig:obc_2_loc}a), we find a single sheet of complex-eigenvalue states localized on the top surface to cover the point-gap region in the complex-energy plane (along with another sheet with eigenstates localized on the bottom surface). We numerically determine the `dispersion' of the surface state as $E(k_x,k_y) \propto (k_x+\mathrm{i}k_y)$ to linear order in $k_x$ and $k_y$. Thus, the system has a \emph{single Fermi point} in the surface Brillouin zone at $k_x=k_y=0$. 

An odd number of Fermi points is impossible in a strictly 2D model due to the non-Hermitian fermion doubling theorem~\cite{Schnyder:2019:1}. This follows because eigenvalues of a 2D Hamiltonian define continuous maps from a two-dimensional BZ torus to the complex plane $\mathbb{C}$. Since the BZ has no boundary, each point in $\mathbb{C}$, including $E=0$, must be the image of an {even} number of momenta in the BZ. The single-sheet covering of the point gap exhibited by the ETI is thus \emph{anomalous} and only possible because it is connected to a 3D bulk spectrum. 

For $\alpha=0$ (Fig.~\ref{fig:obc_2_loc}b), a single exceptional point is found on the surface of the ETI. Likewise, this situation is anomalous since also an odd number of exceptional points cannot be realized in purely 2D according to the non-Hermitian fermion doubling theorem~\cite{Schnyder:2019:1}. In fact, each energy in the point-gap region is covered exactly once, reminiscent of the previously discussed $\alpha=\pi/2$ case.

Interpolating between $\alpha=0$ and $\alpha=\pi/2$ changes the surface spectrum by moving the exceptional point on the surface out of the point gap into the spectral region of the bulk states. This is analogous to surface states of a conventional 3D TI, where the topological surface Dirac cone can either be found within the bulk energy gap, or can be `buried' in the bulk energy bands~\cite{ChenBi2Te3:2009,HasanBiSe:2009}, leaving a single band sheet on the surface (see {Supplementary Note 3}). We find an exact bulk-boundary correspondence between invariant $w_\textrm{3D}$ and the number of point-gap covering surface-state sheets by considering a Hermitian doubled Hamiltonian~\cite{Slager:2020} that corresponds to two ETI copies related by Hermitian conjugation (see Methods).

Another generic property of the ETI emerges when considering OBC in {two} directions (while keeping PBC in the third direction), where a \emph{surface skin effect}~\cite{Lee:2019} localizes order $N$ states exponentially at the hinges. This is related to a higher order skin effect~\cite{Kawabata:2020}. One particular difference between the two values of $\alpha$ can be seen in the localization of the modes, for $\alpha=\pi/2$, the spectral weight of the surface states concentrates on two opposite corners (Fig.~\ref{fig:obc_2_loc}c). By contrast, for $\alpha=0$ the surface exceptional point localizes all the surface states exponentially towards one side (Fig.~\ref{fig:obc_2_loc}d).\\

\noindent
\textbf{Berry flux.} The anomalous boundary states of an ETI are tied to the nontrivial bulk topological invariant~\eqref{eq:Ueda} via the bulk-boundary correspondence. To substantiate this claim we reformulate the invariant as follows. Recall that a non-Hermitian Hamiltonian with a point-gap can be continuously deformed into a unitary matrix while preserving locality and band topology~\cite{Ueda:2018}. Such a deformed Hamiltonian has orthogonal eigenstates with eigenvalues lying on a unit circle, $E(\mathbf{{k}})\to e^{\mathrm{i}\varepsilon(\mathbf{{k}})}$, suggesting an interpretation as a periodically driven (i.e.~`Floquet') quantum system~\cite{Sato:2020}. In this context, $\varepsilon$ is referred to as quasi-energy. 
Then (see {Supplementary Note 4} or Ref.~\onlinecite{Sato:2020}),
\begin{equation}
w_{\mathrm{3D}} = C_{\textrm{FS}(\mu)},
    \label{eqn:second form of invairant}
\end{equation}
where the right-hand side denotes the Chern number of the Fermi surface at (arbitrary) quasi-energy $\mu$, i.e., $\textrm{FS}(\mu)=\{\mathbf{{k}}\in\textrm{BZ}\,|\,\exists a: \varepsilon^a(\mathbf{{k}}) =\mu \}$ with $a$ the band index. Equation~\eqref{eqn:second form of invairant} allows to interpret a nonzero $w_{\mathrm{3D}}$ as counting the quanta of Berry flux that circulate around the point gap in the complex-energy plane~\cite{Sun:2018,Higashikawa:2019}. 

\begin{figure}
    \includegraphics[width=\linewidth]{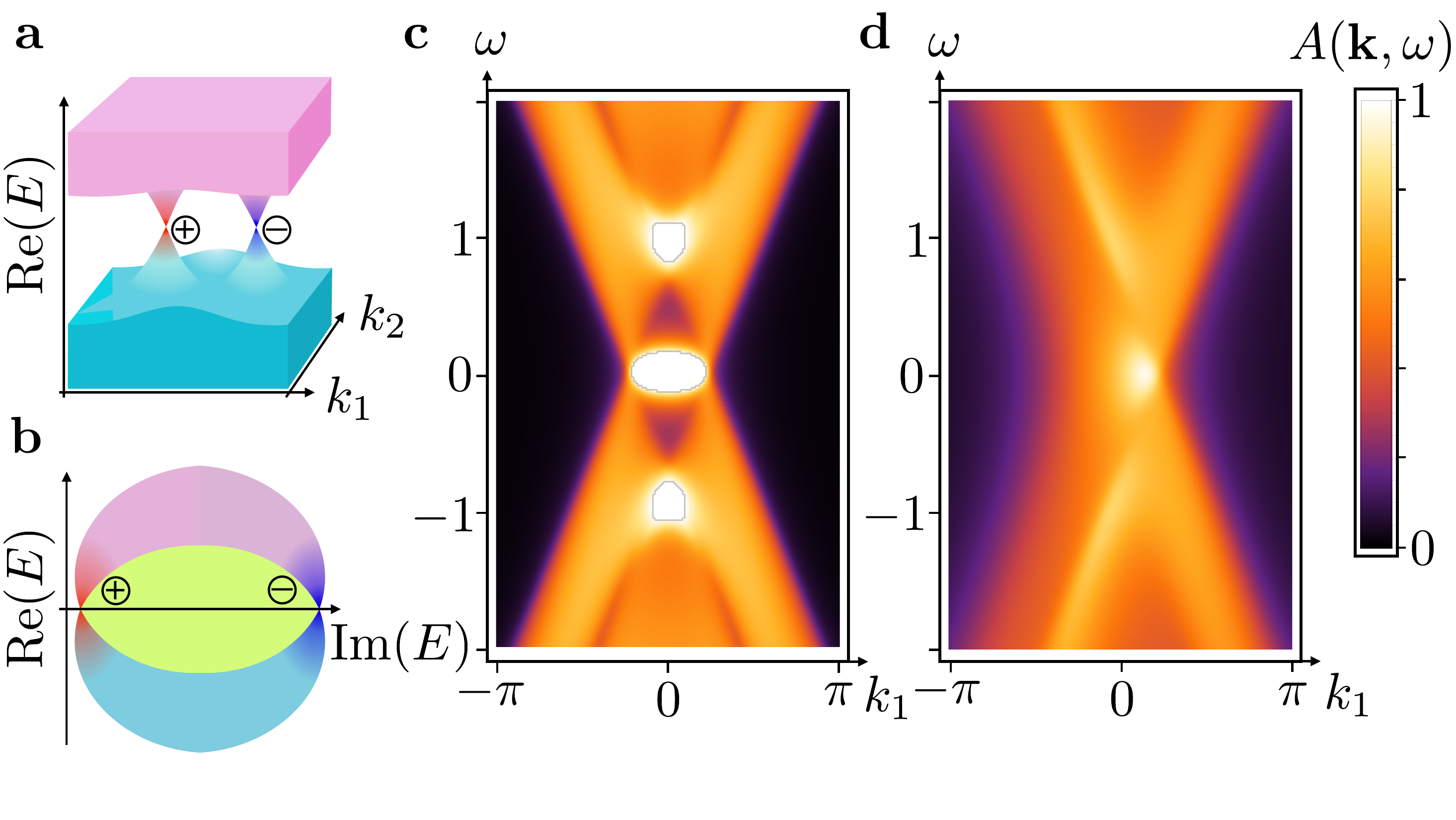}
    \caption{\label{fig:Greens} \textsf{\textbf{Inducing an ETI in a Hermitian Weyl semimetal.} \textbf{{a}}~Schematic band structure of a Weyl semimetal with two Weyl cones of opposite chiral charge (red vs.~blue) connecting bulk valence (cyan) and conduction (pink) bands. \textbf{{b}}~An ETI with a point gap (green) arises if we assign different lifetimes to the two chiral fermions. The bulk states carry a quantum of Berry flux around the point gap. \textbf{{c}}~and~\textbf{{d}}~Spectral density derived from the surface Green's function~\cite{vonOppen17} $G(\mathbf{{k}}, \omega)$ via $A(\mathbf{{k}}, \omega) = {-1/\pi \, \mathrm{Im\, Tr \,} G(\mathbf{{k}}, \omega+\mathrm{i} \epsilon)}$, along momenta $k_{1,2} = (k_x\pm k_y)/2$ and for frequency $\omega$. We set the smearing factor to $\epsilon=0.1$ and the non-Hermitian term to $\delta = 0$ (c) resp. to~$\delta = -1/2$ (d), for Hamiltonian parameters $M=3$, $\alpha = \pi/2$, and  $B=1/2$. We added a term $ \mathrm{i} \delta \tau_0 \sigma_0$ to the Hamiltonian in order for all eigenvalues to have non-positive imaginary part.
    }}
\end{figure}

The Berry flux can be related to the ETI surface states as follows. Consider the system with OBC in $z$ direction and PBC otherwise and assume a series of cuts through the 2D surface BZ. Each cut represents a fictitious 2D non-Hermitian system with OBC in one direction. In 2D, isolated bands can be characterized by an integer Chern number $C$~\cite{Matthew:2019,Schen:2018}. If $|C|$ is nonzero, the corresponding number of edge-localized topological modes connect the band with the other bands inside the complex plane, similar to Hermitian Chern insulators that describe the integer quantum Hall effect in lattice systems~\cite{Thouless:1982}. 

We visualize the construction for the model~\eqref{eq: Herm Hamiltonian} in Fig.~\ref{fig:obc_2_loc}e along the cuts shown in the insets. We observe that the four bands of the model project to three distinct regions in the complex plane. Panels (e-ii), (e-iv), and (e-vi) indicate critical cuts when the bands exchange Chern numbers, while in the intermediate regions the bands exhibit fixed values $C=0$ or $C=\pm1$. We observe in the sequence of cuts that a Chern dipole is formed on the left side of the point gap, by transferring a Chern number $+1$ (in clockwise direction) between the two bands that touch in the second panel. The topological charge $C=+1$ is then transferred between the upper left and the upper right band, and finally back to the lower band. Thus, a quantum $+1$ of Berry flux is pumped clockwise around the point gap, in accordance with the rewriting of $w_{\mathrm{3D}}$ in Eq.~\eqref{eqn:second form of invairant}. Note that Fig.~\ref{fig:obc_2_loc}e also indicates the boundary-localized edge modes which at each stage connect bands with opposite Chern number. As such, they necessarily swipe over the entire point-gap region during the pumping process, thus forming the protected topological surface state of the ETI. Consequently, the Chern number flow around the point gap corresponds to a rotation of the chiral edge state, which leads to the anomalous net surface chirality. Combining the series of cuts with a momentum resolution in Fig.~\ref{fig:obc_2_loc}f provides a visual connection between the numerically obtained surface spectra and the topological invariant.

Based on the relation between $w_\textrm{3D}$ and the Berry flux, we expect ETIs to arise generically out of critical 3D band structures, in particular from Weyl semimetals~\cite{WeylRev:2017}. Note that for $\delta=0$ and $\alpha=\pi/2$, Hamiltonian~\eqref{eq: Herm Hamiltonian} is precisely a Weyl semimetal  with a pair of Weyl points at $\mathbf{{k}}=\pm \mathbf{{B}}$. These are a source and a sink of unit Berry flux\cite{Wan:2011}. With OBC, the Weyl points are connected by surface Fermi arcs. Upon including $\delta\neq 0$ the eigenvalues of the Weyl points acquire different imaginary parts, such that inevitably a bulk spectrum with a point gap emerges (Fig.~\ref{fig:Greens}a and b). Under OBC, the point gap is filled with Fermi-arc states that connect the Weyl points (green). These are the ETI surface states. In Fig.~\ref{fig:Greens}c and~d we show the surface spectral function for a Weyl semimetal and an ETI. The latter is characterized by only one visible Weyl cone and a broad signal from the surface states that gets sharper closer to the cone. This indicates that the ETI, which is a point-gapped non-Hermitian topological phase, is spectroscopically a semimetal with vanishing density of states at $\mathrm{Re}(E)=0$.\\

\noindent
\textbf{Non-Hermitian Terms.} We finish by discussing how the non-Hermitian term $\mathrm{i}\delta\tau_x\sigma_0$ may arise in a 3D TI and which measurable consequences it has. In principle, a tailored orbital-dependent coupling to a lossy mode would suffice to give rise to such a term. For instance, we can consider an additional impurity, e.g., $f$-orbital at energy $\mu_{\mathrm{f}}$ in the unit cell with no dispersion but finite lifetime $\Gamma$ as a representative source for non-Hermitian contributions to the electronic self-energy. If this short-lived $f$-electron couples to the $s,p$-orbitals of the topological insulator with hopping strength $t_{\mathrm{f}}$, the single-electron Green's function acquires a complex self-energy $\Sigma=\mathrm{i} t_{\mathrm{f}}
^2 / (\Gamma - \mathrm{i} \mu_{\mathrm{f}}) (\tau_0\sigma_0 + \tau_x\sigma_0)$ (see {Supplementary Note 5}). For $\mu_{\mathrm{f}} \ll \Gamma$, the non-Hermitian term dominates and, up to an overall imaginary shift in the spectrum, contributes the desired non-Hermitian term in Eq.~\eqref{eq: Herm Hamiltonian}. Alternatively, electron-phonon-scattering can act as a source for the non-Hermitian terms in the self-energy, likewise leading to the topological features of an ETI. 

Classical analogues to quantum mechanical topological states can be constructed in a variety of platforms, including phononic~\cite{Lubesky:2014,Susstrunk:2015}, photonic~\cite{Raghu:2008,Wang:2009,PhysRevLett.118.045701} and electrical~\cite{leech2018} metamaterials. The ETI is no exception to this. For realizations {by design}, a two-band model~\cite{Terrier:2020}, which, however, requires a more complicated anti-Hermitian term than the four-band model of Eq.~\eqref{eq: Herm Hamiltonian}, may be more amenable (see {Supplementary Note 6}).\\

{\section{Discussion}}
We introduced 3D ETIs, a phase of matter governed by a local non-Hermitian (Hamiltonian) operator with a point gap and topological surface states.
In particular, we show that a Weyl semimetal with two Weyl nodes at the Fermi energy generically becomes an ETI under a non-Hermitian perturbation that opens a point gap. 
Besides the realization of ETIs in (meta)materials, open questions for future research include: How are ETI phases further differentiated by the addition of symmetries? What are alternative representations of the topological invariants in terms of symmetry indicators? What role do interactions play for the stability of ETIs? Our findings are the first step towards a microscopic understanding of such non-Hermitian topological matter.\\

{\section{Methods}}
\noindent
\textbf{ETI tight-binding model.} As an illustrative model for the ETI phase, we consider in the main text a cubic lattice with two orbitals $s$ ($\gamma = 0$) and $p$ ($\gamma = 1$) and spin $\uparrow, \downarrow$ per site. The lattice is spanned by the unit vectors $\mathbf{{e}}_i$, $i = x, y, z$, giving rise to the tight-binding Hamiltonian
\begin{eqnarray}
    H &=& -M \sum_{\mathbf{{r}},\gamma} (-1)^{\gamma}\, c_{\mathbf{{r}},\gamma}^{\dagger}\sigma_0 c_{\mathbf{{r}},\gamma}^{}\nonumber\\
    &\phantom{=}&+\frac{1}{2} \sum_{\mathbf{{r}},\gamma} \sum_{i = x,y,z} (-1)^{\gamma}\, c_{\mathbf{{r}}+\mathbf{{e}}_i,\gamma}^{\dagger}\sigma_0 c_{\mathbf{{r}},\gamma}^{}+ \mathrm{h.c.}\nonumber\\
    &\phantom{=}&+ \frac{\lambda}{2\mathrm{i}} \sum_{\mathbf{{r}},\gamma} \sum_{i = x,y,z}\, c_{\mathbf{{r}}+\mathbf{{e}}_i,\gamma+1}^{\dagger} \sigma_i^{} c_{\mathbf{{r}},\gamma}^{}+ \mathrm{h.c.}\\
    &\phantom{=}&+ \sum_{\mathbf{{r}},\gamma} \sum_{i = x,y,z}B_i\left[(-1)^{\gamma}\cos(\alpha) + \sin(\alpha)\right]\, c_{\mathbf{{r}},\gamma}^{\dagger} \sigma_i^{} c_{\mathbf{{r}},\gamma}^{} \nonumber\\
    &\phantom{=}&+ \mathrm{i}\delta \sum_{\mathbf{{r}},\gamma}\,c_{\mathbf{{r}},\gamma+1}^{\dagger}\sigma_0 c_{\mathbf{{r}},\gamma}^{},\nonumber
\end{eqnarray}
where we use $\gamma$ modulo 2 and the Pauli matrices as $\sigma_\mu$, $\mu = 0, x, y, z$ with the 0-th Pauli matrix as the 2$\times$2 identity. The operator $c_{\mathbf{{r}},\gamma}^{\dagger} = \left(c_{\mathbf{{r}},\gamma,\uparrow}^{\dagger}\, c_{\mathbf{{r}},\gamma,\downarrow}^{\dagger}\right)$ then creates an electron in orbital $\gamma$ at lattice site $\mathbf{{r}}$ with the respective spin orientation.\\

\noindent
\textbf{Topological invariants.} Two important invariants exist for 3D non-Hermitian systems in the absence of any symmetry. First, specific directions in momentum space are equipped with a weak integer invariant~\cite{Sato-PRL:2020},
\begin{equation}
    \label{eq:1D invariant}
    w_{\mathrm{1D},j} = -\mathrm{i} \int_{\mathrm{BZ}}  \frac{d^3\mathbf{{k}}}{(2 \pi)^3} \Tr[Q_{j}(\mathbf{{k}})],
\end{equation}
with $Q_{j}(\mathbf{{k}}) = [H(\mathbf{{k}})-E]^{-1} \partial_{k_{j}} [H(\mathbf{{k}})-E]$, $E$ is any complex value in the point gap, $j=x,y,z$, and $\mathrm{BZ}=[-\pi,\pi]^3$ denotes the 3D Brillouin zone. A non-zero $w_{\mathrm{1D},j}$ indicates the non-Hermitian skin effect~\cite{Zhang:2020, Sato-PRL:2020}, under which the spectrum collapses upon considering open boundary conditions. Hence, a vanishing $w_{\mathrm{1D},j} = 0$ is required for the observation of topological surface states.

Such a scenario is indicated by an intrinsically 3D integer invariant~\cite{Kawabata:2019,Ueda:2018,Das:2019}
\begin{equation}
    \label{eq:Ueda}
    w_{\mathrm{3D}} = -\int_{\mathrm{BZ}}  \frac{d^3\mathbf{{k}}}{24 \pi^2}\epsilon_{ijk} \Tr[Q_{i}(\mathbf{{k}})Q_{j}(\mathbf{{k}})Q_{k}(\mathbf{{k}})],
\end{equation}
where the summation of repeated indices $i,j,k$ is implied and $\epsilon_{ijk}$ is the Levi-Civita symbol. The physical significance of a non-zero $w_{3\textrm{D}}$ is the ETI phase presented in this manuscript.\\

\noindent
\textbf{Proof of bulk-boundary correspondence.} The ETI bulk-boundary correspondence builds on the fact that the point gap of an ETI necessarily fills up with surface-localized states when open boundary conditions are introduced. These midgap surface states are topologically protected by the nonzero ETI bulk invariant $w_\text{3D}$ without the assumption of additional symmetries, similarly to how the edge states of a Chern insulator are protected by a nonzero Chern number.

We consider an ETI that is described by a Bloch Hamiltonian $Q(\mathbf{{k}})$ with a point gap around the complex energy $E_0$, and a winding number $|w_\text{3D}| = 1$. Since the winding number is quantized, we may equivalently consider any other $E_0$ lying within the same point gap. We now define the Hermitian double
\begin{equation} \label{eq: hermDoublePBC}
\tilde{H}(Q(\mathbf{{k}})-E_0) = \begin{bmatrix} 0 & Q(\mathbf{{k}})-E_0 \\ Q(\mathbf{{k}})^\dagger-E^*_0 & 0\end{bmatrix},
\end{equation}
which describes a topological insulator in Altland-Zirnbauer class AIII with a single surface Dirac cone~\cite{Schnyder3D}. Correspondingly, the Hamiltonian $\tilde{H}_\text{slab}$ obtained by placing the real-space version of $\tilde{H}(\mathbf{{k}})$ in a slab geometry satisfies $\det(\tilde{H}_\text{slab})=0$: for both surfaces, the chiral symmetry of $\tilde{H}_\text{slab}$ enforces a spectral pinning of the surface Dirac crossing to zero energy. We do not resolve any remaining momentum quantum numbers because a surface Dirac cone in class AIII is not pinned to any particular surface momentum. In the slab geometry, the decomposition reads $\tilde{H}_\text{slab}(Q_\text{slab}-E_0)$, where $Q_\text{slab}$ is the non-Hermitian Hamiltonian obtained by placing the real-space version of $Q(\mathbf{{k}})$ in the slab geometry. The surface Dirac cones of $\tilde{H}_\text{slab}$ then imply
\begin{eqnarray}
\det(\tilde{H}_\text{slab}) &=& \det[-(Q_\text{slab}-E_0)(Q_\text{slab}^\dagger-E^*_0)] = 0\nonumber\\
&\phantom{=}&\rightarrow \quad \det(Q_\text{slab}-E_0) = 0,
\end{eqnarray}
from which we deduce that $Q_\text{slab}$ has at least one eigenvalue equal to $E_0$.

After establishing the presence of protected midgap states in the slab spectrum of an ETI, described by a Bloch Hamiltonian $Q(\mathbf{{k}})$, we next derive its unique topological surface characteristic: the surface chirality. By this we mean the accumulated winding of all energies of the states on a given surface around a reference energy $E_0$ (chosen to lie within the point gap), which can be calculated via the formula
\begin{equation} \label{eq: surfacechirality}
\nu(E_0) = \frac{1}{2\pi} \oint_{\gamma (E_0)} \partial_k \left\{\sum_i \mathrm{Arg} [E_i(\mathbf{{k}}_\perp) -E_0] \right\} \mathrm{d}k \in \mathbb{Z},
\end{equation}
where the path of momenta $\gamma (E_0)$ is obtained as the surface Brillouin zone preimage of any connected set of energies $E_i[\gamma (E_0)]$ that encircles $E_0$ counter-clockwise in the complex plane. Note that all $E_i[\gamma (E_0)]$ should lie themselves within the point gap. We will prove that $|\nu(E_0)| = 1$ is nonzero for all choices of $E_0$.

Recall the expression for the surface winding number of a Hermitian topological insulator in class AIII~\cite{Ryu_2010},
\begin{equation} \label{eq: hermwinding}
\nu_\mathrm{Hermitian} = \frac{1}{2\pi} \oint_{\lambda} \mathrm{Im} \, \mathrm{tr} \left[q(\mathbf{{k}}_\perp)^{-1} \partial_k q(\mathbf{{k}}_\perp) \right] \mathrm{d}k \in \mathbb{Z},
\end{equation}
where $q(\mathbf{{k}}_\perp)$ forms the Hermitian surface Hamiltonian $\tilde{H}_\text{surface}(q(\mathbf{{k}}_\perp))$ and $\lambda$ is any (possibly disconnected) path in the surface Brillouin zone that encloses all surface Dirac cones in the spectrum of $\tilde{H}_\text{surface}(q(\mathbf{{k}}_\perp))$ counter-clockwise and only covers surface-localized states. $|\nu_\mathrm{Hermitian}|$ then counts the number of topologically protected surface Dirac cones. 

We now relate the Hermitian surface winding number $\nu_\mathrm{Hermitian}$ to the non-Hermitian surface chirality $\nu$. In the absence of a collapse of the bulk spectrum of $Q(\mathbf{{k}})$ as we open the boundary conditions, we can interpret $[q(\mathbf{{k}}_\perp) - E_0]$ as the effective surface Hamiltonian of the ETI $[Q(\mathbf{{k}}) - E_0]$ [whose Hermitian double is $\tilde{H}(Q(\mathbf{{k}})-E_0)$ in Eq.~\eqref{eq: hermDoublePBC}]. 
Also, as long as $E_0$ lies in the point gap, the nontrivial ETI invariant $|w_\text{3D}| = 1$ implies a single surface Dirac cone for the Hermitian double~\cite{Schnyder3D}, resulting in the equality
\begin{eqnarray} 
\pm 1 &=& \nu_\mathrm{Hermitian} (E_0) \nonumber\\
&=& \frac{1}{2\pi} \oint_{\lambda (E_0)} \partial_k \mathrm{tr} \left\{\mathrm{Im} \log [q(\mathbf{{k}}_\perp)-E_0]\right\} \mathrm{d}k \\
&=& \nu(E_0)\nonumber,
\end{eqnarray}
where we substituted $\lambda \rightarrow \lambda(E_0)$ (the location of the surface Dirac cone varies with $E_0$), and then identified $\gamma (E_0) = \lambda (E_0)$. It remains to be shown that our definitions of $\lambda (E_0)$ and $\gamma (E_0)$ are compatible, that is, that the eigenvalues of $q[\lambda (E_0)]$ wind around $E_0$. This must be so because $\nu(E_0)$ could otherwise not take on non-zero values. Furthermore, since $\nu(E_0)$ is quantized, any other choice of $\gamma (E_0)$ that has the above-mentioned properties is equally valid, thus completing our proof. 

In conclusion, we find that the surface band structure of an ETI is characterized by an anomalous net chirality, which cannot be realized in a purely two-dimensional system, but is instead enabled by the presence of the topologically nontrivial three-dimensional bulk.\\

\noindent
\textbf{Data availability}\\ 
All information needed to evaluate the conclusions in the paper are present in the paper and/or the supplementary {information}. Additional data are available from the corresponding authors upon reasonable request.\\

\noindent
\textbf{{Code availability}}\\ 
{The code used to generate the figures in this work is available upon reasonable request.}\\

\noindent
\textbf{Acknowledgements}\\
This project has received funding from the European Research Council (ERC) under the European Union’s Horizon 2020 research and innovation program (ERC-StG-Neupert-757867-PARATOP). R.~T. is supported by the Deutsche Forschungsgemeinschaft (DFG, German Research Foundation) through project-id 258499086 - SFB 1170 and through the W\"urzburg-Dresden Cluster of Excellence on Complexity and Topology in Quantum Matter –ct.qmat project-id 39085490 - EXC 2147. T.~B. was supported by an Ambizione grant No.~185806 by the Swiss National Science Foundation. \\

\noindent
\textbf{Author contributions}\\
M.M.D., A.S., F.S., T.B. and T.N. contributed equally to the theoretical analysis in this work and wrote the manuscript. F.S. discovered the model used in \eqref{eq: Herm Hamiltonian}. T.B. and T.N. investigated the invariant \eqref{eq:Ueda} and derived the reformulation using Chern numbers. M.M.D., M.H.F. and R.T. investigated the possible origins of the non-Hermitian perturbation in realistic scenarios. All authors discussed and commented on the manuscript.\\

\noindent
\textbf{Competing interests}\\
The authors declare no competing interests.

\end{bibunit}

\bibliography{bibliography}
\bibliographystyle{naturemag}

\end{document}


\clearpage

\begin{bibunit}
\onecolumngrid
\renewcommand\thesection{SUPPLEMENTARY NOTE \arabic{section}:~}
\renewcommand\thesubsection{\Alph{subsection}.~}
\renewcommand{\figurename}{Supplementary Figure}
\renewcommand\bibname{Supplementary References}

\title{Supplementary Information: Exceptional Topological Insulators}

\author{M.~Michael~Denner}
\affiliation{Department of Physics, University of Zurich, Winterthurerstrasse 190, 8057 Zurich, Switzerland}
\author{Anastasiia~Skurativska}
\affiliation{Department of Physics, University of Zurich, Winterthurerstrasse 190, 8057 Zurich, Switzerland}
\author{Frank~Schindler}
\affiliation{Department of Physics, University of Zurich, Winterthurerstrasse 190, 8057 Zurich, Switzerland}
\affiliation{Princeton Center for Theoretical Science, Princeton University, Princeton, NJ 08544, USA}
\author{Mark~H.~Fischer}
\affiliation{Department of Physics, University of Zurich, Winterthurerstrasse 190, 8057 Zurich, Switzerland}
\author{Ronny Thomale}
\affiliation{Institut für Theoretische Physik und Astrophysik, Universität Würzburg, 97074 Würzburg, Germany}
\author{Tom\'{a}\v{s} Bzdu\v{s}ek}
\affiliation{Condensed Matter Theory Group, Paul Scherrer Institute, 5232 Villigen PSI, Switzerland}
\affiliation{Department of Physics, University of Zurich, Winterthurerstrasse 190, 8057 Zurich, Switzerland}
\author{Titus Neupert}
\affiliation{Department of Physics, University of Zurich, Winterthurerstrasse 190, 8057 Zurich, Switzerland}
\date{\today}

\maketitle
\newpage
\onecolumngrid

%

\section{Dirac Theory for the ETI} \label{sec: DiracAppendix}
We here show how the infernal point of the Hamiltonian in Eq. (1) of the main text in presence of open boundary conditions can be derived from a Dirac equation approach, similar to how the gapless surface Dirac fermion of a Hermitian 3D TI can be obtained as a zeromode that is localized at domain walls of a 3D bulk Dirac mass. In comparison to the two zero-energy states of a 3D TI surface Dirac fermion, we find only a single zero-energy state for the ETI, underlining the fact that the ETI surface in some sense realizes ``half" of a Hermitian Dirac cone (recall that upon the introduction of non-Hermitian terms, a 2D Dirac cone generically splits into two exceptional points). 

We start with the Bloch Hamiltonian of a three-dimensional exceptional topological insulator (ETI), corresponding to Eq.~(1) of the main text with the parameter choice $M = 3$, $\lambda=1$, $\mathbf{B} = \bs{0}$, giving
\begin{equation} \label{eq: BlochTheory}
H(\mathbf{k}) = \left(\sum_{i=x,y,z} \cos k_i -3 \right) \tau_z \sigma_0 + \sum_{i=x,y,z} \sin k_i \tau_x \sigma_i + \mathrm{i} \delta \tau_x \sigma_0,
\end{equation}
which describes a four-fold Dirac crossing at $\mathbf{k} = (0,0,0)$ that is gapped out by the non-Hermitian term multiplying $\delta$. Unless otherwise stated, we assume $\delta > 0$. $H(\mathbf{k})$ then has a nontrivial winding number~\cite{Ueda:2018} $w_\text{3D} = 1$.
The long-wavelength limit of the model is given by the Dirac theory
\begin{equation}
H_\text{D} (\mathbf{k}) = \sum_{i=x,y,z} k_i \tau_x \sigma_i + \mathrm{i} \delta \tau_x \sigma_0.
\end{equation}

We want to derive the band structure of the surface with normal in $z$-direction, where $k_x$ and $k_y$ remain as good momentum quantum numbers. For this, we model the interface between an ETI and the vacuum via the Dirac Hamiltonian
\begin{equation} \label{eq: DiracHam}
H_{\text{D},z} (k_x,k_y) = \sum_{i=x,y} k_i \tau_x \sigma_i + (-\mathrm{i} \partial_z) \tau_x \sigma_z + \mathrm{i} \delta \theta(-z) \tau_x \sigma_0 - \mu \theta(z) \tau_z \sigma_0,
\end{equation}
where $\mu > 0$ multiplies a Hermitian mass term that gaps Supplementary Eq.~\eqref{eq: BlochTheory} at $\delta = 0$ into a trivial insulator and $\theta(z)$ is the Heaviside step function. We use the convention $\theta(0) = 0$.

Let us fix $k_x = k_y = 0$ for simplicity -- we have numerically confirmed that the $z$-directed slab spectrum of the Bloch Hamiltonian~\eqref{eq: BlochTheory} away from this point is gapped. Zero-energy states $\Psi(z)$ then satisfy
\begin{equation}
\begin{aligned}
& \left[(-\mathrm{i} \partial_z) \tau_x \sigma_z + \mathrm{i} \delta \theta(-z) \tau_x \sigma_0 - \mu \theta(z) \tau_z \sigma_0 \right] \Psi(z) = 0, \\
& \quad \rightarrow \quad \tau_x \sigma_z \partial_z \Psi(z)= \left[\delta \theta(-z) \tau_x + \mathrm{i} \mu \theta(z) \tau_z \right] \sigma_0 \Psi(z) \\
& \quad \rightarrow \quad \partial_z \Psi(z)= \left[\delta \theta(-z) \tau_0 + \mu \theta(z) \tau_y \right] \sigma_z \Psi(z) \\
& \quad \rightarrow \quad \partial_z \Psi_\lambda(z)= \lambda \left[\delta \theta(-z) \tau_0 + \mu \theta(z) \tau_y \right] \Psi_\lambda(z),
\end{aligned}
\end{equation}
where we have chosen $\Psi_\lambda(z)$ as eigenstates of $\sigma_z$ that satisfy $\sigma_z \Psi_\lambda(z) = \lambda \Psi_\lambda(z)$. 
We next solve the equation on both sides separately. For $z<0$ we obtain
\begin{equation}
\partial_z \Psi^{<}_\lambda(z) = \lambda \delta \tau_0 \Psi^{<}_\lambda(z) \quad \rightarrow \quad \Psi^{<}_{+,1}(z) = \mathcal{N}^{<} e^{\delta z} (1,0,0,0)^\mathrm{T}, \quad \Psi^{<}_{+,2}(z) = \mathcal{N}^{<} e^{\delta z} (0,0,1,0)^\mathrm{T},
\end{equation}
where $\mathcal{N}$ is a normalization factor, and the solutions corresponding to $\lambda = -1$ are not normalizable. For later use, let us note that the following linear combination is an equally valid zero-energy solution for $z<0$:
\begin{equation}
\Psi^{<}_{+}(z) \equiv \frac{\Psi^{<}_{+,2}(z) + \mathrm{i} \Psi^{<}_{+,1}(z)}{\sqrt{2}} = \frac{\mathcal{N}^{<}}{\sqrt{2}} e^{\delta z} (\mathrm{i},0,1,0)^\mathrm{T},
\end{equation}
Likewise, for $z>0$ we obtain
\begin{equation}
\partial_z \Psi^{>}_\lambda(z) = \lambda \mu \tau_y \Psi^{>}_\lambda(z) \quad \rightarrow \quad \Psi^{>}_-(z) = \mathcal{N}^{>} e^{-\mu z} (0,-\mathrm{i},0,1)^\mathrm{T}, \quad \Psi^{>}_+(z) = \mathcal{N}^{>} e^{-\mu z} (\mathrm{i},0,1,0)^\mathrm{T},
\end{equation}
where again the remaining solutions are not normalizable. To obtain a viable zero-energy wavefunction for the entire range of $z$, we need to match solutions across $z = 0$. For simplicity, let us set $\delta = \mu \equiv \Delta$. There is then only a single matchable solution, with wavefunction
\begin{equation}
\Psi(z) = \sqrt{\frac{\Delta}{2}} e^{-\Delta |z|} (\mathrm{i},0,1,0)^\mathrm{T}, \quad H_{\text{D},z} (0,0) \Psi(z) = 0.
\end{equation}

%

\section{Analytical treatment of the ETI tight-binding model}

In this section we analyze the tight-binding model from Eq. (1) of the main text analytically in some detail. We set $B=0$ throughout the discussion, i.e.~we retain isotropy. While this causes certain instabilities in the numerical computation of the spectrum, it also makes the problem more amenable for an analytical treatment. The discussion is organized into subsections as follows. We begin in Supplementary Note~\ref{sec:diagram} with rotating the model to a basis that explicitly reveals its sublattice symmetry. While not necessary to enable the $w_{3\mathrm{D}}$ invariant, the sublattice symmetry greatly simplifies the phase diagram of the model. In Supplementary Note~\ref{sec:slab} we write the analytical form of the Hamiltonian in slab geometry with open boundary conditions. By analyzing the constructed slab Hamiltonian we reveal the existence of the macroscopically defective infernal point at zero surface momentum. Finally, in Supplementary Note~\ref{sec:poly-expansion} we perform a perturbative expansion in the characteristic polynomial of the slab Hamiltonian to infer the dispersion of the surface states around the infernal point. We conclude with analyzing the convergence of our perturbative result by comparing to a numerical computation.

\subsection{Phase diagram and the sublattice symmetry}\customlabel{sec:diagram}{2.A.}

We begin with the model in Eq. (1) with $B=0$, but we rotate the Pauli matrices as $\tau_z \mapsto \tau_x \mapsto \tau_y \mapsto \tau_z$. This brings the Hamiltonian to the form
\begin{equation}
\!H(\mathbf{k};M,\lambda,\delta) \!=\!\!\left(\!\!\begin{array}{cccc}
0
& 0
& \!f(\mathbf{k};M) \!-\! \imi\lambda \sin k_z \!+\! \delta\!
& -\imi \lambda\sin k_x - \lambda\sin k_y \\
0
& 0
& -\imi \lambda\sin k_x + \lambda\sin k_y 
& \!f(\mathbf{k};M) \!+\!\imi \lambda\sin k_z \!+\! \delta \! \\
\!f(\mathbf{k};M) \!+\! \imi \lambda\sin k_z \!-\! \delta \!
& \imi \lambda\sin k_x + \lambda\sin k_y
& 0
& 0\\
 \imi \lambda\sin k_x - \lambda\sin k_y
& \!f(\mathbf{k};M) \!-\!\imi \lambda\sin k_z \!-\! \delta\!
& 0
& 0
\end{array}\!\!\right),\!\label{eqn:ETI-Ham-SLS-basis}
\end{equation}
where $f(\mathbf{k};M) =  \sum_{j=x,y,z} \cos k_j - M$. Note that the rotated basis reveals a previously hidden sublattice symmetry (which is broken for finite $\propto B$). We denote the lower-left (upper-right) block as $h_\textrm{LL}^{\phantom{^\dagger}}$ ($h_\textrm{UR}^\dagger$), where we dropped the dependence on parameters for brevity. Note that because of the non-Hermitian term $\propto \delta$ we have $h_\textrm{LL} \neq h_\textrm{UR}$. Note that the point gap of Hamiltonian $H$ at $E=0$, implies through $0\neq \det H = \det h_\textrm{LL} (\det h_\textrm{UR})^*$ that each of the blocks respects the point gap too (the asterisk `$^*$' indicates complex conjugation), meaning that both matrix blocks have a well-defined inverse.

Recall from Eq.~(5) in the main text that the winding number $w_{3\textrm{D}}$ for a point gap at $E=0$ is determined from matrices $Q_i = H^{-1}\partial_i H$ where $\partial_i \equiv \partial_{k_i}$. It follows from the block-off-diagonal form of the Hamiltonian that $H^{-1}$ is block-off-diagonal with $(h_\textrm{UR}^{\dagger})^{-1}$ in the lower-left and $h_\textrm{LL}^{-1}$ upper-right position, such that 
\begin{equation}
Q_i = \left(\begin{array}{cc} 
h_\textrm{LL}^{-1} \partial_i h_\textrm{LL}^{\phantom{-1}} & \triv \\
\triv & (h_\textrm{UR}^\dagger)^{-1} \partial_i h_\textrm{UR}^\dagger
\end{array}\right) 
\end{equation}
is block-diagonal. As a consequence, it is possible to write the integrand in Eq.~(5) as a sum of two contributions, which can be through simple manipulations brought into the form 
\begin{equation}
\Tr[Q_iQ_jQ_k] = \Tr[h_{\textrm{LL},i}h_{\textrm{LL},j}h_{\textrm{LL},k}] - \Tr[h_{\textrm{UR},i}h_{\textrm{UR},j}h_{\textrm{UR},k}],
\end{equation}
where $h_{\textrm{block},i} \equiv \partial_{k_i} h_\textrm{block}$. Therefore, the total winding number can be expressed as a difference of winding numbers over the two blocks, 
\begin{equation}
w_{3\textrm{D}} = w_{3\textrm{D}}^{\textrm{LL}} - w_{3\textrm{D}}^{\textrm{UR}}.\label{eqn:sublattice-sum-rule}
\end{equation}
The existence of two integer point-gap invariants in the presence of sublattice symmetry is consistent with the classification tables, such as in Ref.~\onlinecite{Kawabata:2019}. (Note that for non-Hermitian systems the definition of \emph{sublattice symmetry} differs from the definition of \emph{chiral symmetry}! We also remark that for Hermitian systems $w_{3\textrm{D}}^{\textrm{LL}} = w_{3\textrm{D}}^{\textrm{UR}}$, such that $w_{3\textrm{D}}$ vanishes.)

In the following we indicate the the pair of winding numbers as $(w_{3\textrm{D}}^{\textrm{LL}},w_{3\textrm{D}}^{\textrm{UR}})$, and we investigate the phase diagram in the parameter space of $(M,\lambda,\delta)$. Note that the two blocks can be expressed as 
\begin{equation}
h_\textrm{LL}(\mathbf{k};M,\lambda,\delta) = h_0(\mathbf{k};M+\delta,\lambda) \qquad\textrm{and}\qquad  h_\textrm{UR}(\mathbf{k};M,\lambda,\delta) = h_0(\mathbf{k};M-\delta,\lambda)
\end{equation}
where
\begin{equation}
h_0(\mathbf{k};m,\lambda) = \Big(\sum_{j=x,y,z} \cos k_j-m\Big)\sigma_0 + \imi \lambda \sum_{j=x,y,z}\sin k_j \sigma_j,
\end{equation}
such that both topological invariants can be extracted by studying the properties of a single block. The point gap of $h_0$ closes when for some $\mathbf{k}$ the determinant vanishes, which corresponds to a pair of conditions
\begin{equation}
\sum_{j=x,y,z}\cos k_j = m\qquad\quad\textrm{and} \qquad\quad \lambda^2 \Big(\sum_{j=x,y,z} \sin^2 k_j\Big) =0.
\end{equation}
Assuming $\lambda\neq 0$, the latter condition is fulfilled at TRIMs, in which case the first condition can only be satifies for $m\in\{\pm 1,\pm 3\}$. This observation fixes the boundaries of the phase diagram. Computing the winding number $w_{3\textrm{D}}^0$ then reveals the phase diagram displayed in Supplementary Fig.~\ref{fig:phase-diagram}a. The phase diagram of the Hamiltonian in Eq.~(\ref{eqn:ETI-Ham-SLS-basis}) is then obtained easily by substituting for the two blocks $m\mapsto M\pm \delta$, and is displayed in Supplementary Fig.~\ref{fig:phase-diagram}b.

\begin{figure}[t!]
\centering
    \includegraphics[width=0.99\linewidth]{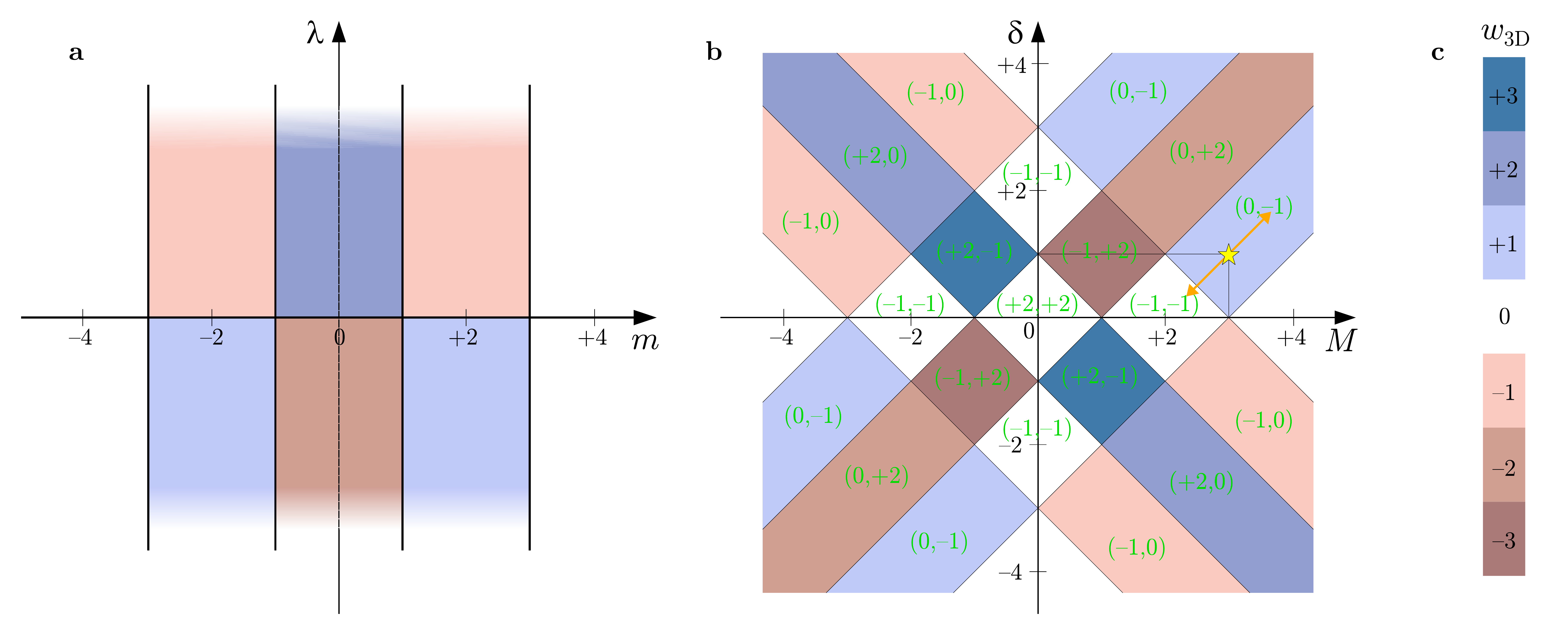}
    \caption{\label{fig:phase-diagram}\textbf{Phase diagrams for the ETI model.} \textbf{a}~Winding number $w_{3\mathrm{D}}$ of the $2\times 2$ block $h_0(\mathbf{k};m,\lambda)$ in the $(m,\lambda)$-parameter space. \textbf{b}~Phase diagram for the model in Eq.~(1) of the main text for $B=0$ and $\lambda>0$ inside the $(M,\delta)$-parameter plane. The pair of green numbers indicate the two invariants $(w_{3\textrm{D}}^{\textrm{LL}},w_{3\textrm{D}}^{\textrm{UR}})$ enabled by the sublattice symmetry, which obey the sum rule in Eq.~(\ref{eqn:sublattice-sum-rule}). All invariants reverse sign for $\lambda < 0$. The star indicates the values $M=+3,\delta=+1$ considered throughout the manuscript, and the orange line indicates the range of parameter treated analytically for $k_x = k_y = 0$ in Supplementary Notes~\ref{sec:slab} and~\ref{sec:poly-expansion} \textbf{c}~Color scheme for $w_{3\textrm{D}}$ utilized in the previous panels.} 
\end{figure}

\subsection{Slab Hamiltonian and the infernal point}\customlabel{sec:slab}{2.B.}

In this section we first present the Hamiltonian for a slab geometry with open boundary condition in $z$-direction. We use the obtained Hamiltonian to reveal the macroscopic defectiveness at zero surface momentum, which we term \emph{infernal point}. Adopting the rotated basis of Eq.~(\ref{eqn:ETI-Ham-SLS-basis}) the ETI slab Hamiltonian can be expressed in a block form as
\begin{equation}
H_\textrm{slab}(k_x,k_y;M,\lambda,\delta;N) = \left(\begin{array}{cccccccc}
H_0 & H_+ & \triv & \triv & \cdots & \triv & \triv & \triv \\
H_- & H_0 & H_+ & \triv & \cdots & \triv & \triv & \triv \\
\triv & H_- & H_0 & H_+  & \cdots & \triv & \triv & \triv \\
\triv & \triv & H_- & H_0 &  \ddots & \triv & \triv & \triv \\
\vdots & \vdots & \vdots & \ddots  & \ddots & \ddots & \vdots & \vdots \\
\triv & \triv & \triv & \triv &  \ddots & H_0 & H_+ & \triv \\
\triv & \triv & \triv & \triv &  \cdots & H_- & H_0 & H_+ \\
\triv & \triv & \triv & \triv &  \cdots & \triv & H_- & H_0 
\end{array}\right)
\end{equation}
where $N$ is the number of layers (blocks), the diagonal blocks are
\begin{equation}
H_0(k_x,k_y;M,\lambda,\delta) = \left(\begin{array}{cccc}
0 
& 0 
& \tilde{f}(k_x,k_y;M)+\delta 
& -\imi \lambda\sin k_x - \lambda\sin k_y \\
0 
& 0 
& -\imi \lambda \sin k_x + \lambda \sin k_y
& \tilde{f}(k_x,k_y;M)+\delta  \\
\tilde{f}(k_x,k_y;M)-\delta 
& \imi \lambda \sin k_x + \lambda \sin k_y 
& 0
& 0 \\
\imi \lambda \sin k_x - \lambda \sin k_y
& \tilde{f}(k_x,k_y;M)-\delta
& 0
& 0 
\end{array}\right)\label{eqn:diag-block}
\end{equation}
where $\tilde{f}(k_x,k_y;M) = \cos k_x + \cos k_y - M$, and the next-to-diagonal blocks are
\begin{equation}
H_+ = \left(\begin{array}{cccc}
0 & 0 & \tfrac{1-\lambda}{2} & 0 \\
0 & 0 & 0 & \tfrac{1+\lambda}{2} \\
\tfrac{1+\lambda}{2} & 0 & 0 & 0 \\
0 & \tfrac{1-\lambda}{2} & 0 & 0 
\end{array}\right)\qquad \textrm{and}\qquad H_- = \left(\begin{array}{cccc}
0 & 0 & \tfrac{1+\lambda}{2} & 0 \\
0 & 0 & 0 &\tfrac{1-\lambda}{2} \\
\tfrac{1-\lambda}{2} & 0 & 0 & 0 \\
0 & \tfrac{1+\lambda}{2} & 0 & 0 
\end{array}\right) = H_+^\top.
\end{equation}
To proceed, we further consider $\lambda = +1$, and $M = 2 + \delta$, which includes the case $M=+3$ and $\delta=+1$ considered in the main text as a special case. (An analogous analytical treatment is also possible for $M = 2 - \delta$.) We keep the number of layers $N$ as a free variable.

For the specified parameters, we aim to derive the spectrum for $k_x = k_y = 0$ analytically by analyzing the characteristic polynomial $\det[H_\textrm{slab}(k_x\!=\!0,\,k_y\!=\!0;\,M\!=\!2\!+\!\delta,\,\lambda\!=\!+1,\,\delta;\,N) - E \mathbbold{1}]$. For these parameters, the diagonal block of $(H_\textrm{slab}-E\mathbbold{1})$ simplifies to
\begin{equation}
H_0 -E \unit = \left(\begin{array}{cccc}
-E & 0 & 0 & 0 \\
0 & -E & 0 & 0 \\
-2\delta & 0 & -E & 0 \\
0 & -2\delta & 0 & -E
\end{array}\right)\qquad\textrm{with inverse}\;\; [H_0 -\lambda \unit]^{-1} = \frac{1}{E^2}\left(\begin{array}{cccc}
-E & 0 & 0 & 0 \\
0 & -E & 0 & 0 \\
2\delta & 0 & -E & 0 \\
0 & 2\delta & 0 & -E 
\end{array}\right),
\end{equation}
and the next-to-diagonal blocks reduce to
\begin{equation}
H_+ = \left(\begin{array}{cccc}
0 & 0 & 0 & 0 \\
0 & 0 & 0 & 1 \\
1 & 0 & 0 & 0 \\
0 & 0 & 0 & 0 
\end{array}\right)\qquad \textrm{and}\qquad H_- = \left(\begin{array}{cccc}
0 & 0 & 1 & 0 \\
0 & 0 & 0 & 0 \\
0 & 0 & 0 & 0 \\
0 & 1 & 0 & 0 
\end{array}\right) = H_+^\top.
\end{equation}
For brevity, we will write $H_\textrm{slab}(0,0,;2+\delta,1,\delta;N)-E\unit \equiv \tilde{H}_\textrm{slab}(N)$. Motivated by Ref.~\onlinecite{Terrier:2020}, we use Schur's determinant identity
\begin{equation}
\det\left(\begin{array}{cc} A & B \\ C & D\end{array}\right) = \det(D)\det(A - B \cdot D^{-1} \cdot C)
\end{equation}
where $A$ and $D$ are square matrices (not necessarily of the same dimension), and it is assumed that $\det D \neq 0$. Setting $D = H_0$ corresponding to the bottom-right $4\times4$ block of $\tilde{H}_\textrm{slab}(N)$, we find  
\begin{equation}
- B  \cdot D^{-1} \cdot C = \left(\begin{array}{ccc}
\ddots & \vdots & \vdots \\
\cdots & \triv  & \triv \\
\cdots & \triv & M
\end{array}\right)\qquad\textrm{with}\;\; M=-H_+\cdot [H_0^{-1} - E\mathbbold{1}]^{-1} \cdot H_- = \left(\begin{array}{cccc}
0 & 0 & 0 & 0 \\
0 & E^{-1} & 0 & 0 \\
0 & 0 & E^{-1} & 0 \\
0 & 0 & 0 & 0 
\end{array}\right)\label{eqn:recur}
\end{equation}
and that 
\begin{equation}
\det \tilde H_\textrm{slab}(N) = E^4 \det [\tilde H_\textrm{slab}(N-1)  - B  \cdot D^{-1} \cdot C].
\end{equation}
The computation unexpectedly simplifies at the next step of applying the Shur's determinant identity. Note that the bottom-right block of $(\tilde H_\textrm{slab}(N-1)  - B  \cdot D^{-1} \cdot C)$ 
is
\begin{equation}
D ' = \left(\begin{array}{cccc}
-E  & 0 & 0 & 0 \\
0 & E^{-1}-E & 0 & 0 \\
-2\delta & 0 & E^{-1}-E & 0 \\
0 & -2\delta & 0 & -E \\
\end{array}\right)\qquad\textrm{with}\;\;\det{D'} = (E^2 - 1)^2,\label{eqn:D-prime}
\end{equation}
and with the same product $-B'\cdot (D')^{-1}\cdot C'$ as obtained in the previous step in Eq.~(\ref{eqn:recur}). We therefore obtain 
\begin{equation}
\det \tilde H_\textrm{slab}(j+2) = (E^2 - 1)^2 \det [\tilde H_\textrm{slab}(j+1)  - B^{(j)}  \cdot (D')^{-1} \cdot C^{(j)}]
\end{equation} 
where 
\begin{equation}
C^{(j)} = \big(\underbrace{\triv\;\,\triv\;\,\cdots\;\, \triv}_{\textrm{$j$ blocks}}\;\, \!\!H_-\big)\qquad\textrm{and}\qquad B^{(j)} = \big(C^{(j)}\big)^\top,
\end{equation}
which remains valid for all $N-3 \geq j \geq 0$. In the last step, knowing that $\tilde H_\textrm{slab}(1) = H_0-E\mathbbold{1}$, we find that $(H_\textrm{slab}(1)  - H_+ \cdot (D')^{-1} H_-) = D'$ with the determinant shown in Eq.~(\ref{eqn:D-prime}). Altogether, we obtain
\begin{equation}
\det \tilde{H}_\textrm{slab}(N) = E^4 (E^2 - 1)^{2(N-1)}.\label{eqn:char-polynomial}
\end{equation}
Therefore, for the chosen parameters ($k_x=k_y=0$, $\lambda = 1$, and $M-\delta=2$), the characteristic polynomial has $2N - 2$ roots at $E=+1$, $2N-2$ roots at $E=-1$, and $4$ roots at $E=0$. 

Knowing the eigenenergies, it is also possible to search for the corresponding eigenstates. Parametrizing a generical eigenvector as 
\begin{equation}
\psi = \left(\ldots \,,\; d^{(j-1)} \;\, | \;\, a^{(j)} \,,\; b^{(j)} \,,\; c^{(j)} \,,\; d^{(j)} \;\, | \;\, a^{(j+1)} \;\,\ldots  \right)^\top\qquad\textrm{where}\quad 1 \leq j \leq N
\end{equation}
the eigentates correspond to solutions of
\begin{equation}\label{eqn:eigenvector-equations}
  \begin{split}
    E a^{(1)} &= 0\\
    E b^{(N)} &= 0 \\
    E c^{(N)} &= -2\delta a^{(N)}\\
    E d^{(1)} &= -2\delta b^{(1)}
  \end{split}
\qquad\textrm{and}\qquad
  \begin{split}
    E a^{(j)} &= c^{(j-1)}\qquad\textrm{for $2\leq j \leq N$}\\
    E b^{(j)} &= d^{(j+1)} \qquad\textrm{for $1\leq j \leq N-1$}\\
    E c^{(j)} &= -2\delta a^{(j)}+ a^{(j+1)}\qquad\textrm{for $1\leq j \leq N-1$}\\
    E d^{(j)} &= b^{(j-1)}-2\delta b^{(j)}\qquad\textrm{for $2\leq j \leq N$}.
  \end{split}
\end{equation}
Assuming $\delta\neq 0$, such that the Hamiltonian is non-Hermitian, there are (not caring about the normalization) the following linearly independent eigenstates:
\begin{itemize}
\item two eigenstates at $E=0$, localized on opposite surfaces, namely $(0,0,0,1|0,\ldots)^\top$ and $(\ldots0|0,0,1,0)^\top$ (both have a single non-vanishing component).
\item a pair of eigenstates at each $E=\pm 1$, also localized on opposite surfaces, namely $(0,\pm 1,0,-2\delta|0,0,0,1| 0\ldots)^\top$ and $(\ldots 0 | 0,0, 1, 0 | \pm 1,0,-2\delta,0)^\top$ (they all have exactly three non-vanishing components).
\end{itemize}
Altogether, the Hamiltonian with dimension $4N$ has only $6$ distinct eigenstates, indicating a huge defectiveness. (We remark that for $\delta=0$ the Hamiltonian becomes Hermitian, and contains $4N$ distinct eigenstates as expected.) 

\subsection{Expansion near the infernal point}~\customlabel{sec:poly-expansion}{2.C.}

\begin{figure}[b!]
\centering
    \includegraphics[width=0.99\linewidth]{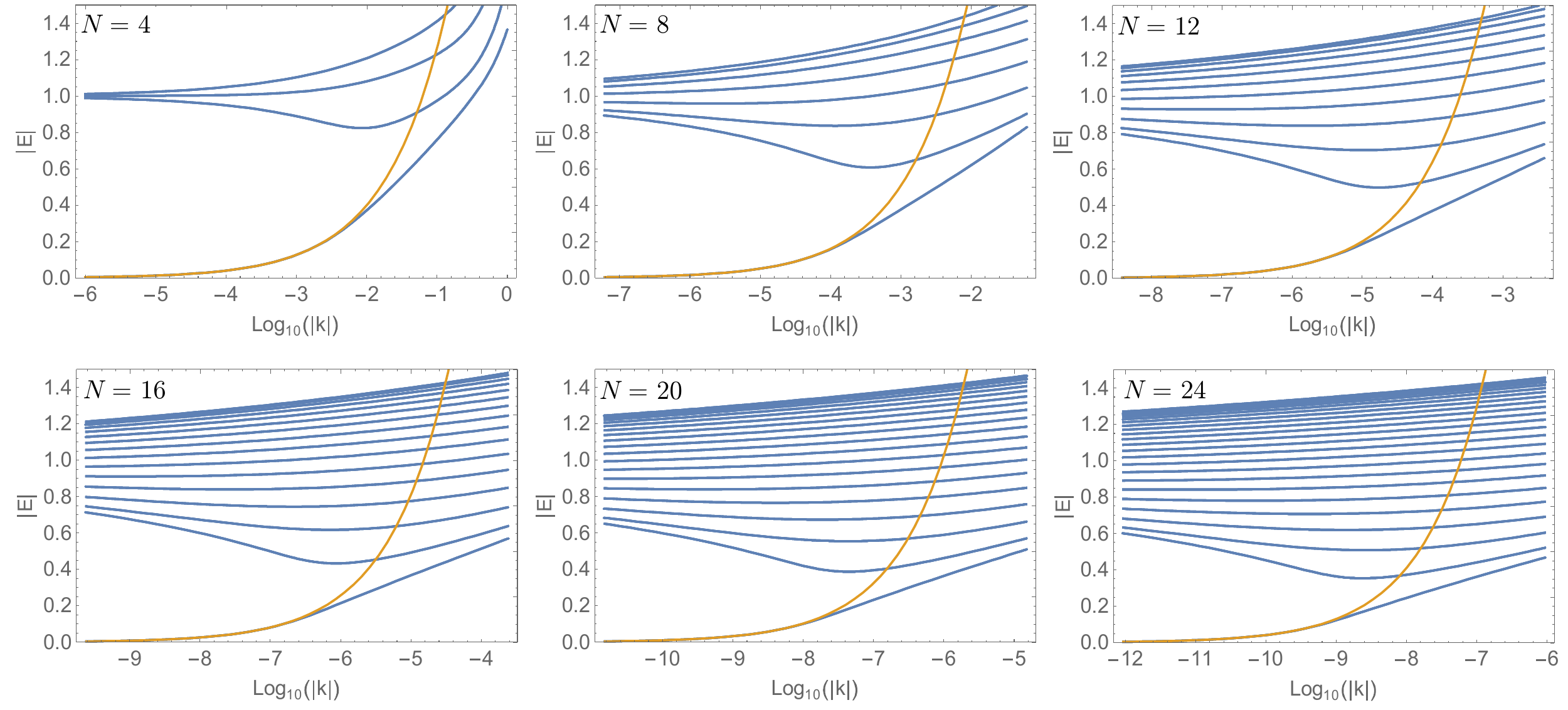}
    \caption{\textbf{Dispersion of the states near the infernal point $k_x\!=\!k_y=0$, $M\!=\!3$, $\delta\!=\!\lambda \!=\!1$ for several small values of $N$.} The blue lines correspond to numerically found eigenvalues, while the yellow line indicates the theoretically predicted dispersion of the zero-energy states in Eq.~(\ref{eqn:sq-root-disp}). We observe that the theoretical result approximates the spectrum well for energies $E\lesssim 0.2$, which corresponds to an exponentially shrinking interval of momenta around the infernal point (notice the logarithmic scale on the horizontal axis). In the thermodynamic limit we expect the spectrum of the slab Hamiltonian with open boundary condition to exhibit a discontinuous jump between $E=0$ and a finite value of energy. }\label{fig:slope} 
\end{figure}

We now extend to a small but finite $k_x$ and $k_y$ by assuming a linear expansion of $\sin k_i \approx k_i$ and $\cos k_i \approx 1$ in Eq.~(\ref{eqn:diag-block}), while keeping $\lambda = 1$ and $M = \delta + 2$. This changes the diagonal block (with substracted $\lambda \unit$) to
\begin{equation}
H_0 -E \unit = \left(\begin{array}{cccc}
-E & 0 & 0 & -\imi k_x - k_y \\
0 & -E & -\imi k_x +k_y& 0 \\
-2\delta & \imi k_x +k_y& -E & 0 \\
\imi k_x - k_y& -2\delta  & 0 & -E 
\end{array}\right).
\end{equation}
At this stage an exact analytic solution becomes infeasible. Nevertheless, one can still deduce some properties about the dispersion near the zero-energy states of the infernal point by considering the expansion of the characteristic polynomial for small values of $k_x$, $k_y$ and $E$. We only need to keep the contributions to the polynomial that are of the lowest order in these variables. These correspond to the following two contributions to the determinant, 
{\tiny \begin{equation}
\left(
\begin{array}{c|c|c|c||c|c|c|c||c|c|c|c||c||c|c|c|c||c|c|c|c||c|c|c|c}
\times  & 
\phantom{\times}  &  
\phantom{\times} & 
\phantom{\times}  &
    \phantom{\times}  &  
    \phantom{\times} & 
    \phantom{\times} &  
    \phantom{\times} &
        \phantom{\times} &  
        \phantom{\times} & 
        \phantom{\times} &   
        \phantom{\times}  
        & \cdots & \
\phantom{\times}  & 
\phantom{\times}  &  
\phantom{\times} & 
\phantom{\times}  &
    \phantom{\times}  &  
    \phantom{\times} & 
    \phantom{\times} &  
    \phantom{\times} &
        \phantom{\times} &  
        \phantom{\times} & 
        \phantom{\times} &   
        \phantom{\times} \\ \hline
\phantom{\times}  & 
\phantom{\times}  &  
\phantom{\times} & 
\phantom{\times}  &
    \phantom{\times}  &  
    \phantom{\times} & 
    \phantom{\times} &  
    \times &
        \phantom{\times} &  
        \phantom{\times} & 
        \phantom{\times} &   
        \phantom{\times}  
        & \cdots & \
\phantom{\times}  & 
\phantom{\times}  &  
\phantom{\times} & 
\phantom{\times}  &
    \phantom{\times}  &  
    \phantom{\times} & 
    \phantom{\times} &  
    \phantom{\times} &
        \phantom{\times} &  
        \phantom{\times} & 
        \phantom{\times} &   
        \phantom{\times}  \\ \hline
\phantom{\times}  & 
\phantom{\times}  &  
\phantom{\times} & 
\phantom{\times}  &
    \times  &  
    \phantom{\times} & 
    \phantom{\times} &  
    \phantom{\times} &
        \phantom{\times} &  
        \phantom{\times} & 
        \phantom{\times} &   
        \phantom{\times}   
        & \cdots & \
\phantom{\times}  & 
\phantom{\times}  &  
\phantom{\times} & 
\phantom{\times}  &
    \phantom{\times}  &  
    \phantom{\times} & 
    \phantom{\times} &  
    \phantom{\times} &
        \phantom{\times} &  
        \phantom{\times} & 
        \phantom{\times} &   
        \phantom{\times} \\ \hline
\phantom{\times}  & 
\phantom{\times}  &  
\phantom{\times} & 
\times &
    \phantom{\times}  &  
    \phantom{\times} & 
    \phantom{\times} &  
    \phantom{\times} &
        \phantom{\times} &  
        \phantom{\times} & 
        \phantom{\times} &   
        \phantom{\times}   
        & \cdots & \
\phantom{\times}  & 
\phantom{\times}  &  
\phantom{\times} & 
\phantom{\times}  &
    \phantom{\times}  &  
    \phantom{\times} & 
    \phantom{\times} &  
    \phantom{\times} &
        \phantom{\times} &  
        \phantom{\times} & 
        \phantom{\times} &   
        \phantom{\times} \\ \hline\hline
\phantom{\times}  & 
\phantom{\times}  &  
\times & 
\phantom{\times}  &
    \phantom{\times}  &  
    \phantom{\times} & 
    \phantom{\times} &  
    \phantom{\times} &
        \phantom{\times} &  
        \phantom{\times} & 
        \phantom{\times} &   
        \phantom{\times}   
        & \cdots & \
\phantom{\times}  & 
\phantom{\times}  &  
\phantom{\times} & 
\phantom{\times}  &
    \phantom{\times}  &  
    \phantom{\times} & 
    \phantom{\times} &  
    \phantom{\times} &
        \phantom{\times} &  
        \phantom{\times} & 
        \phantom{\times} &   
        \phantom{\times} \\ \hline
\phantom{\times}  & 
\phantom{\times}  &  
\phantom{\times} & 
\phantom{\times}  &
    \phantom{\times}  &  
    \phantom{\times} & 
    \phantom{\times} &  
    \phantom{\times} &
        \phantom{\times} &  
        \phantom{\times} & 
        \phantom{\times} &   
        \times  
        & \cdots & \
\phantom{\times}  & 
\phantom{\times}  &  
\phantom{\times} & 
\phantom{\times}  &
    \phantom{\times}  &  
    \phantom{\times} & 
    \phantom{\times} &  
    \phantom{\times} &
        \phantom{\times} &  
        \phantom{\times} & 
        \phantom{\times} &   
        \phantom{\times} \\ \hline
\phantom{\times}  & 
\phantom{\times}  &  
\phantom{\times} & 
\phantom{\times}  &
    \phantom{\times}  &  
    \phantom{\times} & 
    \phantom{\times} &  
    \phantom{\times} &
        \times &  
        \phantom{\times} & 
        \phantom{\times} &   
        \phantom{\times}   
        & \cdots & \
\phantom{\times}  & 
\phantom{\times}  &  
\phantom{\times} & 
\phantom{\times}  &
    \phantom{\times}  &  
    \phantom{\times} & 
    \phantom{\times} &  
    \phantom{\times} &
        \phantom{\times} &  
        \phantom{\times} & 
        \phantom{\times} &   
        \phantom{\times} \\ \hline
\phantom{\times}  & 
\times  &  
\phantom{\times} & 
\phantom{\times}  &
    \phantom{\times}  &  
    \phantom{\times} & 
    \phantom{\times} &  
    \phantom{\times} &
        \phantom{\times} &  
        \phantom{\times} & 
        \phantom{\times} &   
        \phantom{\times}  
        & \cdots & \
\phantom{\times}  & 
\phantom{\times}  &  
\phantom{\times} & 
\phantom{\times}  &
    \phantom{\times}  &  
    \phantom{\times} & 
    \phantom{\times} &  
    \phantom{\times} &
        \phantom{\times} &  
        \phantom{\times} & 
        \phantom{\times} &   
        \phantom{\times} \\ \hline\hline
\phantom{\times}  & 
\phantom{\times}  &  
\phantom{\times} & 
\phantom{\times}  &
    \phantom{\times}  &  
    \phantom{\times} & 
    \times &  
    \phantom{\times} &
        \phantom{\times} &  
        \phantom{\times} & 
        \phantom{\times} &   
        \phantom{\times}  
        & \cdots & \
\phantom{\times}  & 
\phantom{\times}  &  
\phantom{\times} & 
\phantom{\times}  &
    \phantom{\times}  &  
    \phantom{\times} & 
    \phantom{\times} &  
    \phantom{\times} &
        \phantom{\times} &  
        \phantom{\times} & 
        \phantom{\times} &   
        \phantom{\times} \\ \hline
\phantom{\times}  & 
\phantom{\times}  &  
\phantom{\times} & 
\phantom{\times}  &
    \phantom{\times}  &  
    \phantom{\times} & 
    \phantom{\times} &  
    \phantom{\times} &
        \phantom{\times} &  
        \phantom{\times} & 
        \phantom{\times} &   
        \phantom{\times}  
        & \cdots & \
\phantom{\times}  & 
\phantom{\times}  &  
\phantom{\times} & 
\phantom{\times}  &
    \phantom{\times}  &  
    \phantom{\times} & 
    \phantom{\times} &  
    \phantom{\times} &
        \phantom{\times} &  
        \phantom{\times} & 
        \phantom{\times} &   
        \phantom{\times} \\ \hline
\phantom{\times}  & 
\phantom{\times}  &  
\phantom{\times} & 
\phantom{\times}  &
    \phantom{\times}  &  
    \phantom{\times} & 
    \phantom{\times} &  
    \phantom{\times} &
        \phantom{\times} &  
        \phantom{\times} & 
        \phantom{\times} &   
        \phantom{\times}  
        & \cdots & \
\phantom{\times}  & 
\phantom{\times}  &  
\phantom{\times} & 
\phantom{\times}  &
    \phantom{\times}  &  
    \phantom{\times} & 
    \phantom{\times} &  
    \phantom{\times} &
        \phantom{\times} &  
        \phantom{\times} & 
        \phantom{\times} &   
        \phantom{\times} \\ \hline
\phantom{\times}  & 
\phantom{\times} &
\phantom{\times} &
\phantom{\times}  &
    \phantom{\times}  & 
    \times &   
    \phantom{\times} & 
    \phantom{\times} &
        \phantom{\times} &  
        \phantom{\times} & 
        \phantom{\times} &   
        \phantom{\times}  
        & \cdots & \
\phantom{\times}  & 
\phantom{\times}  &  
\phantom{\times} & 
\phantom{\times}  &
    \phantom{\times}  &  
    \phantom{\times} & 
    \phantom{\times} &  
    \phantom{\times} &
        \phantom{\times} &  
        \phantom{\times} & 
        \phantom{\times} &   
        \phantom{\times} \\ \hline\hline
\vdots & \vdots & \vdots & \vdots &
\vdots & \vdots & \vdots & \vdots &
\vdots & \vdots & \vdots & \vdots &
\ddots &
\vdots & \vdots & \vdots & \vdots &
\vdots & \vdots & \vdots & \vdots &
\vdots & \vdots & \vdots & \vdots \\ \hline\hline    
\phantom{\times}  & 
\phantom{\times}  &  
\phantom{\times} & 
\phantom{\times}  &
    \phantom{\times}  &  
    \phantom{\times} & 
    \phantom{\times} &  
    \phantom{\times} &
        \phantom{\times} &  
        \phantom{\times} & 
        \phantom{\times} &   
        \phantom{\times} 
        & \cdots & \
\phantom{\times}  & 
\phantom{\times}  &  
\phantom{\times} & 
\phantom{\times}  &
    \phantom{\times}  &  
    \phantom{\times} & 
    \phantom{\times} &  
    \phantom{\times} &
        \phantom{\times} &  
        \phantom{\times} & 
        \phantom{\times} &   
        \phantom{\times}   \\ \hline
\phantom{\times}  & 
\phantom{\times}  &  
\phantom{\times} & 
\phantom{\times}  &
    \phantom{\times}  &  
    \phantom{\times} & 
    \phantom{\times} &  
    \phantom{\times} &
        \phantom{\times} &  
        \phantom{\times} & 
        \phantom{\times} &   
        \phantom{\times} 
        & \cdots & \
\phantom{\times}  & 
\phantom{\times}  &  
\phantom{\times} & 
\phantom{\times}  &
    \phantom{\times}  &  
    \phantom{\times} & 
    \phantom{\times} &  
    \times &
        \phantom{\times} &  
        \phantom{\times} & 
        \phantom{\times} &   
        \phantom{\times}  \\ \hline
\phantom{\times}  & 
\phantom{\times}  &  
\phantom{\times} & 
\phantom{\times}  &
    \phantom{\times}  &  
    \phantom{\times} & 
    \phantom{\times} &  
    \phantom{\times} &
        \phantom{\times} &  
        \phantom{\times} & 
        \phantom{\times} &   
        \phantom{\times} 
        & \cdots & \
\phantom{\times}  & 
\phantom{\times}  &  
\phantom{\times} & 
\phantom{\times}  &
    \times  &  
    \phantom{\times} & 
    \phantom{\times} &  
    \phantom{\times} &
        \phantom{\times} &  
        \phantom{\times} & 
        \phantom{\times} &   
        \phantom{\times} \\ \hline
\phantom{\times}  & 
\phantom{\times}  &  
\phantom{\times} & 
\phantom{\times}  &
    \phantom{\times}  &  
    \phantom{\times} & 
    \phantom{\times} &  
    \phantom{\times} &
        \phantom{\times} &  
        \phantom{\times} & 
        \phantom{\times} &   
        \phantom{\times} 
        & \cdots & \
\phantom{\times}  & 
\phantom{\times}  &  
\phantom{\times} & 
\phantom{\times}  &
    \phantom{\times}  &  
    \phantom{\times} & 
    \phantom{\times} &  
    \phantom{\times} &
        \phantom{\times} &  
        \phantom{\times} & 
        \phantom{\times} &   
        \phantom{\times}   \\ \hline\hline
\phantom{\times}  & 
\phantom{\times}  &  
\phantom{\times} & 
\phantom{\times}  &
    \phantom{\times}  &  
    \phantom{\times} & 
    \phantom{\times} &  
    \phantom{\times} &
        \phantom{\times} &  
        \phantom{\times} & 
        \phantom{\times} &   
        \phantom{\times} 
        & \cdots & \
\phantom{\times}  & 
\phantom{\times}  &  
\times & 
\phantom{\times}  &
    \phantom{\times}  &  
    \phantom{\times} & 
    \phantom{\times} &  
    \phantom{\times} &
        \phantom{\times} &  
        \phantom{\times} & 
        \phantom{\times} &   
        \phantom{\times}  \\ \hline
\phantom{\times}  & 
\phantom{\times}  &  
\phantom{\times} & 
\phantom{\times}  &
    \phantom{\times}  &  
    \phantom{\times} & 
    \phantom{\times} &  
    \phantom{\times} &
        \phantom{\times} &  
        \phantom{\times} & 
        \phantom{\times} &   
        \phantom{\times}  
        & \cdots & \
\phantom{\times}  & 
\phantom{\times}  &  
\phantom{\times} & 
\phantom{\times}  &
    \phantom{\times}  &  
    \phantom{\times} & 
    \phantom{\times} &  
    \phantom{\times} &
        \phantom{\times} &  
        \phantom{\times} & 
        \phantom{\times} &   
        \times  \\ \hline
\phantom{\times}  & 
\phantom{\times}  &  
\phantom{\times} & 
\phantom{\times}  &
    \phantom{\times}  &  
    \phantom{\times} & 
    \phantom{\times} &  
    \phantom{\times} &
        \phantom{\times} &  
        \phantom{\times} & 
        \phantom{\times} &   
        \phantom{\times}  
        & \cdots & \
\phantom{\times}  & 
\phantom{\times}  &  
\phantom{\times} & 
\phantom{\times}  &
    \phantom{\times}  &  
    \phantom{\times} & 
    \phantom{\times} &  
    \phantom{\times} &
        \times &  
        \phantom{\times} & 
        \phantom{\times} &   
        \phantom{\times}   \\ \hline
\phantom{\times}  & 
\phantom{\times}  &  
\phantom{\times} & 
\phantom{\times}  &
    \phantom{\times}  &  
    \phantom{\times} & 
    \phantom{\times} &  
    \phantom{\times} &
        \phantom{\times} &  
        \phantom{\times} & 
        \phantom{\times} &   
        \phantom{\times} 
        & \cdots & \
\phantom{\times}  & 
\times  &  
\phantom{\times} & 
\phantom{\times}  &
    \phantom{\times}  &  
    \phantom{\times} & 
    \phantom{\times} &  
    \phantom{\times} &
        \phantom{\times} &  
        \phantom{\times} & 
        \phantom{\times} &   
        \phantom{\times}  \\ \hline\hline
\phantom{\times}  & 
\phantom{\times}  &  
\phantom{\times} & 
\phantom{\times}  &
    \phantom{\times}  &  
    \phantom{\times} & 
    \phantom{\times} &  
    \phantom{\times} &
        \phantom{\times} &  
        \phantom{\times} & 
        \phantom{\times} &   
        \phantom{\times} 
        & \cdots & \
\phantom{\times}  & 
\phantom{\times}  &  
\phantom{\times} & 
\phantom{\times}  &
    \phantom{\times}  &  
    \phantom{\times} & 
    \times &  
    \phantom{\times} &
        \phantom{\times} &  
        \phantom{\times} & 
        \phantom{\times} &   
        \phantom{\times}  \\ \hline
\phantom{\times}  & 
\phantom{\times}  &  
\phantom{\times} & 
\phantom{\times}  &
    \phantom{\times}  &  
    \phantom{\times} & 
    \phantom{\times} &  
    \phantom{\times} &
        \phantom{\times} &  
        \phantom{\times} & 
        \phantom{\times} &   
        \phantom{\times}  
        & \cdots & \
\phantom{\times}  & 
\phantom{\times}  &  
\phantom{\times} & 
\phantom{\times}  &
    \phantom{\times}  &  
    \phantom{\times} & 
    \phantom{\times} &  
    \phantom{\times} &
        \phantom{\times} &  
        \times & 
        \phantom{\times} &   
        \phantom{\times}   \\ \hline
\phantom{\times}  & 
\phantom{\times}  &  
\phantom{\times} & 
\phantom{\times}  &
    \phantom{\times}  &  
    \phantom{\times} & 
    \phantom{\times} &  
    \phantom{\times} &
        \phantom{\times} &  
        \phantom{\times} & 
        \phantom{\times} &   
        \phantom{\times}  
        & \cdots & \
\phantom{\times}  & 
\phantom{\times}  &  
\phantom{\times} & 
\phantom{\times}  &
    \phantom{\times}  &  
    \phantom{\times} & 
    \phantom{\times} &  
    \phantom{\times} &
        \phantom{\times} &  
        \phantom{\times} & 
        \times &   
        \phantom{\times}  \\ \hline
\phantom{\times}  & 
\phantom{\times}  &  
\phantom{\times} & 
\phantom{\times}  &
    \phantom{\times}  &  
    \phantom{\times} & 
    \phantom{\times} &  
    \phantom{\times} &
        \phantom{\times} &  
        \phantom{\times} & 
        \phantom{\times} &   
        \phantom{\times}  
        & \cdots & \
\phantom{\times}  & 
\phantom{\times}  &  
\phantom{\times} & 
\phantom{\times}  &
    \phantom{\times}  &  
    \times & 
    \phantom{\times} &  
    \phantom{\times} &
        \phantom{\times} &  
        \phantom{\times} & 
        \phantom{\times} &   
        \phantom{\times}  
\end{array}\right) = +E^4
\end{equation}}
and 
{\tiny \begin{equation}
\left(
\begin{array}{c|c|c|c||c|c|c|c||c|c|c|c||c||c|c|c|c||c|c|c|c||c|c|c|c}
\phantom{\times}  & 
\phantom{\times}  &  
\phantom{\times} & 
\times  &
    \phantom{\times}  &  
    \phantom{\times} & 
    \phantom{\times} &  
    \phantom{\times} &
        \phantom{\times} &  
        \phantom{\times} & 
        \phantom{\times} &   
        \phantom{\times}  
        & \cdots & \
\phantom{\times}  & 
\phantom{\times}  &  
\phantom{\times} & 
\phantom{\times}  &
    \phantom{\times}  &  
    \phantom{\times} & 
    \phantom{\times} &  
    \phantom{\times} &
        \phantom{\times} &  
        \phantom{\times} & 
        \phantom{\times} &   
        \phantom{\times} \\ \hline
\phantom{\times}  & 
\phantom{\times}  &  
\phantom{\times} & 
\phantom{\times}  &
    \phantom{\times}  &  
    \phantom{\times} & 
    \phantom{\times} &  
    \times &
        \phantom{\times} &  
        \phantom{\times} & 
        \phantom{\times} &   
        \phantom{\times}  
        & \cdots & \
\phantom{\times}  & 
\phantom{\times}  &  
\phantom{\times} & 
\phantom{\times}  &
    \phantom{\times}  &  
    \phantom{\times} & 
    \phantom{\times} &  
    \phantom{\times} &
        \phantom{\times} &  
        \phantom{\times} & 
        \phantom{\times} &   
        \phantom{\times}  \\ \hline
\times  & 
\phantom{\times}  &  
\phantom{\times} & 
\phantom{\times}  &
    \phantom{\times}  &  
    \phantom{\times} & 
    \phantom{\times} &  
    \phantom{\times} &
        \phantom{\times} &  
        \phantom{\times} & 
        \phantom{\times} &   
        \phantom{\times}   
        & \cdots & \
\phantom{\times}  & 
\phantom{\times}  &  
\phantom{\times} & 
\phantom{\times}  &
    \phantom{\times}  &  
    \phantom{\times} & 
    \phantom{\times} &  
    \phantom{\times} &
        \phantom{\times} &  
        \phantom{\times} & 
        \phantom{\times} &   
        \phantom{\times} \\ \hline
\phantom{\times}  & 
\times  &  
\phantom{\times} & 
\phantom{\times}  &
    \phantom{\times}  &  
    \phantom{\times} & 
    \phantom{\times} &  
    \phantom{\times} &
        \phantom{\times} &  
        \phantom{\times} & 
        \phantom{\times} &   
        \phantom{\times}   
        & \cdots & \
\phantom{\times}  & 
\phantom{\times}  &  
\phantom{\times} & 
\phantom{\times}  &
    \phantom{\times}  &  
    \phantom{\times} & 
    \phantom{\times} &  
    \phantom{\times} &
        \phantom{\times} &  
        \phantom{\times} & 
        \phantom{\times} &   
        \phantom{\times} \\ \hline\hline
\phantom{\times}  & 
\phantom{\times}  &  
\times & 
\phantom{\times}  &
    \phantom{\times}  &  
    \phantom{\times} & 
    \phantom{\times} &  
    \phantom{\times} &
        \phantom{\times} &  
        \phantom{\times} & 
        \phantom{\times} &   
        \phantom{\times}   
        & \cdots & \
\phantom{\times}  & 
\phantom{\times}  &  
\phantom{\times} & 
\phantom{\times}  &
    \phantom{\times}  &  
    \phantom{\times} & 
    \phantom{\times} &  
    \phantom{\times} &
        \phantom{\times} &  
        \phantom{\times} & 
        \phantom{\times} &   
        \phantom{\times} \\ \hline
\phantom{\times}  & 
\phantom{\times}  &  
\phantom{\times} & 
\phantom{\times}  &
    \phantom{\times}  &  
    \phantom{\times} & 
    \phantom{\times} &  
    \phantom{\times} &
        \phantom{\times} &  
        \phantom{\times} & 
        \phantom{\times} &   
        \times  
        & \cdots & \
\phantom{\times}  & 
\phantom{\times}  &  
\phantom{\times} & 
\phantom{\times}  &
    \phantom{\times}  &  
    \phantom{\times} & 
    \phantom{\times} &  
    \phantom{\times} &
        \phantom{\times} &  
        \phantom{\times} & 
        \phantom{\times} &   
        \phantom{\times} \\ \hline
\phantom{\times}  & 
\phantom{\times}  &  
\phantom{\times} & 
\phantom{\times}  &
    \times  &  
    \phantom{\times} & 
    \phantom{\times} &  
    \phantom{\times} &
        \phantom{\times} &  
        \phantom{\times} & 
        \phantom{\times} &   
        \phantom{\times}   
        & \cdots & \
\phantom{\times}  & 
\phantom{\times}  &  
\phantom{\times} & 
\phantom{\times}  &
    \phantom{\times}  &  
    \phantom{\times} & 
    \phantom{\times} &  
    \phantom{\times} &
        \phantom{\times} &  
        \phantom{\times} & 
        \phantom{\times} &   
        \phantom{\times} \\ \hline
\phantom{\times}  & 
\phantom{\times}  &  
\phantom{\times} & 
\phantom{\times}  &
    \phantom{\times}  &  
    \times & 
    \phantom{\times} &  
    \phantom{\times} &
        \phantom{\times} &  
        \phantom{\times} & 
        \phantom{\times} &   
        \phantom{\times}  
        & \cdots & \
\phantom{\times}  & 
\phantom{\times}  &  
\phantom{\times} & 
\phantom{\times}  &
    \phantom{\times}  &  
    \phantom{\times} & 
    \phantom{\times} &  
    \phantom{\times} &
        \phantom{\times} &  
        \phantom{\times} & 
        \phantom{\times} &   
        \phantom{\times} \\ \hline\hline
\phantom{\times}  & 
\phantom{\times}  &  
\phantom{\times} & 
\phantom{\times}  &
    \phantom{\times}  &  
    \phantom{\times} & 
    \times &  
    \phantom{\times} &
        \phantom{\times} &  
        \phantom{\times} & 
        \phantom{\times} &   
        \phantom{\times}  
        & \cdots & \
\phantom{\times}  & 
\phantom{\times}  &  
\phantom{\times} & 
\phantom{\times}  &
    \phantom{\times}  &  
    \phantom{\times} & 
    \phantom{\times} &  
    \phantom{\times} &
        \phantom{\times} &  
        \phantom{\times} & 
        \phantom{\times} &   
        \phantom{\times} \\ \hline
\phantom{\times}  & 
\phantom{\times}  &  
\phantom{\times} & 
\phantom{\times}  &
    \phantom{\times}  &  
    \phantom{\times} & 
    \phantom{\times} &  
    \phantom{\times} &
        \phantom{\times} &  
        \phantom{\times} & 
        \phantom{\times} &   
        \phantom{\times}  
        & \cdots & \
\phantom{\times}  & 
\phantom{\times}  &  
\phantom{\times} & 
\phantom{\times}  &
    \phantom{\times}  &  
    \phantom{\times} & 
    \phantom{\times} &  
    \phantom{\times} &
        \phantom{\times} &  
        \phantom{\times} & 
        \phantom{\times} &   
        \phantom{\times} \\ \hline
\phantom{\times}  & 
\phantom{\times}  &  
\phantom{\times} & 
\phantom{\times}  &
    \phantom{\times}  &  
    \phantom{\times} & 
    \phantom{\times} &  
    \phantom{\times} &
        \times &  
        \phantom{\times} & 
        \phantom{\times} &   
        \phantom{\times}  
        & \cdots & \
\phantom{\times}  & 
\phantom{\times}  &  
\phantom{\times} & 
\phantom{\times}  &
    \phantom{\times}  &  
    \phantom{\times} & 
    \phantom{\times} &  
    \phantom{\times} &
        \phantom{\times} &  
        \phantom{\times} & 
        \phantom{\times} &   
        \phantom{\times} \\ \hline
\phantom{\times}  & 
\phantom{\times} &
\phantom{\times} &
\phantom{\times}  &
    \phantom{\times}  & 
    \phantom{\times} &   
    \phantom{\times} & 
    \phantom{\times} &
        \phantom{\times} &  
        \times & 
        \phantom{\times} &   
        \phantom{\times}  
        & \cdots & \
\phantom{\times}  & 
\phantom{\times}  &  
\phantom{\times} & 
\phantom{\times}  &
    \phantom{\times}  &  
    \phantom{\times} & 
    \phantom{\times} &  
    \phantom{\times} &
        \phantom{\times} &  
        \phantom{\times} & 
        \phantom{\times} &   
        \phantom{\times} \\ \hline\hline
\vdots & \vdots & \vdots & \vdots &
\vdots & \vdots & \vdots & \vdots &
\vdots & \vdots & \vdots & \vdots &
\ddots &
\vdots & \vdots & \vdots & \vdots &
\vdots & \vdots & \vdots & \vdots &
\vdots & \vdots & \vdots & \vdots \\ \hline\hline    
\phantom{\times}  & 
\phantom{\times}  &  
\phantom{\times} & 
\phantom{\times}  &
    \phantom{\times}  &  
    \phantom{\times} & 
    \phantom{\times} &  
    \phantom{\times} &
        \phantom{\times} &  
        \phantom{\times} & 
        \phantom{\times} &   
        \phantom{\times} 
        & \cdots & \
\phantom{\times}  & 
\phantom{\times}  &  
\phantom{\times} & 
\phantom{\times}  &
    \phantom{\times}  &  
    \phantom{\times} & 
    \phantom{\times} &  
    \phantom{\times} &
        \phantom{\times} &  
        \phantom{\times} & 
        \phantom{\times} &   
        \phantom{\times}   \\ \hline
\phantom{\times}  & 
\phantom{\times}  &  
\phantom{\times} & 
\phantom{\times}  &
    \phantom{\times}  &  
    \phantom{\times} & 
    \phantom{\times} &  
    \phantom{\times} &
        \phantom{\times} &  
        \phantom{\times} & 
        \phantom{\times} &   
        \phantom{\times} 
        & \cdots & \
\phantom{\times}  & 
\phantom{\times}  &  
\phantom{\times} & 
\phantom{\times}  &
    \phantom{\times}  &  
    \phantom{\times} & 
    \phantom{\times} &  
    \times &
        \phantom{\times} &  
        \phantom{\times} & 
        \phantom{\times} &   
        \phantom{\times}  \\ \hline
\phantom{\times}  & 
\phantom{\times}  &  
\phantom{\times} & 
\phantom{\times}  &
    \phantom{\times}  &  
    \phantom{\times} & 
    \phantom{\times} &  
    \phantom{\times} &
        \phantom{\times} &  
        \phantom{\times} & 
        \phantom{\times} &   
        \phantom{\times} 
        & \cdots & \
\times  & 
\phantom{\times}  &  
\phantom{\times} & 
\phantom{\times}  &
    \phantom{\times}  &  
    \phantom{\times} & 
    \phantom{\times} &  
    \phantom{\times} &
        \phantom{\times} &  
        \phantom{\times} & 
        \phantom{\times} &   
        \phantom{\times} \\ \hline
\phantom{\times}  & 
\phantom{\times}  &  
\phantom{\times} & 
\phantom{\times}  &
    \phantom{\times}  &  
    \phantom{\times} & 
    \phantom{\times} &  
    \phantom{\times} &
        \phantom{\times} &  
        \phantom{\times} & 
        \phantom{\times} &   
        \phantom{\times} 
        & \cdots & \
\phantom{\times}  & 
\times  &  
\phantom{\times} & 
\phantom{\times}  &
    \phantom{\times}  &  
    \phantom{\times} & 
    \phantom{\times} &  
    \phantom{\times} &
        \phantom{\times} &  
        \phantom{\times} & 
        \phantom{\times} &   
        \phantom{\times}   \\ \hline\hline
\phantom{\times}  & 
\phantom{\times}  &  
\phantom{\times} & 
\phantom{\times}  &
    \phantom{\times}  &  
    \phantom{\times} & 
    \phantom{\times} &  
    \phantom{\times} &
        \phantom{\times} &  
        \phantom{\times} & 
        \phantom{\times} &   
        \phantom{\times} 
        & \cdots & \
\phantom{\times}  & 
\phantom{\times}  &  
\times & 
\phantom{\times}  &
    \phantom{\times}  &  
    \phantom{\times} & 
    \phantom{\times} &  
    \phantom{\times} &
        \phantom{\times} &  
        \phantom{\times} & 
        \phantom{\times} &   
        \phantom{\times}  \\ \hline
\phantom{\times}  & 
\phantom{\times}  &  
\phantom{\times} & 
\phantom{\times}  &
    \phantom{\times}  &  
    \phantom{\times} & 
    \phantom{\times} &  
    \phantom{\times} &
        \phantom{\times} &  
        \phantom{\times} & 
        \phantom{\times} &   
        \phantom{\times}  
        & \cdots & \
\phantom{\times}  & 
\phantom{\times}  &  
\phantom{\times} & 
\phantom{\times}  &
    \phantom{\times}  &  
    \phantom{\times} & 
    \phantom{\times} &  
    \phantom{\times} &
        \phantom{\times} &  
        \phantom{\times} & 
        \phantom{\times} &   
        \times  \\ \hline
\phantom{\times}  & 
\phantom{\times}  &  
\phantom{\times} & 
\phantom{\times}  &
    \phantom{\times}  &  
    \phantom{\times} & 
    \phantom{\times} &  
    \phantom{\times} &
        \phantom{\times} &  
        \phantom{\times} & 
        \phantom{\times} &   
        \phantom{\times}  
        & \cdots & \
\phantom{\times}  & 
\phantom{\times}  &  
\phantom{\times} & 
\phantom{\times}  &
    \times  &  
    \phantom{\times} & 
    \phantom{\times} &  
    \phantom{\times} &
        \phantom{\times} &  
        \phantom{\times} & 
        \phantom{\times} &   
        \phantom{\times}   \\ \hline
\phantom{\times}  & 
\phantom{\times}  &  
\phantom{\times} & 
\phantom{\times}  &
    \phantom{\times}  &  
    \phantom{\times} & 
    \phantom{\times} &  
    \phantom{\times} &
        \phantom{\times} &  
        \phantom{\times} & 
        \phantom{\times} &   
        \phantom{\times} 
        & \cdots & \
\phantom{\times}  & 
\phantom{\times}  &  
\phantom{\times} & 
\phantom{\times}  &
    \phantom{\times}  &  
    \times & 
    \phantom{\times} &  
    \phantom{\times} &
        \phantom{\times} &  
        \phantom{\times} & 
        \phantom{\times} &   
        \phantom{\times}  \\ \hline\hline
\phantom{\times}  & 
\phantom{\times}  &  
\phantom{\times} & 
\phantom{\times}  &
    \phantom{\times}  &  
    \phantom{\times} & 
    \phantom{\times} &  
    \phantom{\times} &
        \phantom{\times} &  
        \phantom{\times} & 
        \phantom{\times} &   
        \phantom{\times} 
        & \cdots & \
\phantom{\times}  & 
\phantom{\times}  &  
\phantom{\times} & 
\phantom{\times}  &
    \phantom{\times}  &  
    \phantom{\times} & 
    \times &  
    \phantom{\times} &
        \phantom{\times} &  
        \phantom{\times} & 
        \phantom{\times} &   
        \phantom{\times}  \\ \hline
\phantom{\times}  & 
\phantom{\times}  &  
\phantom{\times} & 
\phantom{\times}  &
    \phantom{\times}  &  
    \phantom{\times} & 
    \phantom{\times} &  
    \phantom{\times} &
        \phantom{\times} &  
        \phantom{\times} & 
        \phantom{\times} &   
        \phantom{\times}  
        & \cdots & \
\phantom{\times}  & 
\phantom{\times}  &  
\phantom{\times} & 
\phantom{\times}  &
    \phantom{\times}  &  
    \phantom{\times} & 
    \phantom{\times} &  
    \phantom{\times} &
        \phantom{\times} &  
        \phantom{\times} & 
        \times &   
        \phantom{\times}   \\ \hline
\phantom{\times}  & 
\phantom{\times}  &  
\phantom{\times} & 
\phantom{\times}  &
    \phantom{\times}  &  
    \phantom{\times} & 
    \phantom{\times} &  
    \phantom{\times} &
        \phantom{\times} &  
        \phantom{\times} & 
        \phantom{\times} &   
        \phantom{\times}  
        & \cdots & \
\phantom{\times}  & 
\phantom{\times}  &  
\phantom{\times} & 
\phantom{\times}  &
    \phantom{\times}  &  
    \phantom{\times} & 
    \phantom{\times} &  
    \phantom{\times} &
        \times &  
        \phantom{\times} & 
        \phantom{\times} &   
        \phantom{\times}  \\ \hline
\phantom{\times}  & 
\phantom{\times}  &  
\phantom{\times} & 
\phantom{\times}  &
    \phantom{\times}  &  
    \phantom{\times} & 
    \phantom{\times} &  
    \phantom{\times} &
        \phantom{\times} &  
        \phantom{\times} & 
        \phantom{\times} &   
        \phantom{\times}  
        & \cdots & \
\phantom{\times}  & 
\phantom{\times}  &  
\phantom{\times} & 
\phantom{\times}  &
    \phantom{\times}  &  
    \phantom{\times} & 
    \phantom{\times} &  
    \phantom{\times} &
        \phantom{\times} &  
        \times & 
        \phantom{\times} &   
        \phantom{\times}  
\end{array}\right) = (2\delta)^{2N} (k_x^2+k_y^2).
\end{equation}}
As all the other terms are of higher order in either $E$ or $(k_x^2+k_y^2)$ than the two listed terms, there exists a neighborhood around $k_x=k_y=0$ where these two terms are dominant. In this neighborhood, the dispersion of the previously identified zero-energy states is obtained by solving for $\{E\}_{i=1}^4$ in the equation 
\begin{equation}
(E-E_1)(E-E_2)(E-E_3)(E-E_4) = E^4 + (2\delta)^{2N}(k_x^2 + k_y^2).
\end{equation}
The solution is 
\begin{equation}
E_{1,2,3,4} = \pm \sqrt{\pm\imi} (2\delta)^{N/2}\sqrt{k} \qquad\textrm{where}\quad k=\sqrt{k_x^2 + k_y^2}.  \label{eqn:sq-root-disp}
\end{equation}
Note that the coefficient in the derived square-root dispersion grows exponentially with the number of layers, and becomes \emph{infinitely steep} in the thermodynamic limit $N\to \infty$.

%

\section{Higher-order exceptional point on the surface of ETI}\customlabel{sec:surface-EPs}{3.}

We showed in the main text that by tuning the $g$-factor angle $\alpha$ of the model in Eq.~(1) in the range $[0,\pi/2]$, the surface states filling the point-gap region continuously evolve from a single-sheet covering (for $\alpha = \pi/2$) to a pair of bands connected with an exceptional point (EP) at $E=0$ (for $\alpha = 0$) (cf.~Fig.~2a and b). Rather than encountering some sharp transition, this is a simple consequence of the EP moving/leaving the point-gap region from/to the bulk energy bands. In this section, we show that the evolution of the surface states with $\alpha$ is, in fact, \emph{even richer}, as it exhibits a collision and bouncing of two elementary (so-called \emph{second-order}) EPs, with the critical point corresponding to a \emph{third-order} EP~\cite{Demange:2012}. For brevity, we refer to these two types of band degeneracy as EP2, resp.~EP3. Recall that an $n^\textrm{th}$ order exceptional point (EP$n$) is a singularity where $n$ bands become defective, and disperse as $(k_x + i k_y)^{1/n}$.

\begin{figure}[b!]
    \includegraphics[width=0.82\linewidth]{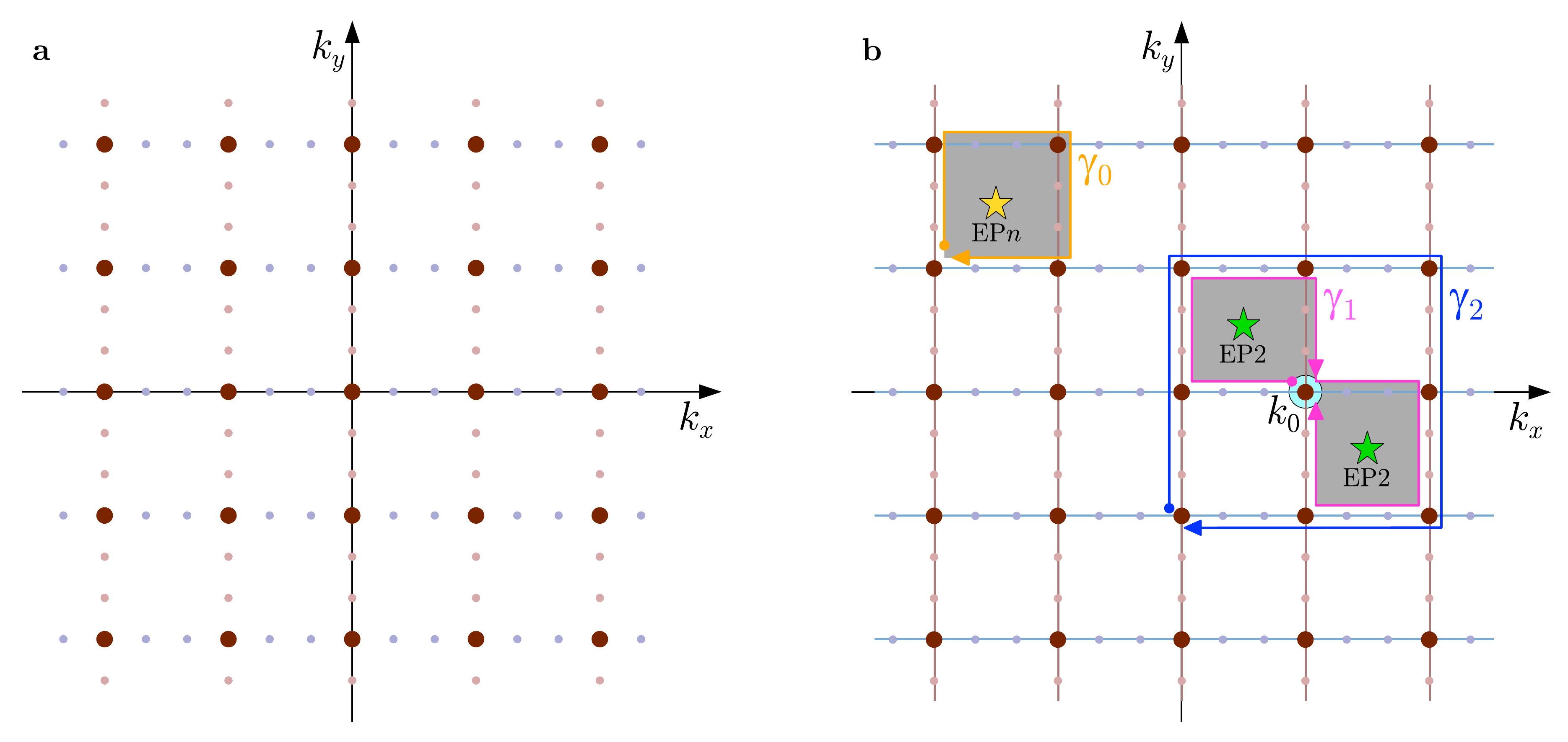}
    \caption{\label{fig:grids-and-paths}\textbf{Momentum grid and paths for the surface spectra in this work.} \textbf{a}~The grid of points in momentum space for which we plot the surface spectra in Fig.~2a and b of the main text and in Supplementary Fig.~\ref{fig:EP_move} below. (\textbf{b}, {\color{orange}orange}) Assuming the presence of an $n^\textrm{th}$ order exceptional point (EP$n$, yellow star), one needs to circumvent it $n$ times until the energy band connects back onto itself. As a consequence, the surface-states spectrum along the square path $\gamma_0$ looks like a polygon with $4n$ sides in the complex energy plane. (\textbf{b}, {\color{magenta}magenta}) If there are two nearby second-order exceptional points (EP2s, green stars), one could enclose each one of them in sequence by following the magenta path $\gamma_1$. The energy spectrum along $\gamma_1$ reveals whether the two EPs are formed by \emph{the same} pair of bands, or by two \emph{different} pairs of adjacent bands. (\textbf{b}, {\color{blue}blue}) If three bands participate in the formation of the two adjacent EP2s, the spectrum along $\gamma_2$ is equivalent to the spectrum along $\gamma_0$ for $n=3$. Therefore, the merging of two EP2s results in an EP3. For a detailed discussion see Supplementary Note~\ref{sec:surface-EPs}}
\end{figure}

To reveal the presence of the EPs, we plot in Supplementary Fig.~\ref{fig:EP_move} the surface states across the critical point in a way similar to Fig~2 a) and b) of the main text. More specifically, we consider a slab geometry with $N=20$ layers, and we take a regular square grid of momentum points with $\Delta k_x = \Delta k_y = 2\pi/800$ (we find the region $\abs{k_{x,y}}<2\pi/10$ to be sufficiently large to include all the surface states filling the point-gap region) inside the two-dimensional Brillouin zone. We indicate all the identified eigenstates as points inside the complex energy plane, and color them blue to red according to their localization (following their inverse participation ratio $\textrm{IPR}[\left|{\psi}\right>] = \sum_i \abs{\psi(i)}^4/\sum_i \abs{\psi(i)}^2$ where $\psi(i)$ are components of the right eigenstate $\left|\psi\right>$). 

Crucially, we also consider Cartesian lines in $\mathbf{k}$-space by including two additional points along each bond of the square lattice, following the schematics of Supplementary Fig.~\ref{fig:grids-and-paths}a. The image of these Cartesian lines is visible in the surface-state energies of Fig.~2a and b and of Supplementary Fig.~\ref{fig:EP_move} as a deformed square grid. From the deformation of the square grid one can deduce how a piece of the surface Brillouin zone is pasted onto the complex energy plane. Besides continuous deformations, one observes that the square grid occasionally exhibits disclination defects. We argue that these indicate various types of exceptional points on the surface of ETI at the corresponding complex energy.

\begin{figure}[t!]
    \includegraphics[width=\linewidth]{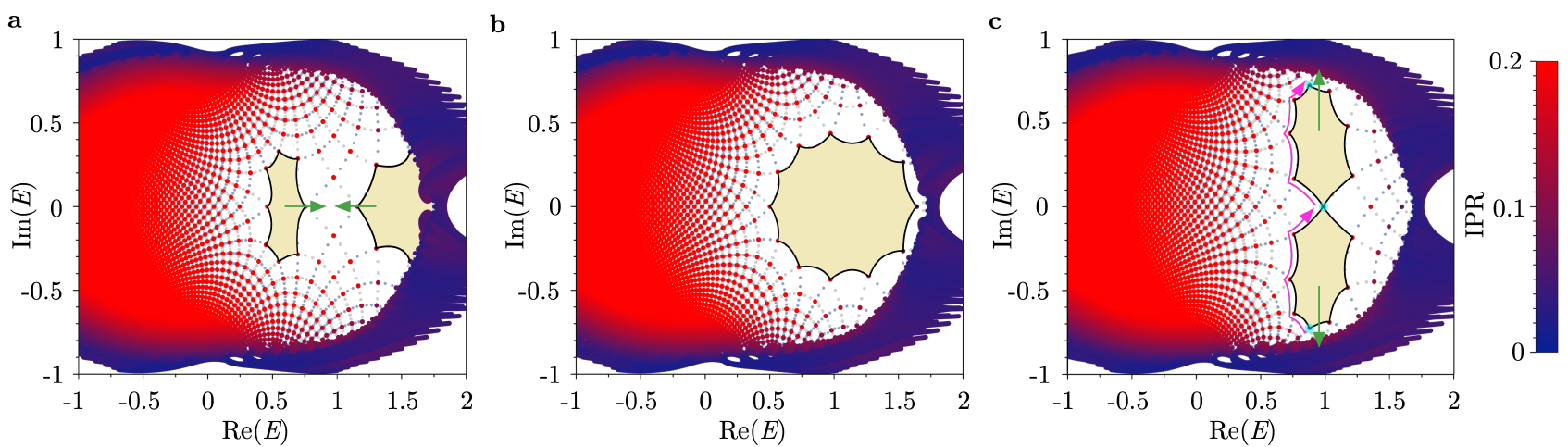}
    \caption{\label{fig:EP_move} \textbf{Surface states for (a-c) $g$-factor angles $\alpha = 0.57$, $0.64$ and $0.72$.} We indicate surface states at momentum resolution $\delta k = 2\pi/800$ in both $k_{x,y}$-directions in red-to-blue tones according to their inverse participation ratio (IPR). The smaller light-gray points with a finer resolution $\delta k = 2\pi/2400$ represent the momentum mesh; more specifically, they indicate the band energies along lines parallel with either $k_x$ or $k_y$ axis (cf.~Supplementary Fig.~\ref{fig:grids-and-paths}a). The presence of `ordinary' (second-order) exceptional points (EP2s) is revealed by an octagon shape with eight right angles [light yellow in a and c], and their motion with increasing $\alpha$ is indicated by the green arrows. Two EPs collide and bounce at $\alpha \approx 0.64$, while forming a third-order exceptional point (EP3) at the collision, revealed by a dodecagon shape with twelve right angles [light yellow in b]. The meaning of the magenta arrows, of the cyan dots in panel c, and the detailed reasoning behind the occurrence of the polygons are clarified in the text of Supplementary Note~\ref{sec:surface-EPs}}
\end{figure}

Let us first consider an isolated $n^\textrm{th}$ order exceptional point (EP$n$) occurring inside one of the squares of the momentum grid [Supplementary Fig.~\ref{fig:grids-and-paths}(b, {\color{orange}orange})]. Then the bands form a Riemann sheet that wind $n$ times around the central singularity until it connects back onto itself. Therefore, one needs to go $n$ times along the edges of the square (i.e.~make $4n$ right turns along the orange path $\gamma_0$) before returning to the original state. This is revealed as an octagon (i.e.~$2\pi$ disclination) for EP2, and as a dodecagon (i.e.~$4\pi$ disclination) for EP3. 

Since both an octagon and a dodecagon are observed in the panels of Supplementary Fig.~\ref{fig:EP_move}, one interprets the evolution of the surface states with increasing $\alpha$ as follows. First (Fig.~2b, $\alpha=0$) there is a single EP2 at $E=0$. With increasing $\alpha$ the EP2 moves to higher energies, while at the same time (Supplementary Fig.~\ref{fig:EP_move}a, $\alpha = 0.57$) another EP2 enters the point-gap region from the bulk states at $E\approx 1.6$. The two EP2s collide (Supplementary Fig.~\ref{fig:EP_move}b, $\alpha = 0.64$) at $E\approx 1.1$, apparently forming an EP3 at the critical point. The EP2s then bounce and depart symmetrically from the real energy axis (Supplementary Fig.~\ref{fig:EP_move}c, $\alpha = 0.72$), until they both leave the point-gap region, leaving behind only a single-sheet covering surface state~(Fig.~2a, $\alpha = \pi/2$). [We remark that without additional analysis, our plots cannot be directly used to tell the momentum $(k_x,k_y)$ at which the EPs are formed.]

To analyze the situation near the critical point (Supplementary Fig.~\ref{fig:EP_move}b, $\alpha = 0.64$) in more detail, we consider a pair of EP2s in two nearby infinitesimal squares of the momentum grid [Supplementary Fig.~\ref{fig:grids-and-paths}(b, {\color{magenta}magenta})]. We construct the magenta path $\gamma_1$ that starts at $\mathbf{k}_0$ (cyan dot) between the EP2s, then (\emph{i}) encloses the first EP2, then (\emph{ii}) returns to $\mathbf{k}_0$, then (\emph{iii}) encloses the second EP2, until finally (\emph{iv}) it returns to $\mathbf{k}_0$ again. We argue that the spectrum along $\gamma_1$ signals whether the two EPs are formed by the same pair of energy bands, or by three contiguous energy bands. To understand this, note that circumventing an EP2 results in a permutation of two band indices, symbolically $(\overline{1,2}) \mapsto (2,1)$. If the two EP2s were formed by the same pair of bands, then the two permutations along $\gamma_1$ act, in sequence, as $(\overline{1,2}) \mapsto (\overline{2,1}) \mapsto (1,2)$ (the overline indicates the two bands which are permuted at the next step), meaning that the total permutation is trivial, and that the initial and the final state encountered along $\gamma_1$ at $\mathbf{k}_0$ are the same. 

We apply this reasoning to the pair of EP2s visible in Supplementary Fig.~\ref{fig:EP_move}c after the collision. Starting at the cyan point on the bottom, and departing in the indicated direction, we follow the magenta path in Supplementary Fig.~\ref{fig:grids-and-paths}b (performing in sequence: three right turns, one left turn, and three right turns). The three cyan points in Supplementary Fig.~\ref{fig:EP_move}c correspond to the same momentum $\mathbf{k}_0$, but lie at three distinct energies. Manifestly, the initial and the final state are different. This implies that there are \emph{three surface bands} at play, and that the two EP2s are formed by different pairs of bands.

We finally explain why the merging of two EP2s results in an EP3. For that, note that the magenta path $\gamma_1$ is continuously deformable into the square blue path $\gamma_2$ [Supplementary Fig.~\ref{fig:grids-and-paths}(b, {\color{blue}blue})]. The total permutation of bands along the paths $\gamma_1$ and $\gamma_2$ is the same, namely $(\overline{1,2},3) \mapsto (2,\overline{1,3}) \mapsto (2,3,1)$. Note that this is a cyclic permutation, exactly as expected for an EP3. Therefore, the spectrum along the square path $\gamma_2$ that encircles two EP2 formed by three adjacent bands looks the same as expected for the square path $\gamma_0$ encircling an EP3. This correspondence suggests the formation of an exact EP3 at the critical point~\cite{Demange:2012}.

%

\section{Bulk invariant}

In this section, we relate the bulk invariant $w_\textrm{3D}$ of a 3D non-Hermitian Hamiltonian with a point gap [Eq.~(5) of the main text] to the flow of Chern numbers around the point gap [Eq.~(2) of the main text]. Note that the same result has been obtained in Ref.~\onlinecite{Sato:2020}. Our argument is divided into four parts, distributed over the four subsection below. First, in Supplementary Note~\ref{sec:proof-part-1} we deform the non-Hermitian Hamiltonian into a Floquet unitary form by employing the polar decomposition while preserving the topological invariants. Second, in Supplementary Note~\ref{sec:two-band}, we depart from the unitary form and review the proof from Ref.~\onlinecite{Sun:2018} which derives the equivalence of Eqs.~(5) and~(2) for the special case of two-band models with vanishing $w_\textrm{1D}$. Finally, by appealing to the topological nature of the integrals in Eqs.~(5) and~(2) we show that the correspondence generalizes to models with arbitrarily many bands [Supplementary Note~\ref{sec:many-band-with-w1D}] and with arbitrary values of $w_\textrm{1D}$ along the three directions of the BZ torus [Supplementary Note~\ref{sec:two-band-with-w1D}].

\subsection{Mapping the non-Hermitian Hamiltonian to a Floquet operator via polar decomposition}~\customlabel{sec:proof-part-1}{4.A.}

We first review the arguments presented in Appendix G of Ref.~\onlinecite{Ueda:2018}, which relate a non-Hermitian Hamiltonian with a point-gap to a unitary Floquet operator. 
To do so, recall that a non-singular matrix $H(\mathbf{k})-E\mathbbold{1}$ (i.e.~with a point gap at energy $E\in\mathbb{C}$) has a polar decomposition. Setting for simplicity $E=0$, this corresponds to
\begin{equation}\label{eq:polar_decompose}
H(\mathbf{k}) = U(\mathbf{k})P(\mathbf{k}),
\end{equation}
where $P(\mathbf{k}) = [H(\mathbf{k})^\dagger H(\mathbf{k})]^{1/2}$ is a positive-definite self-adjoint matrix, and $U(\mathbf{k}) = H(\mathbf{k})P(\mathbf{k})^{-1}$ is a unitary matrix. The decomposition into $U(\mathbf{k})$ and $P(\mathbf{k})$ is unique and depends continuously on $\mathbf{k}$ when $H(\mathbf{k})$ is continuous (which is the case for any physical system).

Importantly, if $A$ and $B$ are positive-definite self-adjoint matrices, then the linear combination $tA+(1-t)B$ for $t\in[0,1]$ is also a positive-definite matrix. Setting these matrices specifically to $A(\mathbf{k}) = \mathbbold{1}$ and $B = P(\mathbf{k})$, we find that
\begin{equation}
F(\mathbf{k},t) = U(\mathbf{k}) \left[t \mathbbold{1} + (1-t)P(\mathbf{k})\right] \label{eqn:homotopy}
\end{equation}
is a \emph{homotopy} that continuously deforms $F(\mathbf{k},0) = H(\mathbf{k})$ to $F(\mathbf{k},1) = U(\mathbf{k})$ while preserving the point-gap at all $\mathbf{k}$ and $t$ [i.e.~$\det F(\mathbf{k},t) \neq 0$]. Furthermore, since we did not assume any symmetry of $H(\mathbf{k})$, we are not required to check for symmetries of $U(\mathbf{k})$. As a consequence, $H(\mathbf{k})$ and $U(\mathbf{k})$ share all point-gap invariants. 

We conclude that for each non-Hermitian $H(\mathbf{k})$ with a point gap there is a canonically associated unitary $U(\mathbf{k})$. The continuous deformation that evolves the first into the latter changes the eigenstates and normalizes eigenvalues to absolute value $1$, but it preserves all point-gap invariants. Besides relating non-Hermitian and Floquet topology, the obtained equivalence is particularly useful for analytical manipulations. Note that a unitary matrix is never defective, meaning that it always has a complete set of eigenstates. In particular, the generic band singularity of $H(\mathbf{k})$ is an exceptional line, while that of $U(\mathbf{k})$ is a Weyl point. This implies that $U(\mathbf{k})$ has the usual spectral decomposition at each $\mathbf{k}$, which is impossible for non-Hermitian Hamiltonians at exceptional degeneracies. 

\subsection{Relating \texorpdfstring{$w_{3\textrm{D}}$}{w3D} to Chern numbers in two-band models with $w_\textrm{1D}=0$}~\customlabel{sec:two-band}{4.B.}

For a Floquet unitary $U(\mathbf{k})$, the eigenvalues become $\mathsf{U}(1)$ phases, i.e.~$\lambda_{\mathbf{k}}^a = \mathrm{e}^{\imi \varepsilon_{\mathbf{k}}^a}$. We call below the complex argument $\varepsilon_{\mathbf{k}}^a$ the \emph{quasi-energy} of $a^\textrm{th}$ band, and we refer to the (counter-clockwise) direction along the unitary circle as the (positive) \emph{quasi-energy direction}. It is possible to consider all the $\mathbf{k}$-points with fixed quasi-energy, 
\begin{equation}
\textrm{FS}(\mu) = \{\mathbf{k}\in\textrm{BZ}\,|\, \exists a: \varepsilon_{\mathbf{k}}^a=\mu \;(\textrm{mod}\, 2\pi)\},
\end{equation}
which corresponds to a Fermi surface of $U(\mathbf{k})$ at quasi-energy $\mu$. The orientation of the Fermi surface $\textrm{FS}(\mu)$ is defined to be along the increasing quasi-energy. In the remainder of the present subsection we consider two-band models, i.e., $U(\mathbf{k}):\textrm{BZ}\to \mathsf{U}(2)$. Furthermore, we assume that $w_\textrm{1D}$ is trivial for all directions along the Brillouin zone torus, in which case the Floquet unitary can be continuously deformed into $\mathsf{SU}(2)$. For cases when these two conditions are fulfilled, it was shown by Ref.~\onlinecite{Sun:2018} that 
\begin{equation}
w_\textrm{3D} = C_{\textrm{FS}(\mu)},\label{eqn:invariant-equivalence}
\end{equation}
i.e.~the winding number of $U(\mathbf{k})$ is equal to the Chern number on the Fermi surface $\textrm{FS}(\mu)$ for an arbitrary choice of Fermi quasi-energy $\mu$ (i.e., equivalently, to the flow of Berry curvature in the quasi-energy direction around the point gap). Below, we review the arguments of Ref.~\onlinecite{Sun:2018} that lead to the validity of Eq.~(\ref{eqn:invariant-equivalence}) in two-band models. These arguments are generalized to the case of many bands and of non-vanishing $w_\textrm{1D}$ in the subsequent Supplementary Notes~\ref{sec:many-band-with-w1D} and~\ref{sec:two-band-with-w1D}

Starting with a generic two-band unitary, note that $\mathsf{U}(2) = \mathsf{U}(1)\ltimes \mathsf{SU}(2)$ (semi-direct product). The decomposition of the Floquet unitary $U(\mathbf{k})$ into the two components is not unique: while the $\mathsf{U}(1)$-component of $U(\mathbf{k})$ is simply $\det[U(\mathbf{k})]$, the $\mathsf{SU}(2)$-component $\tilde{U}(\mathbf{k}) = U(\mathbf{k})/\sqrt{\det[U(\mathbf{k})]}$ has a $\pm$ sign ambiguity. Therefore, we choose to first prove Eq.~(\ref{eqn:invariant-equivalence}) for the class of unitaries $U(\mathbf{k})$ that do not wind non-trivially along the $\mathsf{U}(1)$ component (i.e., for which $w_\textrm{1D}$ is trivial for all directions around the Brillouin-zone torus). In such cases it is possible to choose the branch of the square root globally, implying that $\tilde{U}(\mathbf{k})$ is a global and smooth function on the entire Brillouin zone. Furthermore, the assumption also allows us to continuously deform the $\mathsf{U}(1)$-component to a constant map [e.g., to $1\in\mathsf{U}(1)$], in which case the Floquet unitary operator $U(\mathbf{k})$ is continuously deformed to \emph{special} unitary $\tilde{U}(\mathbf{k})$. Because of the continuity, such a deformation preserves both the left- and the right-hand side of Eq.~(\ref{eqn:invariant-equivalence}).

The general $\mathsf{SU}(2)$ matrix can be decomposed using the identity matrix and the Pauli matrices as
\begin{equation}
\tilde{U}(\mathbf{k}) = n_0(\mathbf{k})\unit -\imi\left[n_x(\mathbf{k})\sigma_x + n_y(\mathbf{k})\sigma_y + n_z(\mathbf{k})\sigma_z \right]\label{eqn:SU2-decomp}
\end{equation}
with norm
\begin{equation}
||n||^2 = n_0^2 + n_x^2 +n_y^2 + n_z^2 = 1\label{eqn:SU2-norm}
\end{equation}
at each $\mathbf{k}$, indicating the well-known property that as a topological space $\mathsf{SU}(2)\cong S^3$. We use this simple expression for a general $\mathsf{SU}(2)$ matrix to simplify the formula for $w_\textrm{3D}$. 

Noting that $\tilde{U}^\dagger(\mathbf{k})$ is obtained from $\tilde{U}(\mathbf{k})$ simply by flipping the sign $-\imi\mapsto +\imi$, one easily checks that
\begin{eqnarray}
\tilde{U}^{-1}\partial_k\tilde{U}=\left({\Sigma}_{a\in\{0,x,y,z\}} n_a\partial_k n_a\right)\unit &+& \imi \left[(\partial_k n_0) n_x - (\partial_k n_x) n_0 - (\partial_k n_y) n_z + (\partial_k n_z) n_y\right] \sigma_x \nonumber \\
&+& \imi \left[(\partial_k n_0) n_y - (\partial_k n_y) n_0 - (\partial_k n_z) n_x + (\partial_k n_x) n_z\right] \sigma_y \nonumber \\
&+& \imi \left[(\partial_k n_0) n_z - (\partial_k n_z) n_0 - (\partial_k n_x) n_y + (\partial_k n_y) n_x\right] \sigma_z ,
\end{eqnarray}
where we dropped the $\mathbf{k}$-arguments for brevity. Note that the obtained prefactor of $\unit$ is simply $\tfrac{1}{2}\partial_k ||n(\mathbf{k})||^2 = 0$ according to Eq.~(\ref{eqn:SU2-norm}), and the term drops out. We thus rewrite 
\begin{equation}
\tilde{U}^{-1}\partial_k\tilde{U} = \sum_{a\in{\{x,y,z\}}} \imi\, [(\partial_k n_0) n_a - (\partial_k n_a) n_0 - \epsilon_{abc}(\partial_k n_b)n_c]\sigma_a. \label{eqn:U-one-der}
\end{equation}
Furthermore, since $\tilde{U}^{-1}\tilde{U}=\unit$, we can rewrite $\tilde{U}^{-1}\partial_i U = -(\partial_i \tilde{U}^{-1})\tilde{U}$ for one of the derivatives in Eq.~(5) of the main text, such that the formula for the 3D winding number reduces to 
\begin{eqnarray}
w_\textrm{3D} &=& +\frac{1}{24\pi^2}\int_{\textrm{BZ}}d^3\mathbf{k}\epsilon_{ijk}\Tr[(\partial_i \tilde{U}^{-1})\tilde{U} \tilde{U}^{-1}(\partial_j \tilde{U})\tilde{U}^{-1}(\partial_k \tilde{U})] \\
&=& +\frac{1}{24\pi^2}\int_{\textrm{BZ}}d^3\mathbf{k}\epsilon_{ijk}\Tr[(\partial_i \tilde{U}^{-1})(\partial_j \tilde{U})\tilde{U}^{-1}(\partial_k \tilde{U})]\label{eqn:w3D-after-PP}
\end{eqnarray}
with implicit summation over $i,j,k\in\{k_x,k_y,k_z\}$. We further use that 
\begin{equation}
(\partial_i \tilde{U}^{-1})(\partial_j \tilde{U}) =  \sum_{a\in\{0,x,y,z\}}(\partial_i n_a)(\partial_i n_j)\unit+\sum_{a\in\{x,y,z\}}(-\imi)\left[(\partial_i n_0)(\partial_j n_a) - (\partial_i n_a)(\partial_j n_0)-\epsilon_{ade}(\partial_i n_d)(\partial_j n_e)\right]\sigma_a.\label{eqn:U-two-der}
\end{equation}
We now substitute expressions~(\ref{eqn:U-one-der}) and~(\ref{eqn:U-two-der}) into Eq.~(\ref{eqn:w3D-after-PP}). Since Pauli matrices have zero trace, the only combinations of terms that contribute to Eq.~(\ref{eqn:w3D-after-PP}) are those that combine two copies of the same Pauli matrix (i.e., $\Tr[\sigma_a\cdot\sigma_a]=2$). Therefore, the expression behind the integral of Eq.~(\ref{eqn:w3D-after-PP}) equals
\begin{equation}
2\epsilon_{ijk}[(\partial_i n_0)(\partial_j n_a) - (\partial_i n_a)(\partial_j n_0) - \epsilon_{ade}(\partial_i n_d) (\partial_j n_e) ][(\partial_k n_0)n_a - (\partial_k n_a)n_0 - \epsilon_{abc}(\partial_k n_b) n_c ],\label{eqn:to-simplify}
\end{equation}
where $i,j,k$ is summed over $\{k_x,k_y,k_z\}$, and $a,b,c$ over $\{x,y,z\}$. 

We would now like to expand the product of the square brackets in Eq.~(\ref{eqn:to-simplify}). One readily observes that many of the resulting terms vanish for symmetry reasons. On the one hand, combining any two terms without the $\epsilon_{a\ldots}$ Levi-Civita symbol results in expressions containing either $\epsilon_{ijk}(\partial_i n_0)(\partial_k n_0)=0$ or 
$\epsilon_{ijk}(\partial_i n_a)(\partial_k n_a)=0$ (or equivalent). The cancellation arises after summation over the $i,k$ indices due to taking the product of objects symmetric [i.e., $(\partial_i n_a)(\partial_k n_a)$] and antisymmetric (i.e., $\epsilon_{ijk}$) in $i \leftrightarrow k$. Similarly, the combination involving $\epsilon_{abc}\epsilon_{ade} = (\delta_{bd}\delta_{ce}-\delta_{be}\delta_{ce})$ is also easily shown to combine the antisymmetric tensor $\epsilon_{ijk}$ with a combination of derivatives symmetric in $i\leftrightarrow k$ or $j\leftrightarrow k$, and therefore also vanishes. In summary, only the products that contain \emph{exactly one} of the $\epsilon_{a\ldots}$ symbols are non-vanishing after summation, which (after also renaming $d\mapsto b$ and $e\mapsto c$) results in 
\begin{equation}
2 \epsilon_{ijk}\epsilon_{abc}[-(\partial_i n_0)(\partial_j n_a)(\partial_k n_b)n_c + (\partial_i n_a)(\partial_j n_0)(\partial_k n_b)n_c - (\partial_i n_b)(\partial_j n_c)(\partial_k n_0) n_a + (\partial_i n_b)(\partial_j n_c)(\partial_k n_a) n_0].
\end{equation}
After slight reordering of the terms and permuting the indices, the same expression is equivalently written as
\begin{equation}
2\epsilon_{ijk}\epsilon_{abc} [n_0 (\partial_i n_a)(\partial_j n_b)(\partial_j n_c) - n_a(\partial_i n_0)(\partial_j n_b)(\partial_k n_c) + n_a(\partial_i n_b)(\partial_j n_0)(\partial_k n_c) - n_a(\partial_i n_b)(\partial_j n_c)(\partial_k n_0)]
\end{equation}
which one should recognize as
\begin{eqnarray}
&\phantom{=}& 2\epsilon_{ijk}[\epsilon_{0bcd} n_0 (\partial_i n_b)(\partial_j n_c)(\partial_j n_d) + \epsilon_{a0cd} n_a(\partial_i n_0)(\partial_j n_c)(\partial_k n_d) \nonumber \\&\phantom{=}& +\, \epsilon_{ab0d} n_a(\partial_i n_b)(\partial_j n_0)(\partial_k n_d) + \epsilon_{abc0} n_a(\partial_i n_b)(\partial_j n_c)(\partial_k n_0)] \nonumber \\
 &=& 2\epsilon_{ijk}\epsilon_{abcd}n_a (\partial_i n_b) (\partial_j n_c)(\partial_k n_d).
\end{eqnarray}
where $a,b,c,d$ now range over the enlarged set $\{0,x,y,z\}$. Therefore, one obtains
\begin{equation}
w_\textrm{3D}=\frac{1}{12\pi^2}\int_\textrm{BZ}d^3\mathbf{k}\epsilon_{ijk}\epsilon_{abcd}n_a(\partial_i n_b)(\partial_j n_c)(\partial_k n_d) = \frac{1}{2\pi^2}\int_\textrm{BZ}d^3\mathbf{k}\epsilon_{abcd}n_a(\partial_{k_x} n_b)(\partial_{k_y} n_c)(\partial_{k_z} n_d),\label{eqn:S3-wrapping number}
\end{equation}
which is recognized as the wrapping number of the Brillouin zone torus around the $3$-sphere. (Note that the normalization is also correct because $2\pi^2$ is exactly the surface area of the unit $3$-sphere.)

The integrand in Eq.~(\ref{eqn:S3-wrapping number}) is interpreted as the oriented area of the unit $3$-sphere swiped by $\mathbf{n}(\mathbf{k}) = (n_0,n_x,n_y,n_z)$ as $\mathbf{k}$ is varied over the infinitesimal volume $d^3\mathbf{k}$. The orientation is important, namely it means that the area is being ``covered'' resp.~``uncovered'' depending on the sign of the integrand. A generic point $x_0\in S^3$ is covered (uncovered) by the map $\tilde{U}(\mathbf{k}):\textrm{BZ}\mapsto S^3$ a number of times which we denote $\nu_+(x_0)$ [$\nu_-(x_0)$]. The geometric nature of the wrapping number (\ref{eqn:S3-wrapping number}) implies that 
\begin{equation}
w_{\textrm{3D}} = \nu_+(x_0) - \nu_-(x_0),\label{eqn:un-covering-number}
\end{equation}
i.e., each $x_0\in S^3$ is wrapped the same number of times if the orientation is taken into account. In a more technical language~\cite{Kennedy:2016}, the equal dimensionality of $\mathsf{SU}(2)\cong S^3$ and of $\textrm{BZ}=T^3$ implies that the Pontryagin manifold of $\tilde{U}:\textrm{BZ}\to \mathsf{SU}(2)$ [i.e.~the framed pre-image $\tilde{U}^{-1}(x_0)$ of any point $x_0\in\mathsf{SU}(2)$, with framing defined by pulling back a basis of tangent space $T_{x_0}\mathsf{SU}(2)$] is generically a set of discrete points $\{\mathbf{k}_i\}_{i\in\mathcal{I}}$ that depends on $\tilde{U}$ and $x_0$, each point carrying positive or negative orientation $\mathfrak{o}(\mathbf{k}_i)$ [the handedness of the pulled-back frame]. It follows from the Pontryagin-Thom construction that the index of the map $\tilde{U}$ is 
\begin{equation}
w_\textrm{3D} = \sum_{i} \mathfrak{o}(\mathbf{k}_i)\label{eqn:index-formula}
\end{equation}
and that it is invariant under cobordisms representing continuous deformations of the map $\tilde{U}$ and the choice of $x_0$. [In particular, the cobordism theory also implies that the invariant cannot change in fine-tuned (i.e., not generic) situations where the pre-images are not point-like or where several point-like pre-images with the same orientation are brought on top of one another.] A pair of pre-images with opposite orientation (i.e., one being ``covered'' and the other ``uncovered'') can pairwise annihilate, while keeping the right-hand side of Eq.~(\ref{eqn:un-covering-number}) invariant. 

The relation in Eq.~(\ref{eqn:un-covering-number}) is particularly useful when used to study Weyl points in the spectrum of $\tilde{U}(\mathbf{k})$. The eigenvalues of the special unitary in Eq.~(\ref{eqn:SU2-decomp}) are $\mathrm{e}^{\pm\imi\theta}$ where $n_0 = \cos\theta$ with $\theta\in[0,\pi]$. Note that the eigenvalues become degenerate (i.e.,~the bands of $\tilde{U}(\mathbf{k})$ exhibit a Weyl point) only when $\theta=0$ [corresponding to the ``north pole'' of the 3-sphere] or $\theta=\pi$ [the ``south pole'' of the 3-sphere]. To prove the relation between $w_\textrm{3D}$ and $C_{\textrm{FS}(\mu)}$, we now study the relation between $w_\textrm{3D}$ and the Weyl points of $\tilde{U}(\mathbf{k})$, whence Eq.~(\ref{eqn:invariant-equivalence}) will be seen as a straightforward consequence.

The described band nodes are generically Weyl points, and they come in four species: They could occur at quasi-energy $0$ or $\pi$, and they can carry left-handed or right-handed chirality. We indicate the number of these four species of Weyl points as $N_\textrm{L}^0$, $N_\textrm{R}^0$, $N_\textrm{L}^\pi$ and $N_\textrm{R}^\pi$. It is detailed in Ref.~\onlinecite{Sun:2018} that positive (negative) coverings of $\pm\unit \in \mathsf{SU}(2)$ produce right-handed (left-handed) Weyl points at the corresponding energy and $\mathbf{k}$, i.e., that $N_{L/R}^{0,\pi}$ correspond to $\nu_\pm(\pm\unit)$. [The fact that a right-handed and left-handed Weyl point can pairwise annihilate is seen as a consequence of the index formula~(\ref{eqn:index-formula}).] Since both $\pm \unit\in\mathsf{SU}(2)$ have to be wrapped $w_\textrm{3D}$ times, it follows from Eq.~(\ref{eqn:un-covering-number}) that
\begin{equation}
w_\textrm{3D} = N_\textrm{R}^0 - N_\textrm{L}^0 = N_\textrm{R}^\pi - N_\textrm{L}^\pi.
\end{equation}
In other words, non-vanishing $w_\textrm{3D}$ implies an imbalance between right-handed and left-handed Fermions (i.e., violating the Nielsen-Ninomiya theorem\cite{Sun:2018,Higashikawa:2019}) both at quasi-energy $0$ and $\pi$. It is well known (and confirmed by explicit calculation of a linearized model around the degeneracies) that right-/left-handed Weyl points inject/remove one qunatum of Berry curvature flow into the higher (quasi-)energy band. It therefore follows from the Weyl-point imbalance in two-band models with vanishing $w_\textrm{1D}$ that Eq.~(\ref{eqn:invariant-equivalence}) is valid for all $\mu$ with non-degenerate Fermi surfaces (i.e., when the Chern number on the right-hand side can be meaningfully computed). We can thus interpret the right-hand side of Eq.~(\ref{eqn:invariant-equivalence}) as the \emph{flow} of Berry curvature in the quasi-energy direction. For example, for the band with quasi-energy $0<\varepsilon_{\mathbf{k}}<\pi$, the right-handed (left-handed) Weyl points at $\varepsilon=0$ act as sources (sinks) of quanta of Berry curvature, while the right-handed (left-handed) Weyl points at $\varepsilon=\pi$ instead act as sinks (sources) of quanta of Berry curvature.

Since Weyl points only occur at a discrete set of energies [namely at $\mu=0$ and $\mu=\pi$ for $\tilde{U}(\mathbf{k})\in\mathsf{SU}(2)$], one can use continuity to \emph{define} the flow of Chern number on the right-hand side of Eq.~(\ref{eqn:invariant-equivalence}) for arbitrary quasi-energy $\mu$ -- including those coinciding with the Weyl point quasi-energy. Furthermore, because of the topological nature of the integrals on both sides of Eq.~(\ref{eqn:invariant-equivalence}), the presented result generalizes to any two-band unitary map $U:\textrm{BZ}\to \mathsf{U}(2)$ with vanishing $w_\textrm{1D}$, which are obtained from the explicitly discussed $\mathsf{SU}(2)$ models by a continuous deformation. 

\subsection{Relating \texorpdfstring{$w_{3\textrm{D}}$}{w3D} to Chern number in models with arbitrarily many bands and with $w_{\textrm{1D}}=0$}~\customlabel{sec:many-band-with-w1D}{4.C.}

The generalization of the relation in Eq.~(\ref{eqn:invariant-equivalence}) to models with a larger number of bands follows directly by combining the (\emph{i}) topological nature of integrals on both sides of the equation, (\emph{ii}) additivity of both integrals under direct sum of two unitary maps ${U}_1\oplus {U}_2$, and (\emph{iii}) the existence of a continuous deformation (homotopy) between any pair of models with the same number of bands and with the same topological invariant $w_\textrm{3D}\in \pi[T^3,\mathsf{SU}(N)] = \mathbb{Z}$. In particular, it is important that the formula for $w_{\textrm{3D}}$ is \emph{identical} for unitary operators with an arbitrary number $N\geq2$ of bands.

To proceed, first note that two-band models can attain \emph{arbitrary} $w_\textrm{3D}\in \mathbb{Z}$. For example, the special unitary in Eq.~(\ref{eqn:SU2-decomp}) with (before normalization)
\begin{eqnarray}
n_0^W(\mathbf{k}) &=& (\cos k_x + \cos k_y + \cos k_z - 2), \nonumber \\
n_x^W(\mathbf{k}) &=& \textrm{Re}[(\sin k_x + \imi \sin k_y)^W],  \\
n_y^W(\mathbf{k}) &=& \textrm{Im}[(\sin k_x + \imi \sin k_y)^W], \nonumber \\
n_z^W(\mathbf{k}) &=& \sin k_z, \nonumber 
\end{eqnarray}
and with $W\in\mathbb{Z}$ has $w_\textrm{3D} = W$. We will introduce the notation for this particular model, 
\begin{equation}
\frac{1}{||n^W(\mathbf{k})||^2}\left\{n_0^W(\mathbf{k})\unit - \imi [n_x^W(\mathbf{k})\sigma_x + n_y^W(\mathbf{k})\sigma_y + n_z^W(\mathbf{k})\sigma_z]\right\} \equiv \tilde{U}^W(\mathbf{k}).
\end{equation}
To proceed, let us investigate the validity of the equivalence in Eq.~(\ref{eqn:invariant-equivalence}) for the special class of $N$-band models 
\begin{equation}
\tilde{U}^\textrm{W}(\mathbf{k})\oplus \textrm{e}^{\imi \varepsilon_1}\oplus \ldots \oplus \textrm{e}^{\imi \varepsilon_{N-2}}\equiv \tilde{U}^W_N(\mathbf{k})\label{eqn:after-homotopy}
\end{equation}
where $\{\varepsilon_i\}_{i=1}^{N-2}$ are distinct and $\mathbf{k}$-independent phases (i.e., quasi-energies) subject to $\sum_i \varepsilon_i = 0$. We also require each of these phases to be different from $0$ and $\pi$. On the one hand, we use the additivity of $w_\textrm{3D}$ for direct sum of unitaries, combined with the fact that $\mathbf{k}$-independent one-band unitaries have vanishing $w_\textrm{3D}$, to show that the $N$-band model in Eq.~(\ref{eqn:after-homotopy}) has $w_\textrm{3D}=W$. On the other hand, the Chern number is also additive under direct sum of models, and contains a contribution from the two-band unitary $\tilde{U}^W(\mathbf{k})$ (for which it was proved in Supplementary Note~\ref{sec:two-band} that $w_\textrm{3D} = C_{\textrm{FS}(\mu)}$) plus the contributions of the single-band models (which for $\mu\neq \varepsilon_i$ do not form Fermi surface, and therefore their Chern number $C_{\textrm{FS}(\mu)}$ trivially vanishes). Therefore, Eq.~(\ref{eqn:invariant-equivalence}) is valid for models in Eq.~(\ref{eqn:after-homotopy}) for all quasi-energies except $\mu\in\{0,\pi,\varepsilon_1,\ldots,\varepsilon_{N-2}\}$. Since this is a set of measure zero, we \emph{define} the flow of Berry curvature $C_{\textrm{FS}(\mu)}$ at these quasi-energies by continuity from the values where it is well-defined.

We now generalize the result in Eq.~(\ref{eqn:invariant-equivalence}) to \emph{arbitrary} $N$-band models $\tilde{U}:\textrm{BZ}\to \mathsf{SU}(N)$. Recall that equivalence classes of such maps under continuous deformations are fully determined by the homotopy set $\pi[T^3,\mathsf{SU}(N)]\cong \mathbb{Z}$, with the corresponding class determined by the value of $w_\textrm{3D}$. Starting with an arbitrary $N$-band special unitary map $\tilde{U}(\mathbf{k})$ that carries $w_\textrm{3D} = W$, it follows from homotopy theory that there is a continuous deformation into $\tilde{U}^W_N(\mathbf{k})$. The continuous deformation preserves $w_\textrm{3D}$ by definition. Furthermore, such a continuous deformation must also preserve the Chern number at each energy. It is therefore a trivial consequence that Eq.~(\ref{eqn:invariant-equivalence}) is valid for arbitrary special unitary models with arbitrarily many bands. By considering further continuous deformations, our proof trivially extends to all $N$-band models $U:\textrm{BZ}\mapsto \mathsf{U}(N)$ that have $w_\textrm{1D}=0$.

\subsection{Relating \texorpdfstring{$w_{3\textrm{D}}$}{w3D} to Chern number in models with arbitrary $w_\textrm{1D}$}~\customlabel{sec:two-band-with-w1D}{4.D.}

The generalization of the relation in Eq.~(\ref{eqn:invariant-equivalence}) to models with non-trivial $w_\textrm{1D}$ is achieved through similar logic as exercised in the previous subsection when generalizing to arbitrarily many bands. Namely, we first prove the relation for specially crafted simple models -- a representative model for each homotopy equivalence class -- and then we argue (based on the invariant nature of the invariants under continuous deformations and on their additivity under direct sum of models) that the correspondence generalizes to arbitrary unitary models.

Recall that the set of homotopy classes of unitary models in 3D are  $\pi[T^3,\mathsf{U}(N)]=\mathbb{Z}^4$ for any $N\geq 2$, where the four integer invariants correspond to winding numbers $w_{\textrm{3D}}$ and $(w_{\textrm{1D},x},w_{\textrm{1D},y},w_{\textrm{1D},z})\equiv \mathbf{w}_\textrm{1D}$. We therefore consider the following $\mathsf{U}(3)$ models (a representative for each homotopy class),
\begin{equation}
U^{(W,\mathbf{w})}(\mathbf{k}) = \tilde{U}^W(\mathbf{k})\oplus \mathrm{e}^{\imi\mathbf{k}\cdot\mathbf{w}}, \label{eqn:crafted-models-(C)}
\end{equation}
for which it is easily checked that $w_\textrm{3D}=W$ and $\mathbf{w}_\textrm{1D} = \mathbf{w}$. Concerning the Chern number, we already argued in the previous subsections that $W = C_{\textrm{FS}(\mu)}$ for the two-band part $\tilde{U}^W(\mathbf{k})$, while the one-band part $\mathrm{e}^{\imi \mathbf{k}\cdot \mathbf{w}}$ manifestly carries zero Chern number as it encodes a constant (i.e., $\mathbf{k}$-independent) eigenstate $(0,0,1)$. Therefore, one readily finds that Eq.~(\ref{eqn:invariant-equivalence}) is valid for the models in Eq.~(\ref{eqn:crafted-models-(C)}). Because of the continuous nature of the integrals defining $\mathbf{w}_{\textrm{1D}}$, $W_\textrm{3D}$ and $C_{\textrm{FS}(\mu)}$, the relation is readily generalized to \emph{arbitrary} 3-band models $U:\textrm{BZ}\to \mathsf{U}(3)$.

To generalize to models with $N\geq 3$ bands, one can consider [in analogy with Eq.~(\ref{eqn:after-homotopy}) of Supplementary Note~\ref{sec:many-band-with-w1D}] a direct-sum composition of the representative model in Eq.~(\ref{eqn:crafted-models-(C)}) with $N-3$ one-band models corresponding to constant quasi-energies. The proof of Eq.~(\ref{eqn:invariant-equivalence}) for arbitrary $\mathsf{U}(N)$ model with $N\geq 3$ is then trivial to complete. Finally, the generalization to $N=2$ model $U^\textrm{2-band}(\mathbf{k})$ is obtained similarly, with a single $\mathbf{k}$-independent band instead added to the two-band model itself [rather than to the homotopy representative in Eq.~(\ref{eqn:crafted-models-(C)})]. Assuming that $U^\textrm{2-band}(\mathbf{k})$ carries $w_\textrm{3D} = W$ and $\mathbf{w}_\textrm{1D} = \mathbf{w}$, there is a homotopy equivalence 
\begin{equation}
U^\textrm{2-band}(\mathbf{k}) \oplus \mathrm{e}^{\imi \varepsilon_1} \; \sim \; \tilde{U}^W(\mathbf{k})\oplus \mathrm{e}^{\imi \mathbf{k}\cdot\mathbf{w}},
\end{equation}
whence the relation in Eq.~(\ref{eqn:invariant-equivalence}) follows for the two-band model $U^\textrm{2-band}(\mathbf{k})$ from the additivity of invariants in direct-sum models. Therefore, we complete the proof of the equivalence $w_\textrm{3D} = C_{\textrm{FS}(\mu)}$ (for arbitrary quasi-energy $\mu$) for arbitrary unitary map $U:\textrm{BZ} \to \mathsf{U}(N)$ with arbitrary $N \geq 2$. Finally, by recalling the continuity arguments of Supplementary Note~\ref{sec:proof-part-1}, the equivalence generalizes to non-Hermitian point-gapped Hamiltonians whenever the Chern number is well-defined (i.e., whenever the Fermi surface is not intersected by an exceptional degeneracy).

\section{Non-Hermitian terms in electronic systems}

In this section, we discuss scenarios, in which the specific structure of the self-energy discussed in the main text could arise in a condensed matter setting. In general, the emergence of non-Hermitian terms in electronic systems is best understood in the language of Green's functions, where the terms arise through a complex self-energy. The non-interacting Green's function is given by
\begin{equation}
    G^{(0)}(\omega) = (\mathrm{i}\omega - H)^{-1},
\end{equation}
with $H$ being the free Hamiltonian of the (electronic) system. This Green's function describes infinitely long-lived excitations at the eigenenergies of $H$. Interactions can be incorporated into the full Green's function introducing the self-energy $\Sigma(\omega)$,
\begin{equation}
    G(\omega) = \{[G^{(0)}(\omega)]^{-1} - \Sigma(\omega)\}^{-1} = [\mathrm{i}\omega - H - \Sigma(\omega)]^{-1} \equiv [\mathrm{i}\omega - H_{\mathrm{eff}}]^{-1},
\end{equation}
where in the last step, we have dropped the $\omega$ dependence of $\Sigma$. If the self-energy is complex, for example due to finite quasi-particle lifetimes, the effective Hamiltonian of the system $H_{\mathrm{eff}}$ is non-Hermitian.

\subsection{Non-Hermitian terms due to coupling to short-lived electronic excitations}

One scenario to obtain a complex self-energy as discussed in the main text, is to consider an additional orbital---for concreteness, we call it an $f$ electron---with no dispersion and a short lifetime $1/\Gamma$, which couples to the $s,p$ orbitals of the topological insulator. The (effective) Hamiltonian of this additional $f$ orbital is

\begin{equation}
    H_{\mathrm{f}} = \mu_{\mathrm{f}} c_{\mathrm{f}}^{\dag}c_{\mathrm{f}}^{}+i \Gamma c_{\mathrm{f}}^{\dag}c_{\mathrm{f}}^{}.
\end{equation}
Considering a hopping Hamiltonian between the $s$, $p$, and $f$ electrons of the form
\begin{equation}
    H_{\mathrm{sp-f}} = t_{\mathrm{s}} c_{\mathrm{s}}^\dag c_{\mathrm{f}}^{} + t_{\mathrm{p}} c_{\mathrm{p}}^\dag c_{\mathrm{f}}^{}, 
    \label{eq:spf_hopping}
\end{equation}
we can find the full Hamiltonian in block form as
\begin{equation}
    H =
    \begin{pmatrix}
   H_{\mathrm{sp}}  &   H_{\mathrm{sp-f}}\\
    H_{\mathrm{sp-f}}^\dag & H_{\mathrm{f}}  \\
    \end{pmatrix}.
\end{equation}
Since the full Green's function is defined as
\begin{equation}
    \left(i \omega - H \right) G(\omega) = \mathbbold{1},
\end{equation}
we obtain 
\begin{equation}
    \begin{pmatrix}
   i \omega - H_{\mathrm{sp}}  &  -H_{\mathrm{sp-f}}\\
    -H_{\mathrm{sp-f}}^\dag & i \omega - H_{\mathrm{f}}  \\
    \end{pmatrix}
    \begin{pmatrix}
   G_{\mathrm{sp}}(\omega)  &  G_{\mathrm{sp-f}}(\omega)\\
    G_{\mathrm{f-sp}}(\omega) & G_{\mathrm{f}}(\omega)  \\
    \end{pmatrix} = \mathbbold{1}.
\end{equation}
Using
\begin{equation}
    G_{\mathrm{f-sp}}(\omega) = \left(i \omega - H_{\mathrm{f}} \right)^{-1} H_{\mathrm{sp-f}}^\dag G_{\mathrm{sp}}(\omega),
\end{equation}
we can rewrite 
\begin{equation}
\begin{split}
    \left(i \omega - H_{\mathrm{sp}} \right)G_{\mathrm{sp}}(\omega) - H_{\mathrm{sp-f}} G_{\mathrm{f-sp}}(\omega) &= \left(i \omega - H_{\mathrm{sp}} \right)G_{\mathrm{sp}}(\omega) - H_{\mathrm{sp-f}} \left(i \omega - H_{\mathrm{f}} \right)^{-1} H_{\mathrm{sp-f}}^\dag G_{\mathrm{sp}}(\omega) \\
    &=\left(i \omega - H_{\mathrm{sp}}- \Sigma(\omega) \right)G_{\mathrm{sp}}(\omega) = 1.
    \end{split}
\end{equation}
This gives a correction to the $s$- and $p$-electron self energy
\begin{equation}
    \Sigma(\omega) = (t_{\mathrm{s}}\; t_{\mathrm{p}})^T \left(i \omega - H_{\mathrm{f}} \right)^{-1} (t_{\mathrm{s}}\; t_p) = \frac{1}{i\omega - \mu_{\mathrm{f}} - i\Gamma} \begin{pmatrix} t_{\mathrm{s}}^2 & t_{\mathrm{s}} t_{\mathrm{p}}\\ t_{\mathrm{s}} t_{\mathrm{p}} & t_{\mathrm{p}}^2\end{pmatrix}.
    \label{eq:spf_selfenergy}
\end{equation}

To lowest order in $\omega$, the self energy thus reads
\begin{equation}
    \Sigma \approx 
    \begin{pmatrix}
   \frac{\mathrm{i}t_{\mathrm{s}}^2}{\Gamma - \mathrm{i}\mu_{\mathrm{f}}}  &   \frac{\mathrm{i}t_{\mathrm{s}}t_{\mathrm{p}}}{\Gamma - \mathrm{i}\mu_{\mathrm{f}}}\\
    \frac{\mathrm{i}t_{\mathrm{s}}t_{\mathrm{p}}}{\Gamma - \mathrm{i}\mu_{\mathrm{f}}} & \frac{\mathrm{i}t_{\mathrm{p}}^2}{\Gamma - \mathrm{i}\mu_{\mathrm{f}}}  \\
    \end{pmatrix} = \frac{\mathrm{i}}{2}\frac{t_{\mathrm{s}}^2+t_{\mathrm{p}}^2}{\Gamma - \mathrm{i}\mu_{\mathrm{f}}} \tau_0 + \mathrm{i}\frac{t_{\mathrm{s}}t_{\mathrm{p}}}{\Gamma - \mathrm{i}\mu_{\mathrm{f}}} \tau_x + \frac{\mathrm{i}}{2}\frac{t_{\mathrm{s}}^2-t_{\mathrm{p}}^2}{\Gamma - \mathrm{i}\mu_{\mathrm{f}}} \tau_z,
\end{equation}
which adds to the effective Hamiltonian of $s$ and $p$ electrons. Considering $t_{\mathrm{s}} \approx t_{\mathrm{p}} \equiv t_{\mathrm{f}}$, the contributions proportional to $\tau_z$ vanish. Including spins of $s,p$ orbitals, we obtain the self-energy
\begin{equation}
    \label{eq:non_herm_f_electron}
    \Sigma = \mathrm{i}\frac{t_{\mathrm{f}}^2}{\Gamma - \mathrm{i} \mu_{\mathrm{f}}}(\tau_0 + \tau_x)\sigma_0.
\end{equation}
If the $f$ electron sits close to the chemical potential, in other words $\mu_{\mathrm{f}} \ll \Gamma$, the non-Hermitian term dominates in Eq.~(\ref{eq:non_herm_f_electron}) and gives rise to the physics discussed in the main text.

\subsection{Non-Hermitian terms due to electron-phonon scattering}

As an alternative scenario, we consider the scattering of electrons and phonons, one of the most fundamental interactions in solids. For simplicity, we describe this coupling with a toy model, the Holstein Hamiltonian, which in second quantization reads
\begin{align}
    H &= H_{\mathrm{el}} + H_{\mathrm{ph}} + H_{\mathrm{el-ph}}\notag \\
    &= \sum_{\mathbf{k}} c_{\mathbf{k},\nu}^{\dagger} \left(\mathbf{k} \cdot \Gamma-\mu\right)_{\nu,\nu'} c_{\mathbf{k},\nu'}^{} + \omega_0 \sum_{\mathbf{q}} \left(b_{\mathbf{q}}^{\dagger}b_{\mathbf{q}}^{} + \frac{1}{2}\right) + \sum_{\mathbf{k},\mathbf{q},\nu, \nu'} \frac{g_{\nu, \nu'}}{\sqrt{N}} \left(b_{\mathbf{q}}^{\dagger} + b_{-\mathbf{q}}^{}\right) c_{\mathbf{k}+\mathbf{q},\nu}^{\dagger}c_{\mathbf{k},\nu'}^{}.
\end{align}
Here, we consider non-interacting electrons with a 3D Dirac dispersion $\mathbf{k} \cdot \Gamma$, with $\Gamma_i = \tau_x \sigma_i$, and model the phonons as a single branch of Einstein phonons of frequency $\omega_0$ and momentum $\mathbf{q}$. The coupling strength $g_{\nu,\nu'}$ is taken to be small and momentum independent but we allow electrons to scatter from $\nu$ to $\nu'$. Here and in the following, we use $\nu$ as a combined index of orbital and spin.

The interaction with the phonons leads to a non-trivial self-energy in the electron Green's function
\begin{equation}
    G(\mathbf{k}, \mathrm{i}\omega_{m})_{\nu, \nu'}^{-1} =  G^{(0)}(\mathbf{k}, \mathrm{i}\omega_{m})_{\nu, \nu'}^{-1} + \Sigma_{\mathrm{el-ph}}(\mathbf{k}, \mathrm{i}\omega_m)_{\nu, \nu'}
    \label{eq:el_greens}
\end{equation}
with the bare (non-interacting) electron Green's function given by 
\begin{equation}
    \label{eq:Greens_el}
    G^{(0)}(\mathbf{k}, \mathrm{i}\omega_{m})_{\nu, \nu'} = \left(\mathrm{i}\omega_{m} - H_{\mathrm{el}}\right)_{\nu, \nu'}^{-1}
\end{equation}
and $\omega_m = (2m+1)\pi T$ ($m$ an integer) are the fermionic Matsubara frequencies. In the following, we use a lowest-order approximation to the electron-phonon problem~\cite{Giustino:2017}, since we are only interested in the \emph{structure} of the resulting self-energy.
The self-energy can be calculated as
\begin{equation}
    \Sigma_{\mathrm{el-ph}}(\mathbf{k}, \mathrm{i}\omega_m)_{\nu, \nu'} = - \frac{1}{N \beta} \sum_{\mathbf{k}',m'} \sum_{\nu'',\nu'''} g_{\nu,\nu''} G(\mathbf{k}',\mathrm{i}\omega_{m'})_{\nu'',\nu'''} g_{\nu''',\nu'} D(\mathbf{q},\mathrm{i}(\omega_m-\omega_{m'})),
    \label{eq:full_selfenergy}
\end{equation}
where we have relied on Migdal-Eliashberg theory to neglect vertex corrections~\cite{Migdal:1958,Eliashberg:1960}. This is valid even for the case of Dirac materials~\cite{Roy:2014}, but only correct in the weak-coupling regime $g/\omega_0 \ll 1$~\cite{Alexandrov:2001, Bauer:2011}. 
It is also customary to replace the full electron and phonon Green's functions in Eq.~\eqref{eq:full_selfenergy} to lowest order with the bare Green's function in Eq.~\eqref{eq:Greens_el} and the bare phonon Green's function~\cite{Migdal:1958,Eliashberg:1960}
\begin{equation}
    \label{eq:Greens_ph}
    D^{(0)}(\mathbf{q}, \mathrm{i}\Omega_{n}) = \left[\omega_{0}^2 + \Omega_n^2\right]^{-1},
\end{equation}
with $\Omega_n = 2n \pi T$ the bosonic Matsubara frequencies.
This corresponds to the processes of emitting and subsequently absorbing a phonon with momentum $\mathbf{q}$ ($\mathbf{-q}$), see Supplementary Fig.~\ref{fig:feyn_eph}.

\begin{figure}[t]
  \centering
  \includegraphics[width=0.7\linewidth]{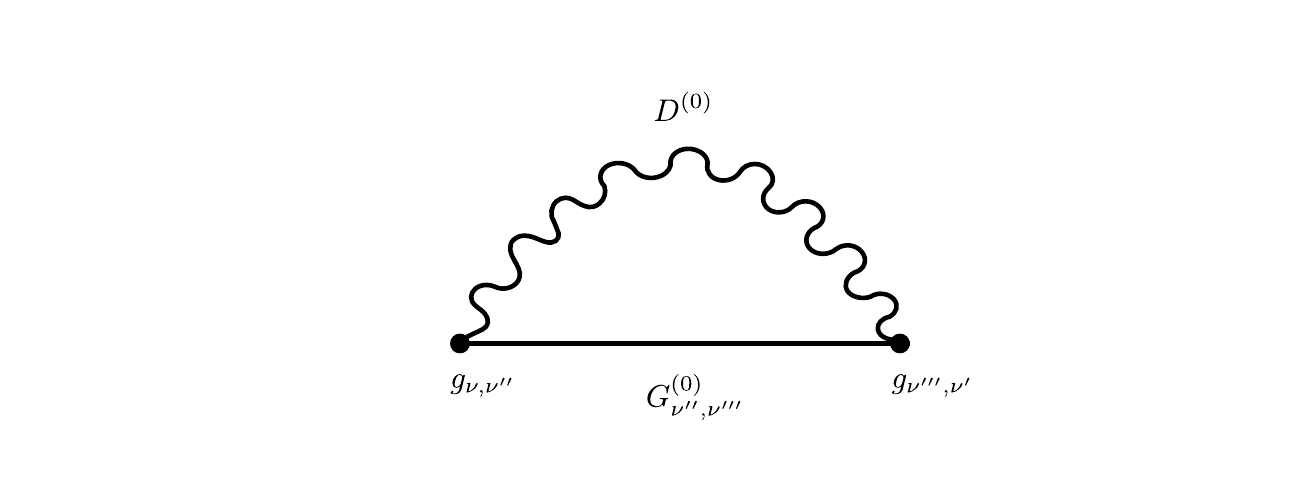}
  \caption{\label{fig:feyn_eph} Lowest-order Feynman diagram of the electron self-energy from the electron-phonon coupling.}
\end{figure}

The electron self-energy thus simplifies to
\begin{equation}
    \Sigma_{\mathrm{el-ph}}(\mathbf{k}, \mathrm{i}\omega_m)_{\nu, \nu'} =- \frac{1}{N \beta} \sum_{\mathbf{k}',m'} \left[ g \frac{(\mathrm{i}\omega_{m'} +\mu)\mathbbold{1} + \mathbf{k}' \cdot \Gamma}{(\mathrm{i}\omega_{m'} +\mu)^2-|\mathbf{k}'|^2} g \frac{1}{\left[(\omega_{0}^2 + (\omega_m-\omega_{m'})^2)\right]}\right]_{\nu, \nu'},
\end{equation}
where, for simplicity, we have neglected the matrix indices of $g_{\nu,\nu'}$ and the element $\nu, \nu'$ is taken after matrix multiplication.
The term linear in $k'$ vanishes upon integration, such that, performing the summation over the Matsubara frequencies, we find
\begin{equation}
\begin{split}
    \frac{1}{\beta}\sum_{m'} \frac{(\mathrm{i}\omega_{m'} +\mu)}{(\mathrm{i}\omega_{m'} +\mu)^2-|\mathbf{k}'|^2} \frac{1}{\left[(\omega_{0}^2 + (\omega_m-\omega_{m'})^2)\right]} &=-\frac{1}{2}\frac{ n_{\mathrm{F}}(-|\mathbf{k}'|-\mu)}{(\mathrm{i}\omega_m +|\mathbf{k}'| + \mu)^2-\omega_0^2} - \frac{1}{2}\frac{ n_{\mathrm{F}}(|\mathbf{k}'|-\mu)}{(\mathrm{i}\omega_m -|\mathbf{k}'| + \mu)^2-\omega_0^2}\\ & ~~~~~+\frac{1}{2} \frac{(\mathrm{i}\omega_m + \mu +\omega_0) n_{\mathrm{B}}(\omega_0)}{\omega_0(\mathrm{i}\omega_m -|\mathbf{k}'| + \mu + \omega_0)(\mathrm{i}\omega_m +|\mathbf{k}'| + \mu + \omega_0)}\\ 
    &~~~~~+\frac{1}{2} \frac{(\mathrm{i}\omega_m + \mu -\omega_0) (1+n_{\mathrm{B}}(\omega_0))}{\omega_0(\mathrm{i}\omega_m -|\mathbf{k}'| + \mu - \omega_0)(\mathrm{i}\omega_m +|\mathbf{k}'| + \mu - \omega_0)}  \\
    &= F_{\mathrm{F}}(\mathbf{k}',\mathrm{i}\omega_{m}) + F_{\mathrm{B}}(\mathbf{k}',\mathrm{i}\omega_{m}).
\end{split}
\end{equation}
To continue, we have separated the first two fermionic and the last two bosonic terms, and $n_{\mathrm{F}}(x)$, $n_{\mathrm{B}}(x)$ are the fermion and boson distribution functions, respectively. We thus find the self-energy
\begin{equation}
    \Sigma_{\mathrm{el-ph}}(\mathbf{k}, \mathrm{i}\omega_m)_{\nu, \nu'} = - \frac{1}{N} \sum_{\mathbf{k}'} (F_{\mathrm{F}}(\mathbf{k}',\mathrm{i}\omega_{m}) + F_{\mathrm{B}}(\mathbf{k}',\mathrm{i}\omega_{m})) (g^2)_{\nu,\nu'},
\end{equation}
where $(g^2)_{\nu,\nu'} = \sum_{\nu''} g_{\nu,\nu''} g_{\nu'',\nu'}$. In the following, we perform the remaining integration in $\mathbf{k}'$:
\begin{equation}
\begin{split}
    - \frac{1}{N} \sum_{\mathbf{k}'} F_{\mathrm{F}}(\mathbf{k}',\mathrm{i}\omega_{m}) &=  \frac{1}{4 \pi^2} \int_0^{\infty} dk' k'^2\left[\frac{ n_{\mathrm{F}}(-k'-\mu)}{(\mathrm{i}\omega_m +k' + \mu)^2-\omega_0^2} + \frac{ n_{\mathrm{F}}(k'-\mu)}{(\mathrm{i}\omega_m -k' + \mu)^2-\omega_0^2}\right]\\
    &=   \frac{1}{4 \pi^2}\left[\int_{-\infty}^{-\mu} d\hat{k} \frac{(\hat{k}+\mu)^2 n_{\mathrm{F}}(\hat{k})}{(\mathrm{i}\omega_m -\hat{k})^2-\omega_0^2} + \int_{-\mu}^{\infty} d\hat{k} \frac{(\hat{k}+\mu)^2 n_{\mathrm{F}}(\hat{k})}{(\mathrm{i}\omega_m -\hat{k})^2-\omega_0^2}\right]\\
    &=  \frac{1}{4 \pi^2} \int_{-\infty}^{\infty} d\hat{k} \frac{(\hat{k}+\mu)^2 n_{\mathrm{F}}(\hat{k})}{(\mathrm{i}\omega_m -\hat{k})^2-\omega_0^2},
\end{split}
\end{equation}
where, in the second line, we substituted $\hat{k} = -k'- \mu$ in the first and $\hat{k} = k'- \mu$ in the second term. Performing the analytic continuation $\hat{k} \rightarrow z$ yields
\begin{equation}
     \frac{1}{4 \pi^2} \int_{-\infty}^{\infty} f(z) dz = \frac{1}{4 \pi^2} \int_{-\infty}^{\infty} dz \frac{(z+\mu)^2 n_{\mathrm{F}}(z)}{(z-\mathrm{i}\omega_m +\omega_0) (z-\mathrm{i}\omega_m -\omega_0)},
\end{equation}
where $f(z)$ has two poles $z_{1,2}=\mathrm{i}\omega_m \pm \omega_0$ in the upper half of the complex plane. Choosing an integration contour $\mathcal{C}$ of radius $R$ in the upper half surrounding the poles allows to evaluate the integral with the residue theorem. As the arc in the complex plane vanishes in the limit $R \rightarrow \infty$, the fermionic term yields
\begin{equation}
   - \frac{1}{N} \sum_{\mathbf{k}'} F_{\mathrm{F}}(\mathbf{k}',\mathrm{i}\omega_{m}) =- \frac{\mathrm{i}}{4 \pi \omega_0}  \left[(\mathrm{i}\omega_m + \mu - \omega_0)^2 n_{\mathrm{F}}(\mathrm{i}\omega_m-\omega_0) - (\mathrm{i}\omega_m + \mu + \omega_0)^2 n_{\mathrm{F}}(\mathrm{i}\omega_m+\omega_0)\right].
\end{equation}
The bosonic integral 
\begin{equation}
\begin{split}
    - \frac{1}{N} \sum_{\mathbf{k}'} F_{\mathrm{B}}(\mathbf{k}',\mathrm{i}\omega_{m}) &=  -\frac{1}{\omega_0} \frac{1}{4 \pi^2} \int_0^{\infty} dk' k'^2\left[\frac{(\mathrm{i}\omega_m + \mu +\omega_0) n_{\mathrm{B}}(\omega_0)}{(\mathrm{i}\omega_m + \mu + \omega_0)^2 - k'^2}
    + \frac{(\mathrm{i}\omega_m + \mu -\omega_0) (1+n_{\mathrm{B}}(\omega_0))}{(\mathrm{i}\omega_m + \mu - \omega_0)^2 - k'^2} \right]
\end{split}
\end{equation}
is formally divergent due to large and intermediate momentum states. We therefore introduce a regularization parameter $\lambda$, so that singularities appear solely in this quantity. Correspondingly, we limit the momentum integral to the region $|\mathbf{k}'| < \lambda$, in which the description with the effective Dirac Hamiltonian remains valid

\begin{equation}
\begin{split}
    - \frac{1}{N} \sum_{\mathbf{k}'} F_{\mathrm{B}}(\mathbf{k}',\mathrm{i}\omega_{m}) &=  - \frac{1}{4 \pi^2 \omega_0} \int_0^{\lambda} dk' k'^2\left[\frac{(\mathrm{i}\omega_m + \mu +\omega_0) n_{\mathrm{B}}(\omega_0)}{(\mathrm{i}\omega_m + \mu + \omega_0)^2 - k'^2}
    + \frac{(\mathrm{i}\omega_m + \mu -\omega_0) (1+n_{\mathrm{B}}(\omega_0))}{(\mathrm{i}\omega_m + \mu - \omega_0)^2 - k'^2} \right] \\
    &= -\frac{\mathrm{1}}{4 \pi^2 \omega_0}  \left[n_{\mathrm{B}}(\omega_0) (\mathrm{i}\omega_m + \mu + \omega_0)^2 \operatorname {artanh}\left(\frac{\lambda}{\mathrm{i}\omega_m + \mu + \omega_0}\right)-n_{\mathrm{B}}(\omega_0) (\mathrm{i}\omega_m + \mu + \omega_0) \lambda\right.\\
     &\quad\left.+(1+n_{\mathrm{B}}(\omega_0)) (\mathrm{i}\omega_m + \mu - \omega_0)^2 \operatorname {artanh}\left(\frac{\lambda}{\mathrm{i}\omega_m + \mu - \omega_0}\right)-(1+n_{\mathrm{B}}(\omega_0)) (\mathrm{i}\omega_m + \mu - \omega_0) \lambda
    \right].
\end{split}
\label{eq:Boson_term}
\end{equation}

The combination of both terms in the self-energy yields

\begin{equation}
\begin{split}
    \Sigma_{\mathrm{el-ph}}(\mathbf{k}, \mathrm{i}\omega_m)_{\nu, \nu'} &= - \frac{1}{N} \sum_{\mathbf{k}'} (F_{\mathrm{F}}(\mathbf{k}',\mathrm{i}\omega_{m}) + F_{\mathrm{B}}(\mathbf{k}',\mathrm{i}\omega_{m})) (g^2)_{\nu,\nu'}\\
    &= \Phi(\mathrm{i}\omega_{m}) (g^2)_{\nu,\nu'},
\end{split}
\label{eq:self_energy_phi}
\end{equation}
where we combined the integrals in the function $\Phi$. Finally, we can analyze the matrix structure of the self-energy. Since the electron-phonon scattering can in principle scatter between $s$ and $p$ orbitals but is spin preserving, we parametrize $g_{\nu,\nu'}= g_{\rho}\tau_{\rho}\sigma_0$, where $\rho = 0,x,y,z$, $\tau_{\rho}$ are Pauli matrices, and $ \sigma_0$ is the $2\times2$ identity. This yields
\begin{equation}
    (g^2)_{\nu,\nu'} = \Big[\sum_{\rho} g_{\rho}^2\tau_{0}\sigma_0 + 2 \sum_{a}g_{0}g_{a}\tau_{a}\sigma_0 + \mathrm{i}\sum_{a,b}g_{a}g_{b}\epsilon_{abc}\tau_{c}\sigma_0\Big]_{\nu\nu'},
\end{equation}
where $a = x,y,z$. Note that for $g_x \neq 0$, the phonon needs to break inversion symmetry, while $g_y\equiv 0$ for time-reversal-symmetric scattering. Finally, if we assume $g_x \gg g_z$, we obtain
\begin{equation}
    g g \approx (g_0^2 + g_x^2) \tau_0\sigma_0 + 2 g_0 g_x \tau_x\sigma_0,
\end{equation}
which results in a self-energy of the form
\begin{equation}
    \Sigma_{\mathrm{el-ph}}(\mathrm{i}\omega_m) =  \Phi(\mathrm{i}\omega_{m}) \left((g_0^2 + g_x^2) \tau_0\sigma_0 + 2 g_0 g_x \tau_x\sigma_0\right),
\end{equation}
where we removed the momentum argument, since the right hand side does not contain a $\mathbf{k}$-dependence.

At small $\omega_m$, $\Phi(\mathrm{i}\omega_{m}) \approx \Phi(0)$, we can investigate the influence of the cutoff $\lambda$. Being the dominant scale in the problem, we take $\lambda \gg \omega_0$, where $\omega_0 > \mu$. This allows to expand the hyperbolic tangent in Eq. (\ref{eq:Boson_term}),

\begin{equation}
    \operatorname{artanh}\left(\frac{\lambda}{\mu + \omega_0}\right) \approx -\mathrm{i} \frac{\pi}{2}, \qquad \frac{\lambda}{\mu + \omega_0} \gg 1,
\end{equation}

\begin{equation}
    \operatorname{artanh}\left(\frac{\lambda}{\mu - \omega_0}\right) \approx \mathrm{i} \frac{\pi}{2}, \qquad \frac{\lambda}{\mu - \omega_0} \ll -1,
\end{equation}

which yields

\begin{equation}
\begin{split}
    \Phi(0) &= -\frac{1}{N} \sum_{\mathbf{k}'} (F_{\mathrm{F}}(\mathbf{k}',0) + F_{\mathrm{B}}(\mathbf{k}',0)) \\
    &=\frac{\mathrm{i}}{4 \pi \omega_0} \left[(\mu + \omega_0)^2 n_{\mathrm{F}}(\omega_0)-(\mu - \omega_0)^2 (1-n_{\mathrm{F}}(\omega_0)) \right] \\
    &\quad+\frac{\mathrm{i}}{8 \pi \omega_0} \left(4 \mu \omega_0 n_{\mathrm{B}}(\omega_0) - (\mu - \omega_0)^2\right) + \frac{\lambda}{4 \pi^2 \omega_0} \left(2 \mu n_{\mathrm{B}}(\omega_0) + \mu - \omega_0\right).
\end{split}
\end{equation}

The last term constitutes the real part of the self-energy, which can be parametrically small by a suitable combination of $\omega_0,\mu, n_{\mathrm{B}}(\omega_0)$. This means the imaginary part of the self-energy becomes dominant by a suitable choice of the phonon frequency $\omega_0$ and the selective driving via $n_{\mathrm{B}}(\omega_0)$, while $\mu \ll \omega_0$. In such a setting, $ \Phi(0) \approx \mathrm{i} \phi(0)$, yielding

\begin{equation}
    \Sigma_{\mathrm{el-ph}}(0) = \mathrm{i}\beta \tau_0\sigma_0 + \mathrm{i}\delta \tau_x\sigma_0,
    \label{eq:final_self-energy}
\end{equation}
with $\beta = \phi(0)(g_0^2 + g_x^2) $ and $\delta =2\phi(0) g_0 g_x$. The appearance of the term $\beta \tau_0\sigma_0$ shifts the point gap on the imaginary axis, without altering the properties discussed in the main body of the paper. 

The result derived above is only valid for small electron-phonon couplings $g/\omega_0$. For stronger couplings, the self-energy has to be calculated to higher orders in $g$ or even self-consistently. Such a rather tedious calculation requires to take care of the branch cuts in \eqref{eq:self_energy_phi}. Importantly, the overall structure of the self-energy is preserved to the next order in the coupling constant.

%

\section{Discussion of two band models with non-trivial $w_{\mathrm{3D}}$}

Two-band models often serve as minimal models describing the physics of topological phases in case of Hermitian Hamiltonians.
Here we would like to discuss two specific examples of non-Hermitian two-band models with a point gap in their spectrum: non-Hermitian extensions of a Chern insulator and a non-Hermitian Weyl semimetal proposed in Ref.~\onlinecite{Lee:2019-2,Terrier:2020} and point out that these models are not easily regularizable to host a single exceptional point.

\subsection{Non-Hermitian Chern insulator}

First, consider a non-Hermitian variant of Chern insulator in 2D rotated in the complex plane, described by

\begin{equation}
\label{eq:non-Hermitian-Chern}
    H = \frac{(1 + \cos k_x + \cos k_y) \sigma_z + \sin k_x \sigma_x + \sin k_y \sigma_y}{\sqrt{3 + 2 \cos k_x + \cos(k_x - k_y) + 2 \cos k_y + \cos(k_x + k_y) }}  + \left(\frac{1}{2} e^{i k_z} - 1\right)\sigma_0.
\end{equation}

This model possesses a point gap in the bulk spectrum, see Supplementary Fig.~\ref{fig:non-Hermitian-Chern}a. Even though the 3D winding number for this model is non-trivial $w_{3\mathrm{D}} = 1$, open boundaries in $z$-direction yield the non-Hermitian skin effect, with occupations concentrating on one side of the system as shown in Supplementary Fig.~\ref{fig:non-Hermitian-Chern}b. For open boundaries in $x$ and $y$-direction, the point gap regions are filled by edge states. The corresponding weak invariants $w_{1\mathrm{D},i = \{x,y,z\}}$ support this observation, being zero for $x,y$ and one in $z$-direction. The appearance of the non-Hermitian skin effect, signalled by a non-trivial $w_{1\mathrm{D}}$ invariant, hence causes a collapse of the point gap. This disqualifies the model in Eq.(\ref{eq:non-Hermitian-Chern}) for hosting stable surface states like for instance exceptional points.

\begin{figure}[h]
  \centering
  \includegraphics[width=0.7\linewidth]{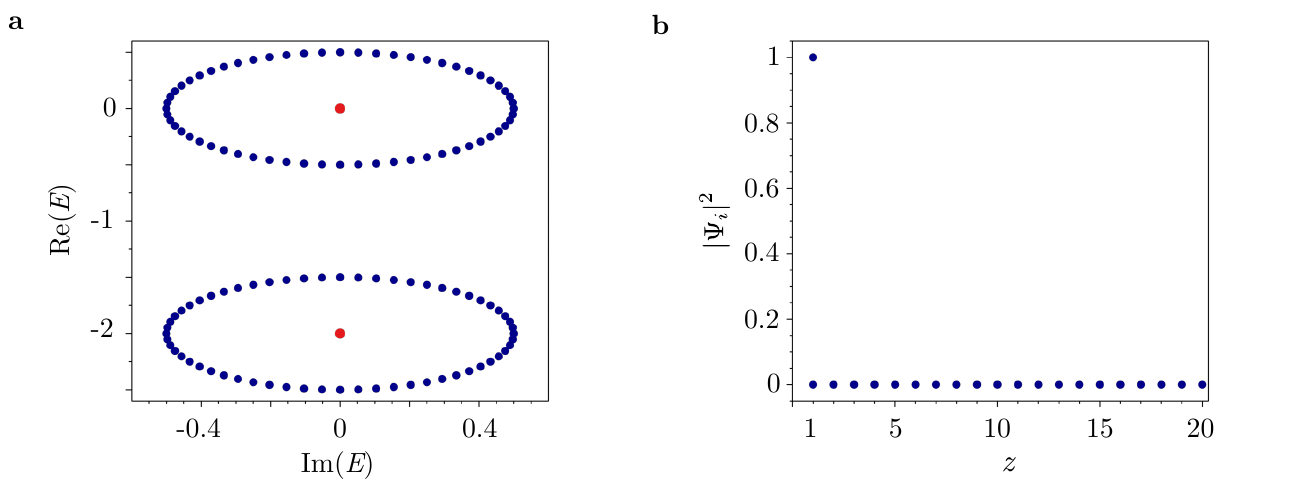}
\caption{\label{fig:non-Hermitian-Chern} \textbf{Energy spectra and localization properties of the non-Hermitian Chern insulator.} \textbf{a}~Bulk spectrum of the non-Hermitian Chern insulator (blue) and open boundary condition spectrum in $z$ direction (red) for a system of 20 layers. \textbf{b}~The occupation of lattice sites for OBC in $z$ direction for a system of 20 layers.}
\end{figure}

\subsection{Non-Hermitian Weyl semimetal}

The two-band model describing a dissipative Weyl semimetal with a point gap in the bulk spectrum has recently been proposed in Ref.~\onlinecite{Lee:2019-2,Terrier:2020}. The non-Hermitian Hamiltonian of the model reads as follows 
\begin{equation}
\label{eq:nhw_model}
H_{\mathrm{NHWS}} = \imi\left(h + \sum_{i\in\{x,y,z\}}\cos k_i\right)\sigma_0 + \sum_{i\in\{x,y,z\}} \sin k_i \sigma_i.
\end{equation}

The determinant of this Hamiltonian is real at every $\mathbf{k}$-point, therefore there are no 1D winding numbers, whereas the 3D winding number is non-trivial. The weak Chern numbers cannot be properly defined due to the occurrence of Weyl points at TRIMs. 

\begin{figure}[h]
  \centering
  \includegraphics[width=0.7\linewidth]{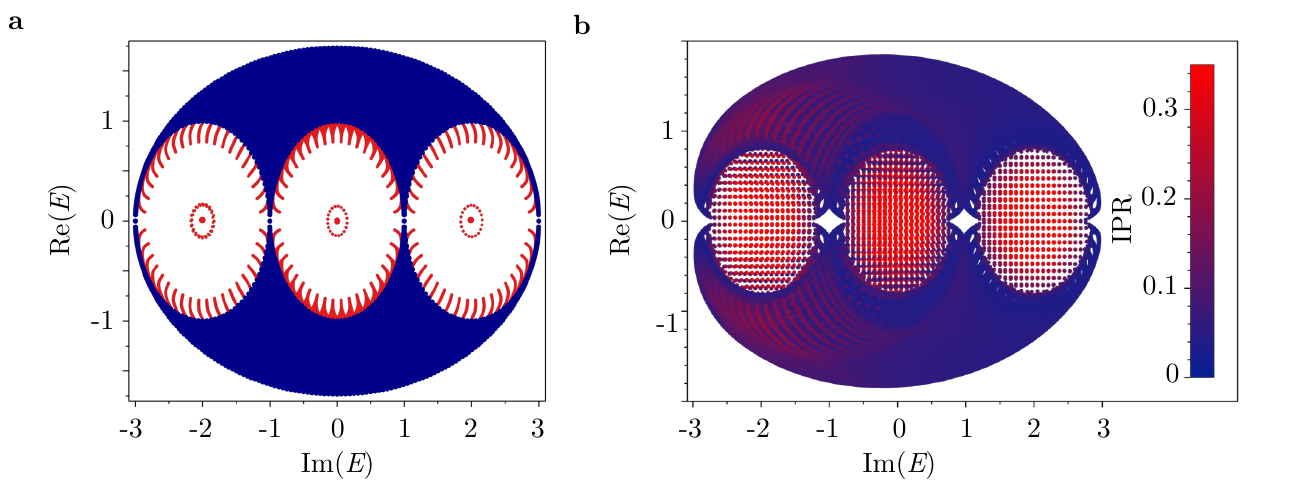}
  \caption{\label{fig:nh_weyl} \textbf{Energy spectrum in the complex plane for the non-Hermitian Weyl semimetal.} \textbf{a}~Bulk energy spectrum (blue) and spectrum for open boundary conditions in $x$ direction (red) for the model in Eq. (\ref{eq:nhw_model}) for $h = 0$ and 20 layers. The analytical solution shows eigenstates at the center of the point gaps. \textbf{b}~Open boundary condition spectrum for the regularization term introduced in Eq. (\ref{eq:reg_term}) ($B=0.1$, 20 layers), resulting in a single surface band covering the point gaps, indicated in red-to-blue tones according to their inverse participation ratio (IPR).}
\end{figure}

The spectrum for open boundaries in the $x$ direction features three analytically calculated eigenvalues at $\epsilon_{1,2,3} = i(h\pm2), ih$ appearing in the point gap. Each of these states is 2N-fold degenerate, where N is the number of layers. The ellipses of states being present around the analytical solutions in Supplementary Fig.~\ref{fig:nh_weyl}a are due to the numerical instability of the model. This resembles the infernal point discussed in the main section of the paper, as the surface state is fine-tuned rather than generic. Similar to the four-band case this surface state can be regularized along certain momentum directions by adding a small perturbation

\begin{equation}
    \label{eq:reg_term}
    H = H_{\mathrm{NHWS}} + B \left[\left(1-\cos k_x\right) \sigma_y + \left(1-\cos k_y\right) \sigma_z + \left(1-\cos k_z\right) \sigma_x\right],
\end{equation}
which yields a single sheet of eigenstates covering the OBC point gap as illustrated in Supplementary Fig.~\ref{fig:nh_weyl}b. The above perturbation moves Weyl cones in the Brillouin zone, ensuring that they do not fall on the same momentum direction when considering open boundary conditions. In a two-band model, this is not possible to ensure with constant regularization terms for surface orientations along cardinal axes. Correspondingly, without a $k$-dependent perturbation, there is always a direction in momentum space along which the OBC system has an infernal point, with this direction being the connection of two Weyl points.

\subsection{Effective theory at a spin-orbit coupled band edge}

In the main text, we focus on how an ETI can arise near the critical Dirac point between a topological and a trivial insulator upon the addition of a suitable non-Hermitian coupling. The rational for considering this case is that the density of states is very low near such a Dirac point, and hence it is the most favorable situation for observing the physical consequences of the ETI point gap topology. However, point gaps with the same topology can also arise in other generic situations. One that corresponds to a reasonably small density of states is the band edge of a generic spin-orbit coupled band.  To demonstrate this, let us analyse the behaviour of the ETI around ${\mathbf k} = (\pi,\pi,\pi)$ by projecting the Hamiltonian onto the states corresponding to the upper-bands. Diagonalizing the original Hamiltonian, Eq.~(1) from the main text,  $H(\pi,\pi,\pi)$ with $B=0$ for simplicity, we obtain two eigenvalues $E_{\pm}~=\pm \sqrt{(3+M-\delta)(3+M+\delta)}\,$ each doubly degenerate with corresponding eigenvectors given by the rows of the matrices $V$ for the upper-band and $W$ for the lower-energy band as follows

\begin{equation}
    V =\frac{1}{\mathcal{N}_v}\begin{pmatrix}
           0&v&0&1 \\
            v&0&1&0\\
     \end{pmatrix}\,,\qquad
    W =\frac{1}{\mathcal{N}_w} 
    \begin{pmatrix}
            0&w&0&1 \\
            w&0&1&0\\
     \end{pmatrix}\,,
\end{equation}
where 
\begin{equation}
\begin{split}
v=&\, -\frac{i(-3-M + \sqrt{(3 + M - \delta)(3 + M + \delta)})}{\delta},\\
w=&\, -\frac{i(3+M + \sqrt{(3 + M - \delta)(3 + M + \delta)})}{\delta}
\end{split}
\end{equation}
are normalization factors
\begin{equation}
\begin{split}
\mathcal{N}_v=&\,\sqrt{1+\frac{(-3-M+\sqrt{(3+M-\delta)(3+M+\delta)})^2}{\delta^2}}
\\
\mathcal{N}_w=&\,\sqrt{1+\frac{(3+M+\sqrt{(3+M-\delta)(3+M+\delta)})^2}{\delta^2}}
.
\end{split}
\end{equation}

We expand $H({\mathbf{k}})$ around ${\mathbf{k}} = (\pi,\pi,\pi)$, shifting momentum $\mathbf{k} \rightarrow \mathbf{p}+(\pi,\pi,\pi)$ and keeping only the terms of the first order in $\mathbf{p}$. The linearized Hamiltonian reads as follows

\begin{equation}
    h({\mathbf{p}}) = (-3-M)\tau_z\sigma_0+i\delta\tau_x\sigma_0-\lambda \sum_{j={x,y,z}}p_j \tau_x\sigma_j.
\end{equation}

We apply degenerate perturbation theory to first order in $\mathbf p$ and to first  order in $\delta$.
The corrections to the energy in the upper-band are going to be the eigenvalues of the following matrix 
\begin{equation}\label{eq: hamiltonian_upper_band}
    h_{\mathrm{eff}}(\mathbf{p}) = V h(\mathbf{p}) V^\dagger - \frac{Vh(\mathbf{p})W^\dagger Wh(\mathbf{p})V^\dagger}{E_+-E_-} = q_1\sigma_0-iq_2 \bs{\xi}\cdot \bs{\sigma}+\mathcal{O}(\mathbf{p}^2)\,,
\end{equation}
where $\bs{\xi} = (-p_x,p_y,p_z)$, $q_1 = \sqrt{(3+M-\delta)(3+M+\delta)}$, and $q_2 = \delta \lambda/(3+M)$.
The resulting linearized Hamiltonian in Eq.(\ref{eq: hamiltonian_upper_band}) is of the same form as the Hamiltonian in Eq. (\ref{eq:nhw_model}) for small ${\mathbf k}$, which has been studied earlier in Ref.~\onlinecite{Terrier:2020}. Thus, an ETI Hamiltonian arises generically at the band edge of a spin-orbit coupled band when subject to a suitable non-Hermitian perturbation. The fact that $q_2$ is linear in $\lambda$ emphasizes the importance of spin-orbit coupling for this result.

\end{bibunit}

\bibliography{bibliography}
\bibliographystyle{naturemag}